\documentclass[twoside]{article}
\usepackage{epsfig}
\usepackage{url}

\newlength{\figurewidth}
\newtheorem{theorem}{Theorem}[section]
\newtheorem{corollary}[theorem]{Corollary}
\newtheorem{definition}[theorem]{Definition}
\newenvironment{proof}{{\bf Proof:}}{\hfill{\rule{2mm}{2mm}}}

\renewcommand{\sectionmark}[1]{\markboth{\textsc{J. Leskovec et al.}}
{\textsc{Graphs Over Time}}}
\renewcommand{\subsectionmark}[1]{\markboth{\textsc{J. Leskovec et al.}}
{\textsc{Graphs Over Time}}}

\newcommand{\UnnumberedFootnote}[1]{{\def\thefootnote{}\footnote{#1}
\addtocounter{footnote}{-1}}}

\newcommand{\hide}[1]{}

\newcommand{\GPL}{{Densification Power Law}}
\newcommand{\GPLplot}{{DPL plot}}
\newcommand{\CCDC}{{Difficulty Constant}}
\newcommand{\CCDF}{{Difficulty Function}}
\newcommand{\CGA}{Community Guided Attachment}

\newcommand{\arxiv}{{arXiv}}

\newcommand{\dbar}{\bar{d}}

\def\fwdprob{p}
\def\bckprob{p_b}
\def\backratio{r}

\newcommand{\dmax}{d_{max}}
\newcommand{\ddslope}{\gamma}

\newcommand{\T}[1]{
\Theta\left({#1}\right) }

\newcommand{\pfrac}[2]{
\left(\frac{{#1}}{{#2}}\right) }

\newcommand{\captionfonts}{\small}
\makeatletter
\long\def\@makecaption#1#2{%
  \vskip\abovecaptionskip
  \sbox\@tempboxa{{\captionfonts #1: #2}}%
  \ifdim \wd\@tempboxa >\hsize
    {\captionfonts #1: #2\par}
  \else
    \hbox to\hsize{\hfil\box\@tempboxa\hfil}%
  \fi
  \vskip\belowcaptionskip}
\makeatother

\begin{document}

\title{
  Graph Evolution:\\Densification and Shrinking Diameters}

\author{
  Jure Leskovec\\
    \textit{\normalsize School of Computer Science, Carnegie Mellon University, Pittsburgh, PA}\\
  Jon Kleinberg\\
    \textit{\normalsize Department of Computer Science, Cornell University, Ithaca, NY} \\
  Christos Faloutsos\\
    \textit{\normalsize School of Computer Science, Carnegie Mellon University, Pittsburgh, PA}
}

\maketitle

\begin{abstract}
How do real graphs evolve over time? What are ``normal'' growth
patterns in social, technological, and information networks? Many
studies have discovered patterns in {\em static graphs}, identifying
properties in a single snapshot of a large network, or in a very
small number of snapshots; these include heavy tails for in- and
out-degree distributions, communities, small-world phenomena, and
others. However, given the lack of information about network
evolution over long periods, it has been hard to convert these
findings into statements about trends over time.

Here we study a wide range of real graphs, and we observe some
surprising phenomena. First, most of these graphs densify over time,
with the number of edges growing super-linearly in the number of
nodes. Second, the average distance between nodes often {\em
shrinks} over time, in contrast to the conventional wisdom that such
distance parameters should increase slowly as a function of the
number of nodes (like $O(\log n)$ or $O(\log(\log n)$).

Existing graph generation models do not exhibit these types of
behavior, even at a qualitative level. We provide a new graph
generator, based on a ``forest fire'' spreading process, that has a
simple, intuitive justification, requires very few parameters (like
the ``flammability'' of nodes), and produces graphs exhibiting the
full range of properties observed both in prior work and in the
present study.

We also notice that the ``forest fire'' model exhibits a sharp
transition between sparse graphs and graphs that are densifying.
Graphs with decreasing distance between the nodes are generated
around this transition point.

Last, we analyze the connection between the temporal evolution of
the degree distribution and densification of a graph. We find that
the two are fundamentally related. We also observe that real
networks exhibit this type of relation between densification and the
degree distribution.
\end{abstract}
\UnnumberedFootnote{Work also appears in ACM Transactions on 
Knowledge Discovery from Data, 1(1), 2007}

\section{Introduction}
\label{sec:intro}

In recent years, there has been considerable interest in graph
structures arising in technological, sociological, and scientific
settings: computer networks (routers or autonomous systems connected
together); networks of users exchanging e-mail or instant messages;
citation networks and hyperlink networks; social networks
(who-trusts-whom, who-talks-to-whom, and so forth); and countless
more~\cite{newman03structure}. The study of such networks has
proceeded along two related tracks: the measurement of large network
datasets, and the development of random graph models that
approximate the observed properties.

Many of the properties of interest in these studies are based on two
fundamental parameters: the nodes' {\em degrees} (i.e., the number
of edges incident to each node), and the {\em distances} between
pairs of nodes (as measured by shortest-path length). The
node-to-node distances are often studied in terms of the {\em
diameter} --- the maximum distance --- and a set of closely related
but more robust quantities including the average distance among
pairs and the {\em effective diameter} (the 90th percentile
distance, a smoothed form of which we use for our studies).

Almost all large real-world networks evolve over time by the addition
and deletion of nodes and edges. Most of the recent models of network
evolution capture the growth process in a way that incorporates two
pieces of ``conventional wisdom:''

\begin{itemize}
\item[(A)] {\em Constant average degree assumption}:
   The average node degree in the network remains constant over time.
   (Or equivalently, the number of edges grows linearly in the number of   nodes.)
\item[(B)] {\em Slowly growing diameter assumption}:
   The diameter is a slowly growing
   function of the network size, as in ``small world'' graphs
   \cite{barabasi99diameter,tomkins00bowtie,milgram67smallworld,watts98collective}.
\end{itemize}

For example, the intensively-studied {\em preferential attachment
model} \cite{barabasi99emergence,newman03structure} posits a network
in which each new node, when it arrives, attaches to the existing
network by a constant number of out-links, according to a
``rich-get-richer'' rule. Recent work has given tight asymptotic
bounds on the diameter of preferential attachment
networks~\cite{bollobas04diameter,chung02average}; depending on the
precise model, these bounds grow
logarithmically~\cite{redner05growth} or even slower than
logarithmically in the number of nodes.

How are assumptions (A) and (B) reflected in data on network growth?
Empirical studies of large networks to date have mainly focused on
{\em static} graphs, identifying properties of a single snapshot or
a very small number of snapshots of a large network. For example,
despite the intense interest in the Web's link structure, the recent
work of Ntoulas et al.~\cite{ntoulas04evolution} noted the lack of
prior empirical research on the evolution of the Web. Thus, while
one can assert based on these studies that, qualitatively, real
networks have relatively small average node degrees and diameters,
it has not been clear how to convert these into statements about
trends over time.

\paragraph*{\bf The present work: Densification laws and shrinking
diameters}
Here we study a range of different networks, from several domains, and
we focus specifically on the way in which fundamental network
properties vary with time. We find, based on the growth patterns of
these networks, that principles (A) and (B) need to be reassessed.
Specifically, we show the following for a broad range of networks
across diverse domains.

\begin{itemize}
\item[($A'$)] {\em Empirical observation: Densification power laws}:
   The networks are becoming {\em denser} over time, with
   the average degree increasing (and hence with the number of edges
   growing super-linearly in the number of nodes).
   Moreover, the densification follows a power-law pattern.
\item[($B'$)] {\em Empirical observation: Shrinking diameters:}
   The effective diameter is, in many cases, actually
   {\em decreasing} as the network grows.
\end{itemize}
We view the second of these findings as particularly surprising:
Rather than shedding light on the long-running debate over exactly
how slowly the graph diameter {\em grows} as a function of the
number of nodes, it suggests a need to revisit standard models so as
to produce graphs in which the effective diameter is capable of
actually {\em shrinking} over time. We also note that, while
densification and decreasing diameters are properties that are
intuitively consistent with one another (and are both borne out in
the datasets we study), they are qualitatively distinct in the sense
that it is possible to construct examples of graphs evolving over
time that exhibit one of these properties but not the other.

We can further sharpen the quantitative aspects of these findings. In
particular, the densification of these graphs, as suggested by ($A'$),
is not arbitrary; we find that as the graphs evolve over time, they
follow a version of the relation
\begin{eqnarray}
    e(t) \propto n(t)^a
    \label{eq:gpl}
\end{eqnarray}

where $e(t)$ and $n(t)$ denote the number of edges and nodes of the
graph at time $t$, and $a$ is an exponent that generally lies
strictly between $1$ and $2$. We refer to such a relation as a {\em
densification power law}, or {\em growth power law}. (Exponent $a =
1$ corresponds to constant average degree over time, while $a = 2$
corresponds to an extremely dense graph where each node has, on
average, edges to a constant fraction of all nodes.)

What underlying process causes a graph to systematically densify,
with a fixed exponent as in Equation (\ref{eq:gpl}), and to
experience a decrease in effective diameter even as its size
increases? This question motivates the second main contribution of
this work: we present two families of probabilistic generative
models for graphs that capture aspects of these properties. The
first model, which we refer to as {\em \CGA\ }
(CGA)~\cite{leskovec05densification}, argues that graph
densification can have a simple underlying basis; it is based on a
decomposition of the nodes into a nested set of communities, such
that the difficulty of forming links between communities increases
with the community size. For this model, we obtain rigorous results
showing that a natural tunable parameter in the model can lead to a
densification power law with any desired exponent $a$. The second
model, which is more sophisticated, exhibits both densification and
a decreasing effective diameter as it grows. This model, which we
refer to as the {\em Forest Fire Model}, is based on having new
nodes attach to the network by ``burning'' through existing edges in
epidemic fashion. The mathematical analysis of this model appears to
lead to novel questions about random graphs that are quite complex,
but through simulation we find that for a range of parameter values
the model exhibits realistic behavior in densification, distances,
and degree distributions. It is thus the first model, to our
knowledge, that exhibits this full set of desired properties.

Accurate properties of network growth, together with models supporting
them, have implications in several contexts.

\begin{itemize}
\item {\em Graph generation:} Our findings form means for assessing
the quality of graph generators. Synthetic graphs are important for
`what if' scenarios, for extrapolations, and for simulations, when
real graphs are impossible to collect (like, {\em e.g.}, a very large
friendship graph between people).

\item {\em Graph sampling:} Datasets consisting
of huge real-world graphs are increasingly available, with sizes
ranging from the millions to billions of nodes. There are many known
algorithms to compute interesting measures (shortest paths,
centrality, betweenness, etc.), but most of these algorithms become
impractical for large graphs. Thus sampling is essential --- but
sampling from a graph is a non-trivial problem since the
goal is to maintain structural properties of the network.
Densification laws can help discard bad sampling methods, by
providing means to reject sampled subgraphs.

A recent work of Leskovec and
Faloutsos~\cite{leskovec06sampling} proposed two views on
sampling from large graphs. For {\em Back-in-time} sampling the goal
is to find a sequence of sampled subgraphs that matches the
evolution of the original graph and thus obey the temporal growth
patterns. On the other hand, {\em Scale-down} sampling aims for a
sample that matches the properties of the original large graph. The
authors considered various sampling strategies, propose evaluation
techniques, and use the temporal graph patterns presented in this
paper to evaluate the quality of the sampled subgraphs.

\item {\em Extrapolations:} For several real graphs, we have a
lot of snapshots of their past. What can we say about their future?
Our results help form a basis for validating scenarios for graph
evolution.

\item {\em Abnormality detection and computer network
management:} In many network settings, ``normal'' behavior will
produce subgraphs that obey densification laws (with a predictable
exponent) and other properties of network growth. If we detect
activity producing structures that deviate significantly from this,
we can flag it as an abnormality; this can potentially help with the
detection of {\em e.g.} fraud, spam, or distributed denial of service
(DDoS) attacks.
\end{itemize}

The rest of the paper is organized as follows:
Section~\ref{sec:related} surveys the related work.
Section~\ref{sec:observations} gives our empirical findings on
real-world networks across diverse domains.
Section~\ref{sec:explanations} describes our proposed models and
gives results obtained both through analysis and simulation.
Section~\ref{sec:ddOverTm} gives the formal and experimental
analysis of the relationship between the degree distribution and the
graph densification over time. We conclude and discuss the
implications of our findings in Section~\ref{sec:conclusion}.

\section{Related Work}
\label{sec:related}

Research over the past few years has identified classes of
properties that many real-world networks obey. One of the main areas
of focus has been on {\em degree power laws}, showing that the set
of node degrees has a heavy-tailed distribution. Such degree
distributions have been identified in phone call
graphs~\cite{Abello98phone}, the Internet
\cite{faloutsos99powerlaw}, the
Web~\cite{barabasi99emergence,huberman99growth,kumar99trawling},
click-stream data~\cite{zhiqiang01dgx} and for a who-trusts-whom
social network~\cite{deepay04rmat}. Other properties include the
``small-world phenomenon,'' popularly known as ``six degrees of
separation'', which states that real graphs have surprisingly small
(average or effective) diameter
(see~\cite{barabasi99diameter,bollobas04diameter,tomkins00bowtie,chung02average,kleinberg01nips,milgram67smallworld,watts98collective,watts02identity}).

In parallel with empirical studies of large networks, there has been
considerable work on probabilistic models for graph generation. The
discovery of degree power laws led to the development of random
graph models that exhibited such degree distributions, including the
family of models based on {\em preferential
attachment}~\cite{barabasi99emergence,abello02handbook,cooper03model},
{\em copying model}~\cite{Kleinberg99cocoon,kumar00stochastic}, and
the related {\em growing network with redirection}
model~\cite{redner01organization}, which produces graphs with
constant diameter and {\em logarithmically} increasing average
degree~\cite{redner05growth}.

Similar to our Forest Fire Model is the work of
Vazquez~\cite{Vazquez01Disordered,Vazquez2003} where ideas
based on random walks and recursive search for generating networks
were introduced. In a random walk model the walk starts at a random
node, follows links, and for each visited node with some probability
an edge is created between the visited node and the new node. It can
be shown that such model will generate graphs with power-law degree
distribution with exponent $\gamma \ge 2$. On the other hand, in the recursive search
model first a new node is added to the network, and the edge to a random
node is created. If an edge is created to a node in the network,
then with some probability $q$ an edge is also created to each of
its 1-hop neighbors. This rule is recursively applied until no
edges are created. The recursive search model is similar to our Forest
Fire Model in a sense that it exploits current network structure to
create new edges. However, there is an important difference that
in recursive search model the average degree scales at most {\em logarithmically} (and
not as a power-law) with the number of nodes in the network. Our
simulation experiments also indicated that the diameter of networks
generated by the recursive search does not decrease over time, but
it either slowly increases or remains constant.

We point the reader to ~\cite{mitzenmacher04brief,newman03structure,li05scalefree} for
overviews of this area. Recent work of Chakrabarti and
Faloutsos~\cite{deepay06survey} gives a survey of the properties
of real world graphs and the underlying generative models for
graphs.

It is important to note the fundamental contrast between one of our
main findings here --- that the average number of out-links per node
is growing polynomially in the network size --- and body of work on
degree power laws. This earlier work developed models that almost
exclusively used the assumption of node degrees that were bounded by
constants (or at most logarithmic functions) as the network grew;
our findings and associated model challenge this assumption, by
showing that networks across a number of domains are becoming {\em
denser} over time.

Dorogovtsev and Mendes in a series of
works~\cite{dorogovtsev01growth,dorogovtsev02lang,dorogovtsev03evolution}
analyzed possible scenarios of nonlinearly growing networks while
maintaining scale-free structure. Among considered hypothetical
scenarios were also those where the number of links grows
polynomially with the number of edges, i.e. \GPL, while maintaining
power-law degree distribution. The authors call this an {\em
accelerated growth} and propose preferential attachment type models
where densification is forced by introducing an additional ``node
attractiveness'' factor that is not only degree-dependent but also
time-dependent. The motivation for their work comes from the fact
that authors~\cite{broder00web,faloutsos99powerlaw} reported the
increase of the average degree over time on the Web and the
Internet. Our work here differs in that it presents
measurements on many time evolving networks to support our findings,
and proposes generative models where densification is an emerging
property of the model. Besides densification we also address the
shrinking diameters and consider models for generating them.

The bulk of prior work on the empirical study of network datasets
has focused on {\em static} graphs, identifying patterns in a single
snapshot, or a small number of network snapshots (see also the
discussion of this point by Ntoulas et
al.~\cite{ntoulas04evolution}). Two exceptions are the very
recent work of Katz~\cite{katz05scale}, who independently
discovered densification power laws for citation networks, and the
work of Redner~\cite{redner04citation}, who studied the
evolution of the citation graph of {\em Physical Review} over the
past century. Katz's work builds on his earlier research on
power-law relationships between the size and the recognition of
professional communities~\cite{katz99self}; his work on
densification is focused specifically on citations, and he does not
propose a generative network model to account for the densification
phenomenon, as we do here. Redner's work focuses on a range of
citation patterns over time which are different from the network
properties we study here.

Our \CGA\ (CGA) model, which produces densifying graphs, is an
example of a hierarchical graph generation model, in which the
linkage probability between nodes decreases as a function of their
relative distance in the
hierarchy~\cite{deepay04rmat,kleinberg01nips,watts02identity,leskovec05densification,leskovec05kronecker,Abello04Hier}.
Again, there is a distinction between the aims of this past work and
our model here; where these earlier network models were seeking to
capture properties of individual snapshots of a graph, we seek to
explain a time evolution process in which one of the fundamental
parameters, the average node degree, is varying as the process
unfolds. Our Forest Fire Model follows the overall framework of
earlier graph models in which nodes arrive one at a time and link
into the existing structure; like the copying model discussed above,
for example, a new node creates links by consulting the links of
existing nodes.  However, the recursive process by which nodes in
the Forest Fire Model creates these links is quite different,
leading to the new properties discussed in the previous section.

This paper extends the work of~\cite{leskovec05densification} by
introducing new measurements and analysis of the evolution of degree
distribution and its connection to \GPL. In a follow-up
paper~\cite{leskovec05kronecker} we introduced {\em Kronecker
Graphs}, a mathematically tractable model of graph growth and
evolution. Kronecker graphs are based on tensor products of graph
adjacency matrices, and exhibit a full set of static and temporal
graph properties. The emphasis of the work on Kronecker graphs is on
the ability to prove theorems about their properties, and not to
provide simple generative models, like our Forest Fire model, that
intrinsically lead to densification and shrinking diameter.

\section{Observations}
\label{sec:observations}

We study the temporal evolution of several networks, by observing
snapshots of these networks taken at regularly spaced points in
time. We use datasets from seven different sources; for each, we
have information about the time when each node was added to the
network over a period of several years --- this enables the
construction of a snapshot at any desired point in time. For each of
datasets, we find a version of the densification power law from
Equation~(\ref{eq:gpl}), $e(t) \propto n(t)^a$; the exponent $a$
differs across datasets, but remains remarkably stable over time. We
also find that the effective diameter decreases in all the datasets
considered.

The datasets consist of two citation graphs for different areas in
the physics literature, a citation graph for U.S. patents, a graph
of the Internet, five bipartite affiliation graphs of authors
with papers they authored, a recommendation network, and an email communication network. Overall, then, we consider 12 different
datasets from 7 different sources.

\begin{figure}[!tp]
\begin{center}
  \begin{tabular}{cc}
    \epsfig{file=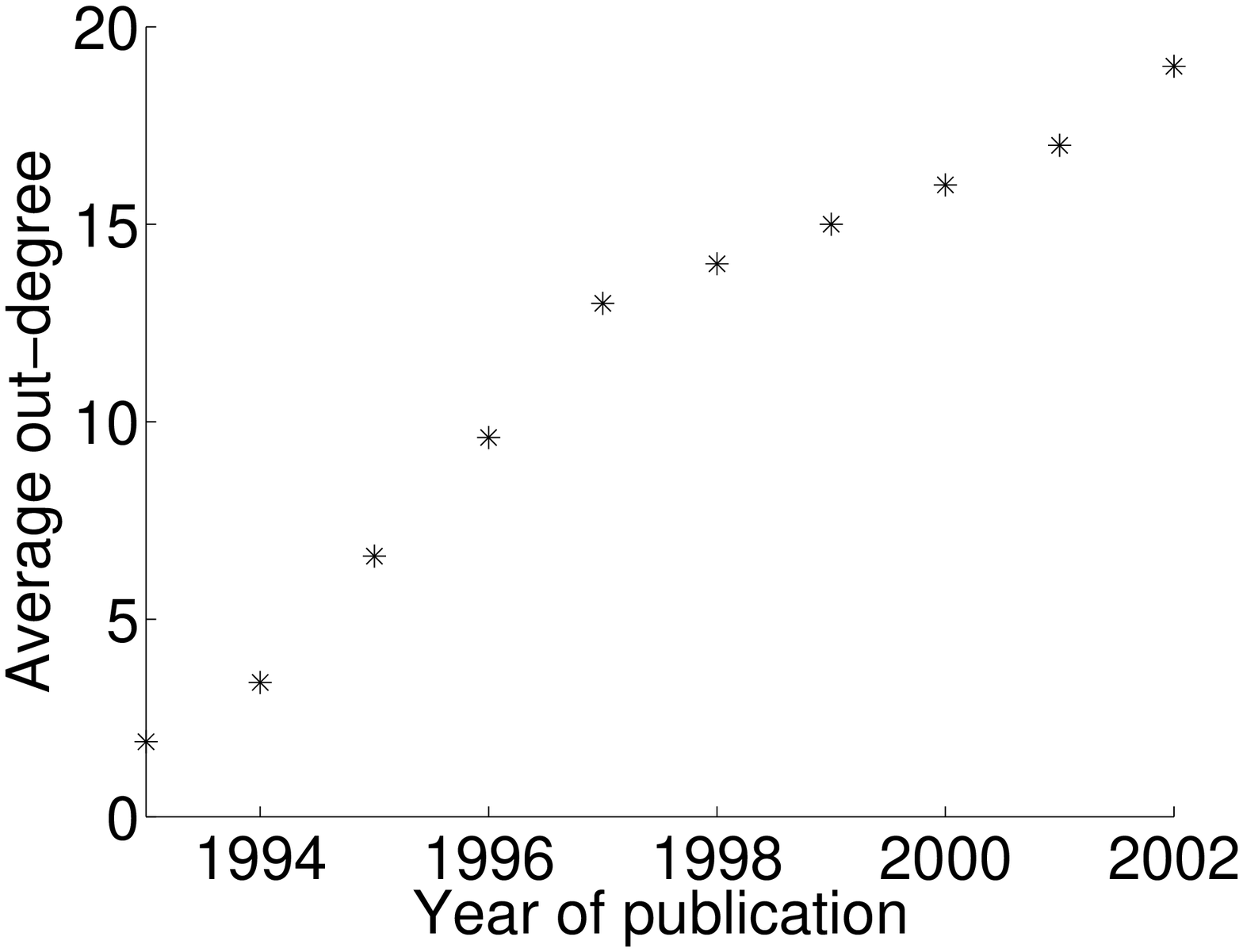, width=2.2in} &
    \epsfig{file=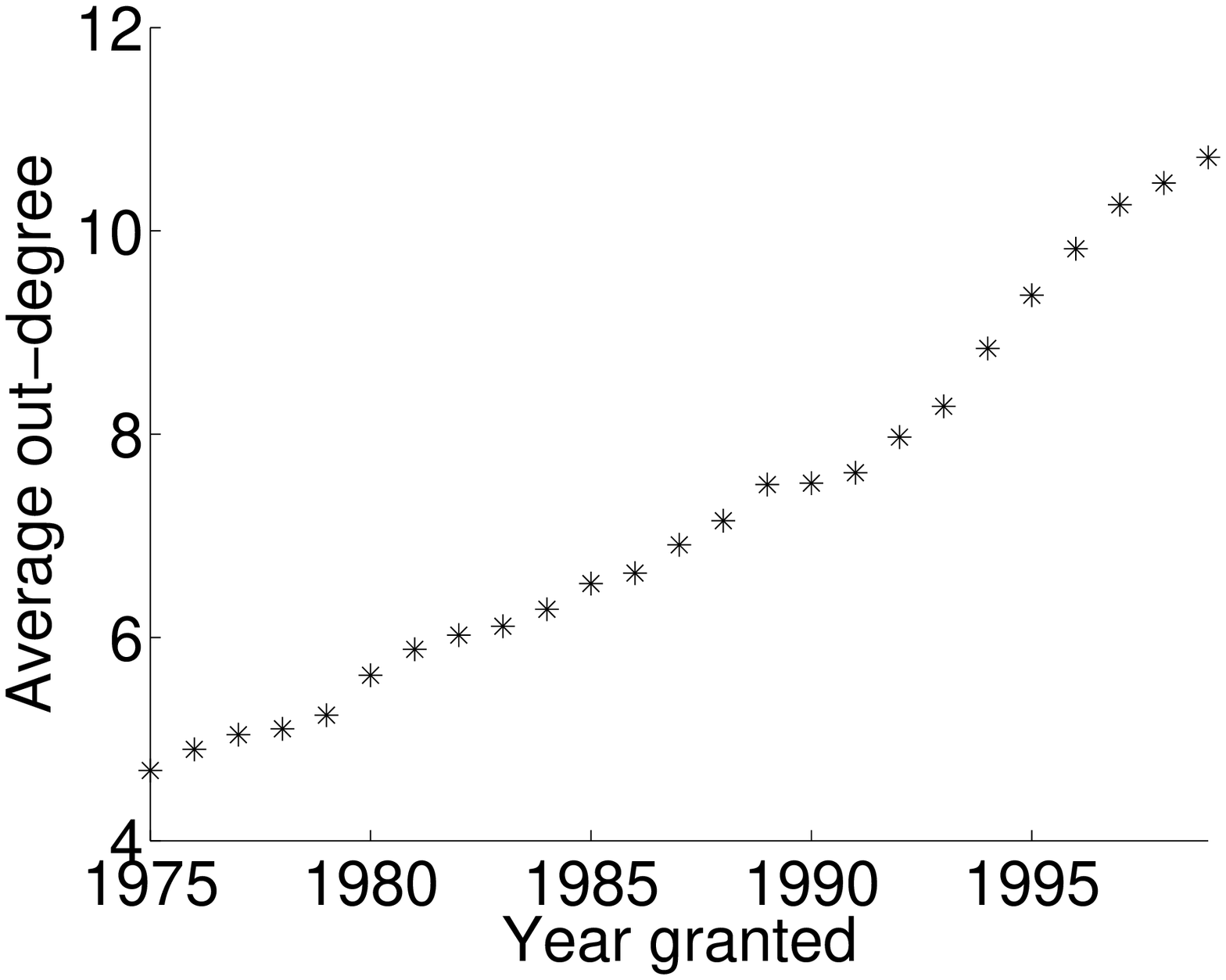, width=2.2in} \\
    (a) arXiv & (b) Patents  \\
    \epsfig{file=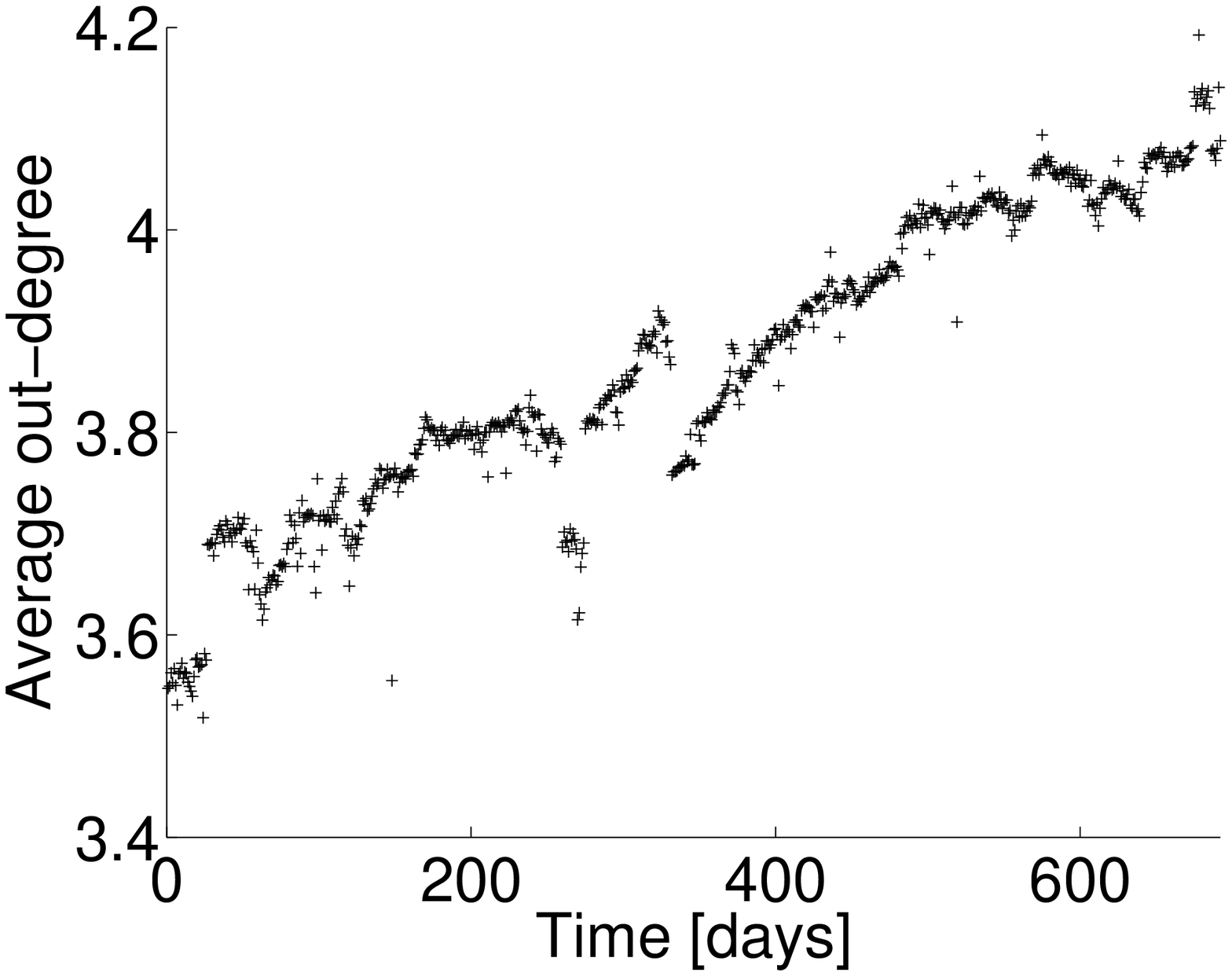, width=2.2in} &
    \epsfig{file=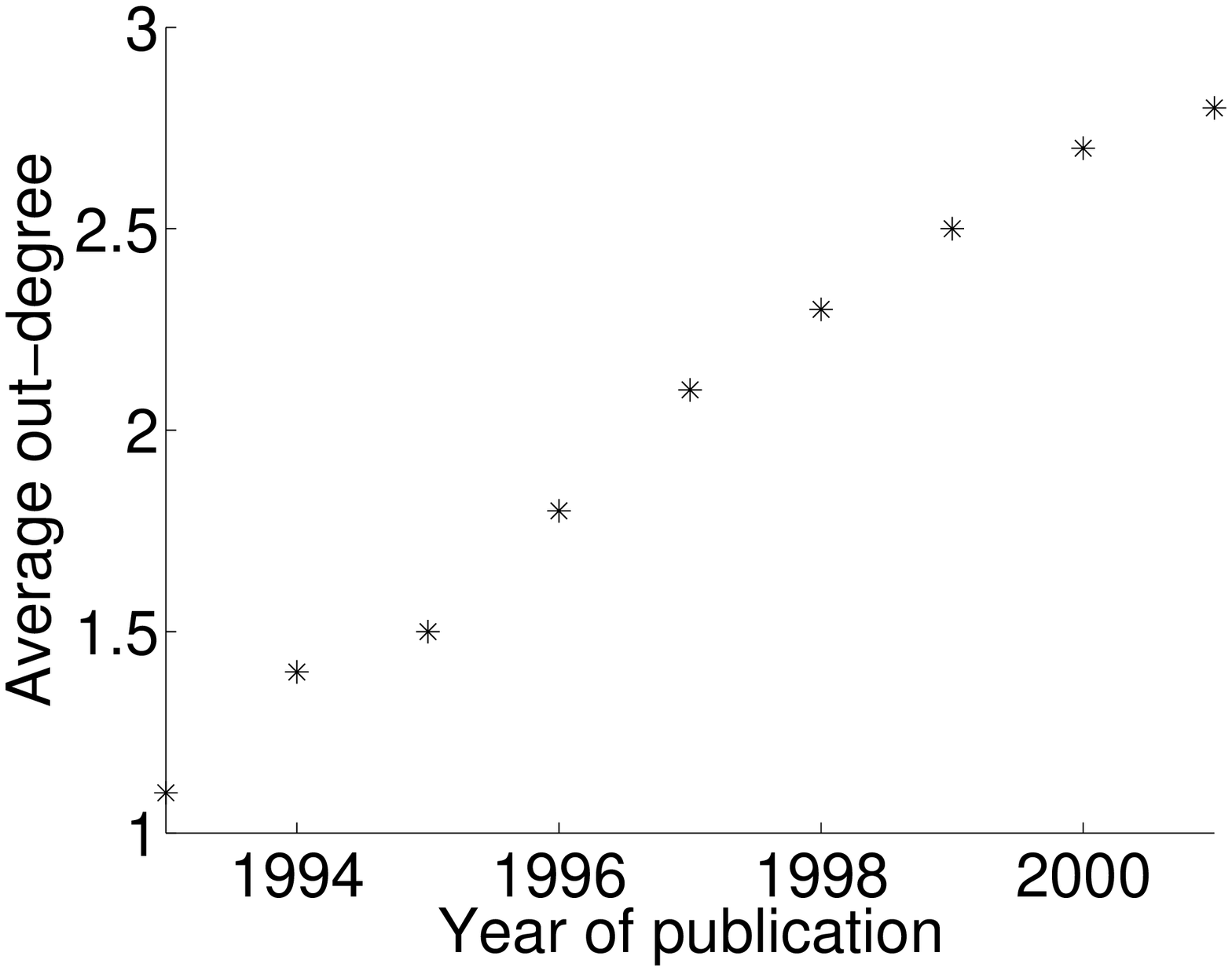, width=2.2in} \\
    (c) Autonomous Systems & (d) Affiliation network \\
  \end{tabular}
  \caption{The average node out-degree over time. Notice that it
  increases, in all 4 datasets. That is, all graphs are {\em
  densifying}.}
  \label{fig:AvgDegGrowth}
\end{center}
\end{figure}

\subsection{Densification Laws}

Here we describe the datasets we used, and our findings related to
densification. For each graph dataset, we have, or can generate,
several
time snapshots, for which we study the number of nodes $n(t)$ and the
number of edges $e(t)$ at each timestamp $t$. We denote by $n$ and $e$
the final number of nodes and edges. We use the term {\em \GPL\ plot}
(or just \GPLplot) to refer to the log-log plot of number of edges
$e(t)$ versus number of nodes $n(t)$.

\subsubsection{ArXiv citation graph}
\label{sec:arxiv}

 We first investigate a citation graph provided as part of
the 2003 KDD Cup~\cite{gehrke03kddcup}. The HEP--TH (high energy
physics theory) citation graph from the e-print arXiv covers all the
citations within a dataset of $n$=29,555 papers with $e$= 352,807
edges. If a paper $i$ cites paper $j$, the graph contains a directed
edge from $i$ to $j$. If a paper cites, or is cited by, a paper
outside the dataset, the graph does not contain any information
about this. We refer to this dataset as {\em arXiv}.

This data covers papers in the period from January 1993 to April 2003
(124 months). It begins within a few months of the inception of the
arXiv, and thus represents essentially the complete history of its
HEP--TH section. For each month $m$ ($1 \le m \le 124$) we
create a citation graph using all papers published up to month $m$.
For each of these graphs, we plot the number of nodes versus the number
of edges on a logarithmic scale and fit a line.

\begin{figure}[tp]
\begin{center}
  \begin{tabular}{cc}
    \epsfig{file=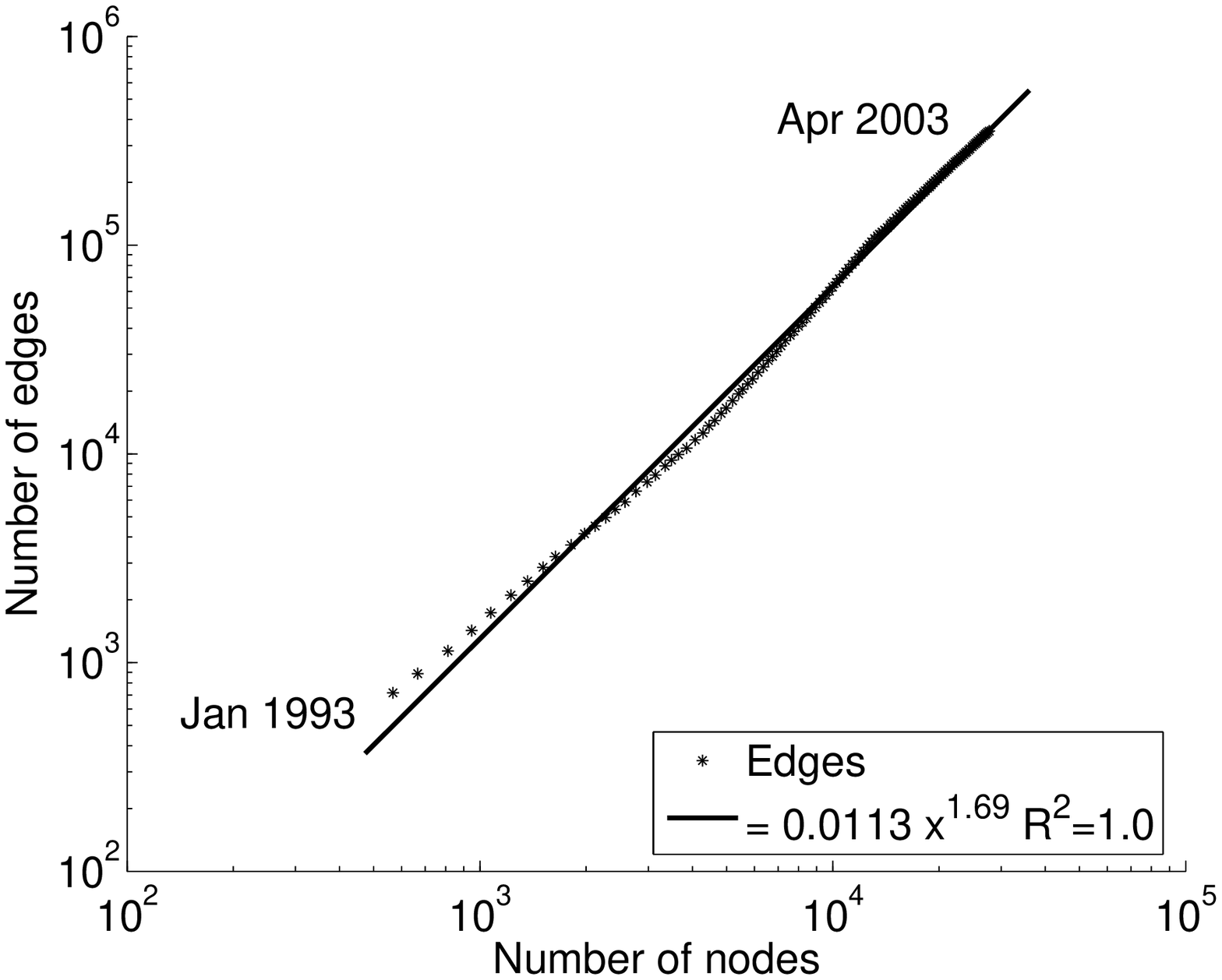, width=2.2in} &
    \epsfig{file=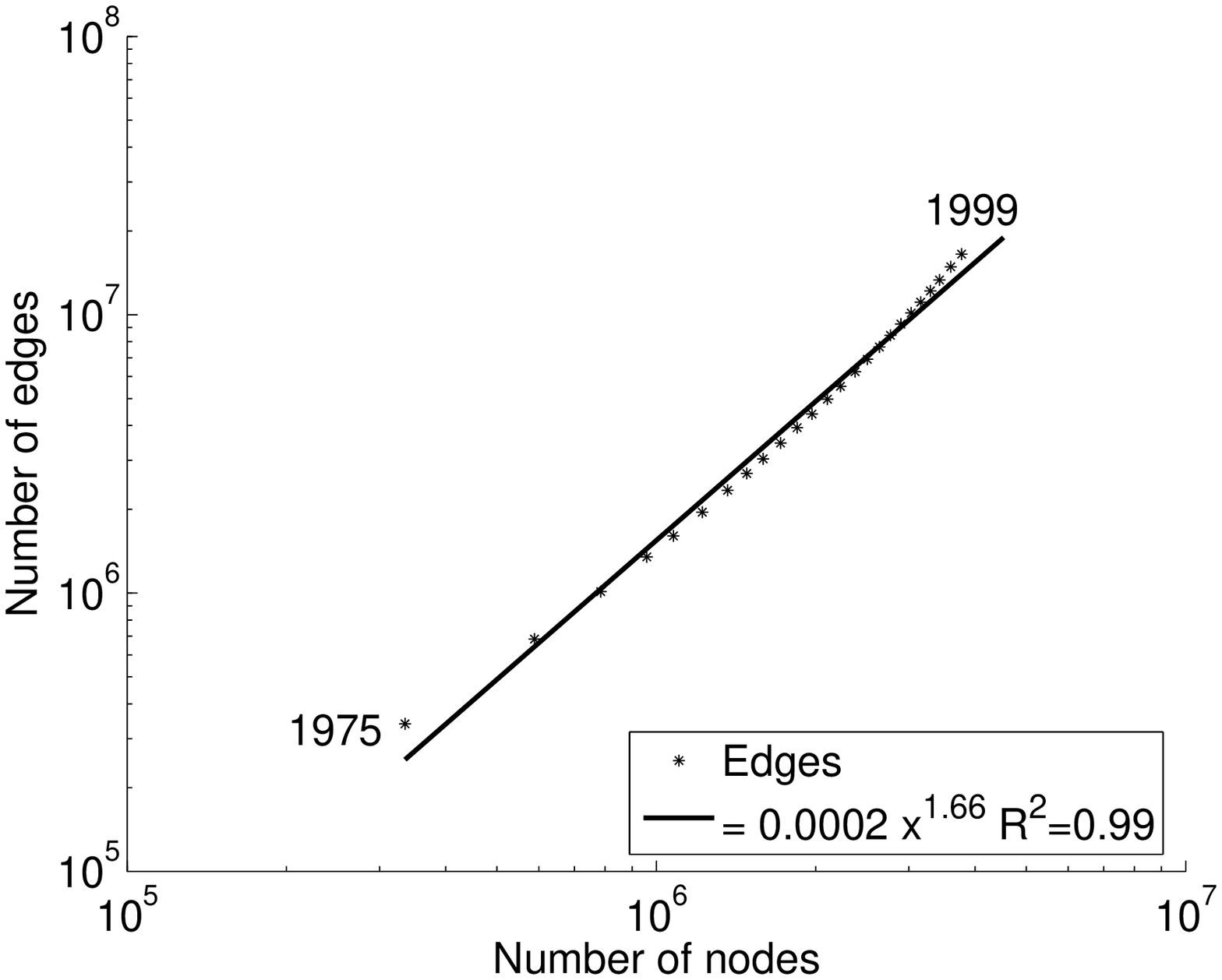, width=2.2in} \\
    (a) arXiv & (b) Patents  \\
    \epsfig{file=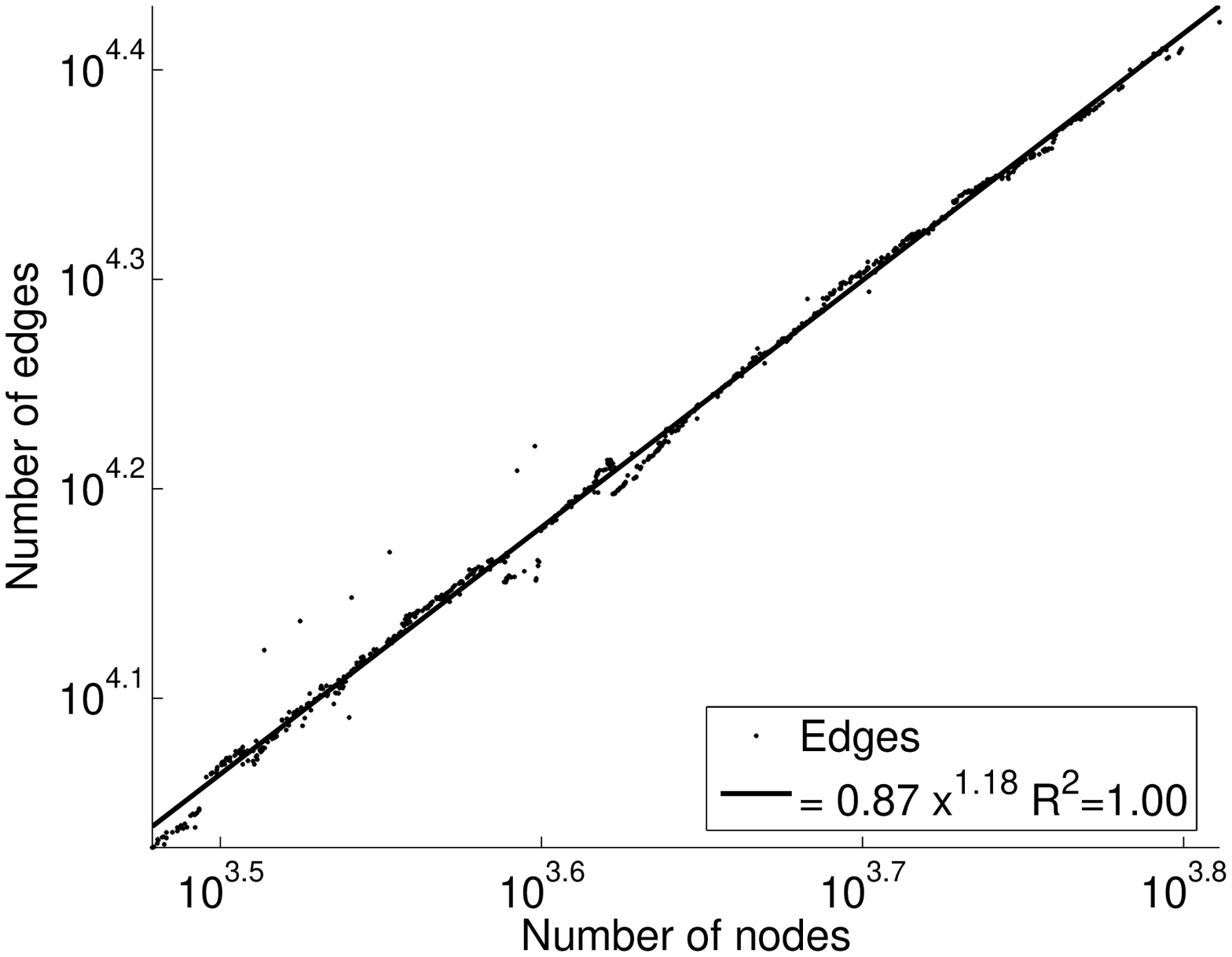, width=2.2in} &
    \epsfig{file=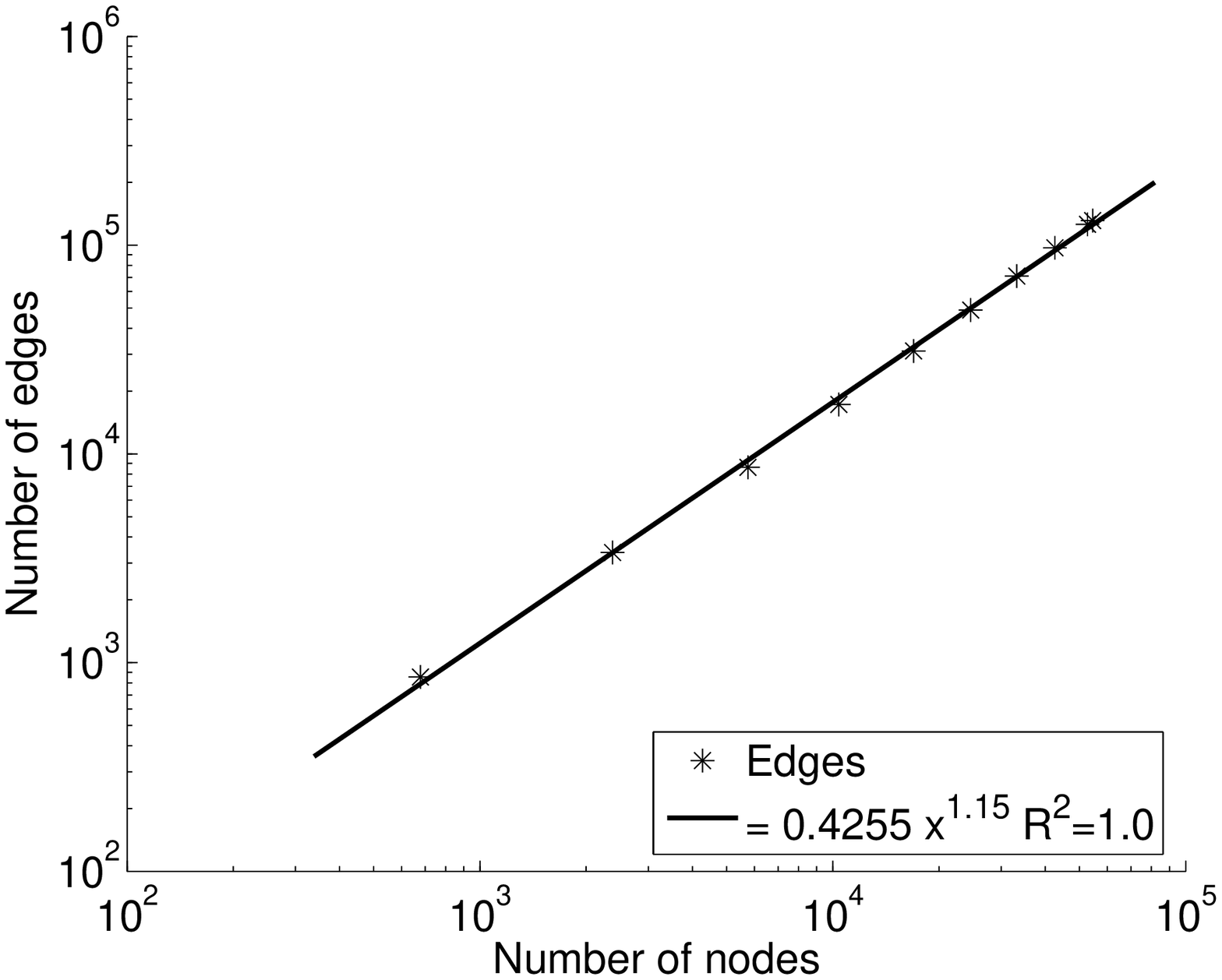 , width=2.2in} \\
    (c) Autonomous Systems & (d) Affiliation network \\
    \epsfig{file=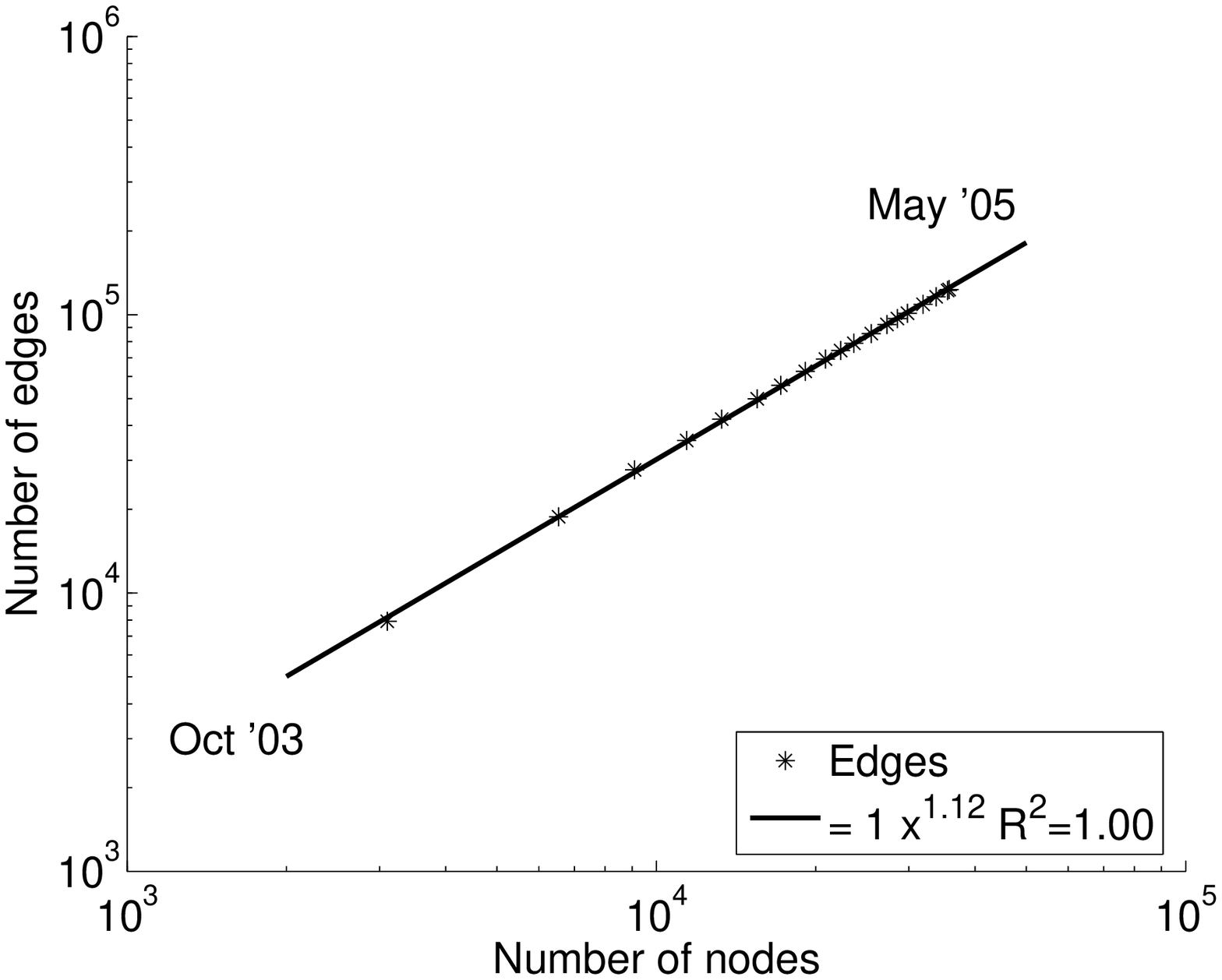, width=2.2in} &
    \epsfig{file=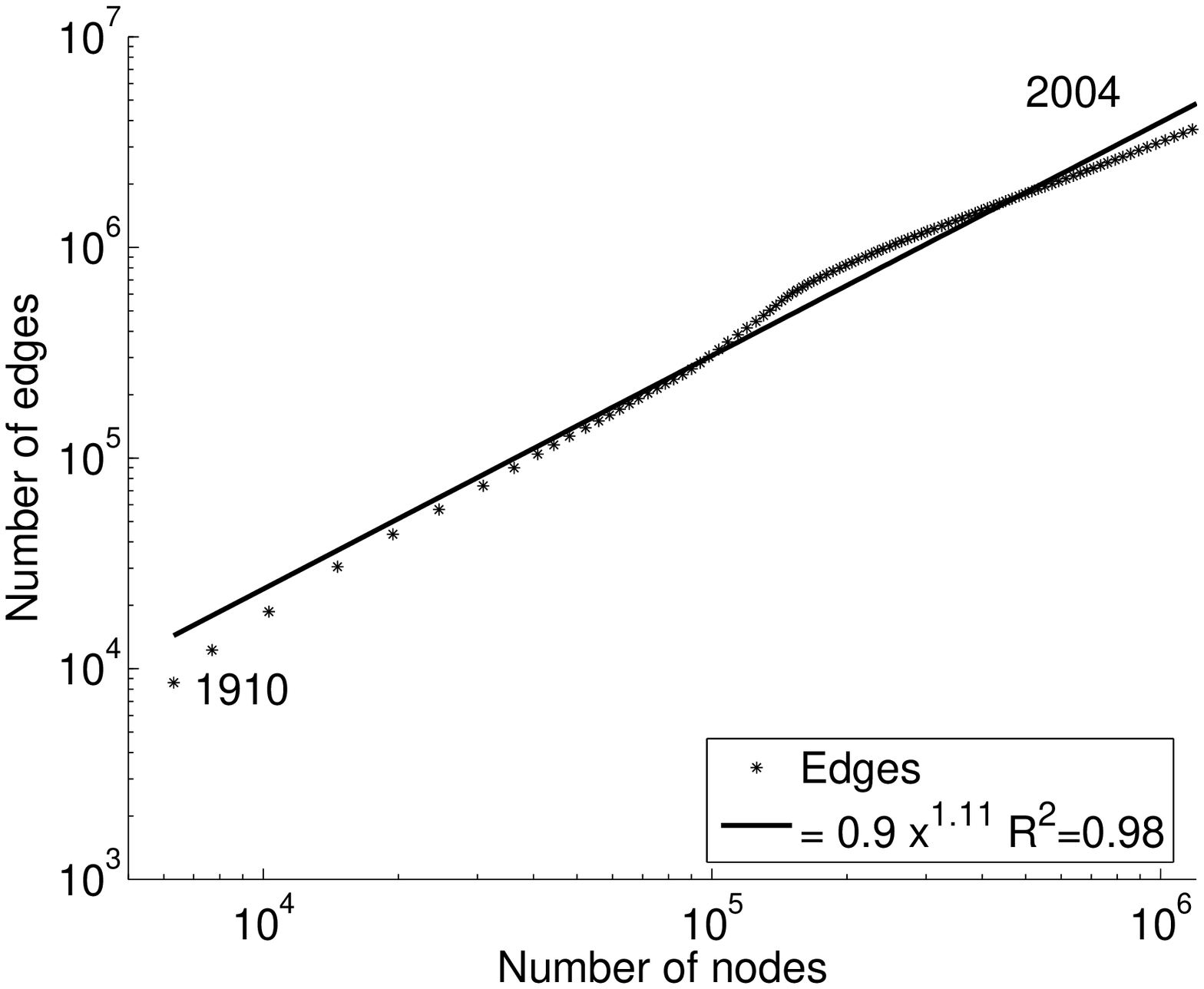, width=2.2in} \\
    (e) Email network & (f) IMDB actors to movies network \\
  \end{tabular}
  \caption{Number of edges $e(t)$ versus number of nodes $n(t)$, in
  log-log scales, for several graphs. All 4 graphs obey the \GPL, with
  a
  consistently good fit. Slopes: $a=$ 1.68, 1.66, 1.18, 1.15,
  1.12, and 1.11 respectively.}
  \label{fig:powerGrowth}
\end{center}
\end{figure}

Figure~\ref{fig:powerGrowth}(a) shows the \GPLplot; the slope is
$a=1.68$ and corresponds to the exponent in the densification law.
Notice that $a$ is significantly higher than 1, indicating a large
deviation from linear growth. As noted earlier, when a graph has $a
> 1$, its average degree increases over time.
Figure~\ref{fig:AvgDegGrowth}(a) exactly plots the average degree
$\dbar$ over time, and it is clear that $\dbar$ increases. This
means that the average length of the bibliographies of papers
increases over time.

There is a subtle point here that we elaborate next: With almost any
network dataset, one does not have data reaching all the way back to
the network's birth (to the extent that this is a well-defined
notion). We refer to this as the problem of the ``{\em missing
past}.'' Due to this, there will be some effect of increasing
out-degree simply because edges will point to nodes prior to the
beginning of the observation period, {\em i.e.} over time less references are pointing to papers outside the dataset. We refer to such nodes as {\em
phantom nodes}, with a similar definition for {\em phantom edges}.
In all our datasets, we find that this effect is relatively minor
once we move away from the beginning of the observation period; on
the other hand, the phenomenon of increasing degree continues
through to the present. For example, in arXiv, nodes over the most
recent years are primarily referencing non-phantom nodes; we observe
a knee in Figure~\ref{fig:AvgDegGrowth}(a) in 1997 that appears to
be attributable in large part to the effect of phantom nodes.
(Later, when we consider a graph of the Internet, we will see a case
where comparable properties hold in the absence of any ``missing
past'' issues.) A similar observation of growing reference lists
over time was also independently made by Krapivsky and
Redner~\cite{redner05growth}.

We also experimented with a second citation graph, taken from the
HEP--PH section of the arXiv, which is about the same size as our
first \arxiv\ dataset. It exhibits the same behavior, with the
densification exponent $a = 1.56$. The plot is omitted for brevity
but we show the summary of results on all 11 datasets we considered
in table~\ref{tab:timeGrowth}.

\subsubsection{Patents citation graph}

Next, we consider a U.S. patent dataset maintained by the National
Bureau of Economic Research~\cite{hall01patents}. The data set spans
37 years (January 1, 1963 to December 30, 1999), and includes all
the utility patents granted during that period, totaling
$n$=3,923,922 patents. The citation graph includes all citations
made by patents granted between 1975 and 1999, totaling
$e$=16,522,438 citations. For the patents dataset there are
1,803,511 nodes for which we have no information about their
citations (we only have the in-links). Because the dataset begins in
1975, it too has a ``missing past'' issue, but again the effect of
this is minor as one moves away from the first few years.

The patents data also contains citations outside the dataset. For
patents outside the dataset the time is unknown. These patents have
zero out-degree and are at some time cited by the patents from within the
dataset. We set the time (grant year) of these out-of-dataset patents to the
year when they were first cited by a patent from the dataset. This is
natural and is equivalent to saying that patents for which grant year
is unknown are in the dataset from the beginning, but when counting,
we count only non-zero degree nodes. So the time when we first count
an unknown patent is when it gets a first link.

We follow the same procedure as with \arxiv. For each year $Y$ from
1975 to 1999, we create a citation network on patents up to year
$Y$, and give the \GPLplot, in Figure~\ref{fig:powerGrowth}(b). As
with the arXiv citation network, we observe a high densification
exponent, in this case $a=1.66$.

Figure~\ref{fig:AvgDegGrowth}(b) illustrates the increasing out-degree
of patents over time. Note that this plot does not incur any of the
complications of a bounded observation period, since the patents in the
dataset include complete citation lists, and here we are simply
plotting
the average size of these as a function of the year.

\subsubsection{Autonomous systems graph}

The graph of routers comprising the Internet can be organized into
sub-graphs called Autonomous Systems (AS). Each AS exchanges traffic
flows with some neighbors (peers). We can construct a communication
network of who-talks-to-whom from the BGP (Border Gateway Protocol)
logs.

We use the {\em Autonomous Systems (AS)} dataset
from~\cite{AsOregon}. The dataset contains 735 daily instances which
span an interval of 785 days from November 8 1997 to January 2 2000.
The graphs range in size from $n$=3,011 nodes and $e$=10,687 edges
to the largest AS graph that has $n$=6,474 nodes and  $e$=26,467
edges.

In contrast to citation networks, where nodes and edges only get added
(not deleted) over time, the AS dataset also exhibits both the addition
and deletion of the nodes and edges over time.

Figure~\ref{fig:powerGrowth}(c) shows the \GPLplot\ for the
Autonomous Systems dataset. We observe a clear trend: Even in the
presence of noise, changing external conditions, and disruptions to
the Internet we observe a strong super-linear growth in the number
of edges over more than 700 AS graphs. We show the increase in the
average node degree over time in Figure~\ref{fig:AvgDegGrowth}(c).
The densification exponent is $a=1.18$, lower than the one for the
citation networks, but still clearly greater than $1$.

\subsubsection{Affiliation graphs}

Using the arXiv data, we also constructed bipartite {\em affiliation
graphs}. There is a node for each paper, a node for each person who
authored at least one arXiv paper, and an edge connecting people to the
papers they authored. Note that the more traditional {\em co-authorship
network} is implicit in the affiliation network: two people are
co-authors if there is at least one paper joined by an edge to each of
them.

We studied affiliation networks derived from the five largest
categories
in the arXiv (ASTRO--PH, HEP--TH, HEP--PH, COND--MAT and GR--QC). We
place a time-stamp on each node: the submission date of each paper, and
for each person, the date of their first submission to the arXiv. The
data for affiliation graphs covers the period from April 1992 to March
2002. The smallest of the graphs (category GR--QC) had 19,309 nodes
(5,855 authors, 13,454 papers) and 26,169 edges. ASTRO--PH is the
largest graph, with 57,381 nodes (19,393 authors, 37,988 papers) and
133,170 edges. It has 6.87 authors per paper; most of the other
categories also have similarly high numbers of authors per paper.

For all these affiliation graphs we observe similar phenomena, and
in particular we have densification exponents between $1.08$ and
$1.15$. We present the complete set of measurements only for
ASTRO--PH, the largest affiliation graph.
Figures~\ref{fig:AvgDegGrowth}(d) and \ref{fig:powerGrowth}(d) show
the increasing average degree over time, and a densification
exponent of $a=1.15$. Table~\ref{tab:timeGrowth} shows the sizes and
{\GPL} exponents for other four affiliation graphs.

\subsubsection{Email network}
\label{sec:email}

We also considered an email network from a large European research
institution. For a period from October 2003 to May 2005 (18 months)
we have anonymized information about all incoming and outgoing email
of the research institution. For each sent or received email message
we know the time, the sender and the recipient of the email. Overall
we have 3,038,531 emails between 287,755 different email addresses.
Note that we have a complete email graph for only 1,258 email
addresses that come from the research institution. Furthermore,
there are 35,756 email addresses that both sent and received email
within the span of our dataset. All other email addresses are
either non-existing, mistyped or spam.

Given a set of email messages we need to create a graph. Since there
can be multiple emails sent between same two addresses (nodes) we
follow the practice of Kossinets and
Watts~\cite{kossinets06email}. Given a set of email messages,
each node corresponds to an email address.  We create an edge
between nodes $i$ and $j$, if they exchanged messages both ways,
{\em i.e.} $i$ sent at least one message to $j$, and $j$ sent at least one
message to $i$.

Similarly to citation networks, we take all email messages up to
particular time $t$ and create a graph using the procedure described
above. So, in the first month we observe 254,080 emails between
38,090 different addresses. Using the
procedure~\cite{kossinets06email} of generating a graph from a set
of emails, we get $n$=6,537 nodes and $e$=18,812 edges. After 18
months, at the end of the dataset, we have $n$=35,756 nodes and
$e$=123,254 edges.

Figure~\ref{fig:powerGrowth}(e) presents the \GPLplot\ for the Email
network. Observe a clear trend: the email network is densifying,
regardless of the fact that it is growing and that new parts of
social network (email address space) are being explored. The
densification exponent is $a=1.12$, lower than the one for the
citation networks but more similar to those from affiliation
networks. Still clearly greater than $1$.

Note that there is one issue with this dataset: we have complete
information about all sent and received emails only for the core of
the network (1258 email addresses from the institution). For the
rest of the addresses, the nodes on the periphery, we only have
their communication (links) with the core of the network.

Regardless of how we look at the email network it always densifies:
If we consider only the core of the network, the densification is
very high. This is expected, since the number of nodes (people at
the research institution) basically remains constant over time and
the edges can only be added, not deleted, and densification
naturally occurs.

The network also densifies if we consider the core
plus the periphery but when determining edges we take a 2 month sliding
window~\cite{kossinets06email}. This means that for every month $m$, we take all email messages between $m-2$ and
$m$, and create a graph, where there is an edge, if nodes exchanged
emails both ways in the last 2 months. This graph also densifies with
densification exponent $a= 1.21$.

Interestingly, the sliding window email network has higher
densification exponent than the full evolving email network. A
possible explanation is that email usage is increasing over time and
not all nodes (email addresses) are active at all times. Over the 18
month time period the size of 2-month sliding window graphs
increases from 7,000 to 10,000 nodes. On the other hand the full
email graph (composed of all nodes up to month $m$) grows from 3,000
to 38,000 nodes over the same time period. This means that there is
a large number of e-mail addresses that are active only for a period
of time. In a moving window graph we observe only active users and
thus more edges since email usage has also increased and people
communicate more. As opposed to the evolution of the full email
network, the moving window graphs do not have to accumulate the
history, i.e sparse graphs from the past, so they densify faster.

\subsubsection{IMDB actors to movies network}
\label{sec:imdb}

The Internet Movie Data Base (IMDB, \url{http://www.imdb.com}) is a
collection of facts about movies and actors. For every movie we know
the year of production, genre, and actor names that appeared in the
movie. From IMDB we obtained data about 896,192 actors and 334,084
movies produced between 1890 and 2004 (114 years).

Given this data we created a bi-partite graph of actors to movies the same way as in the case of affiliation networks.
This means that whenever a new movie appears, it links to all the
actors participating in it. We create a new actor node when the
actor first appears in any movie. This way, when a new movie
appears, we first create a movie node. Then we introduce actor
nodes, but only for actors for whom this was their first appearance
in a movie. Then we link actors and the movie.

In our experiment we started observing the graph in 1910, when the
giant connected component started to form. Before 1910 the largest
connected component consisted of less than 15\% of the nodes. At the
beginning of our observation period the network had $n$=7,690 nodes
(4,219 actors and 3,471 movies) and $e$=12,243 edges. At the end of
the dataset in 2004, we have $n$=1,230,276 nodes and $e$=3,790,667
edges.

We follow the usual procedure: for every year $Y$ we take all the
movies up to year $Y$ and actors that appeared in them. We create a
graph and measure how the number of edges grows with the number of
nodes. Figure~\ref{fig:powerGrowth}(f) presents the \GPLplot\ for
the IMDB actors to movies network. Again, notice the nontrivial
densification exponent of $a=1.11$.

\subsubsection{Product recommendation network}
\label{sec:recnet}

We also report the analysis of~\cite{leskovec06viral}, where they
measured the densification of a large person-to-person
recommendation network from a large on-line retailer. Nodes represent people and edges represent recommendations. The network
generation process was as follows. Each time a person {\em
purchases} a book, music CD, or a movie he or she is given the
option of sending emails recommending the item to friends. Any of
the recipients of the recommendation that makes a purchase can
further recommend the item, and by this propagation of recommendations the network forms.

The network consists of $e$=15,646,121 recommendations made among
$n$=3,943,084 distinct users. The data was collected from June 5
2001 to May 16 2003. In total, 548,523 products were recommended. We
report the \GPL\ exponent $a=1.26$ in table~\ref{tab:timeGrowth}.

\begin{table}
\begin{center}
  \begin{tabular}{l|r|r|r||c}
  {\sc Dataset} & {\sc Nodes} & {\sc Edges} & {\sc Time} & {\sc DPL
  exponent}\\
  \hline
  Arxiv HEP--PH   & 30,501 & 347,268 & 124 months & 1.56\\
  Arxiv HEP--TH   & 29,555 & 352,807 & 124 months & 1.68\\
  Patents         & 3,923,922 & 16,522,438 & 37 years & 1.66\\
  AS              & 6,474 & 26,467 & 785 days & 1.18\\
  Affiliation ASTRO--PH & 57,381 & 133,179 & 10 years & 1.15 \\
  Affiliation COND--MAT & 62,085 & 108,182 & 10 years & 1.10 \\
  Affiliation GR-QC     & 19,309 & 26,169 & 10 years & 1.08 \\
  Affiliation HEP--PH   & 51,037 & 89,163 & 10 years & 1.08 \\
  Affiliation HEP--TH   & 45,280 & 68,695 & 10 years & 1.08 \\
  Email       & 35,756 & 123,254 & 18 months & 1.12\\
  IMDB        & 1,230,276 & 3,790,667 & 114 years & 1.11\\
  Recommendations & 3,943,084 & 15,656,121 & 710 days & 1.26 \\
  \end{tabular}
  \caption{Dataset names with sizes, time lengths
  and Densification Power Law exponents. Notice very high densification
  exponent for citation networks ($\approx 1.6$), around $1.2$ for
  Autonomous Systems and lower (but still significant) densification
  exponent ($\approx 1.1$) for affiliation and collaboration type
  networks.}
  \label{tab:timeGrowth}
\end{center}
\end{table}

\subsection{Shrinking Diameters}

\begin{figure}[!tp]
\begin{center}
  \begin{tabular}{cc}
    \epsfig{file=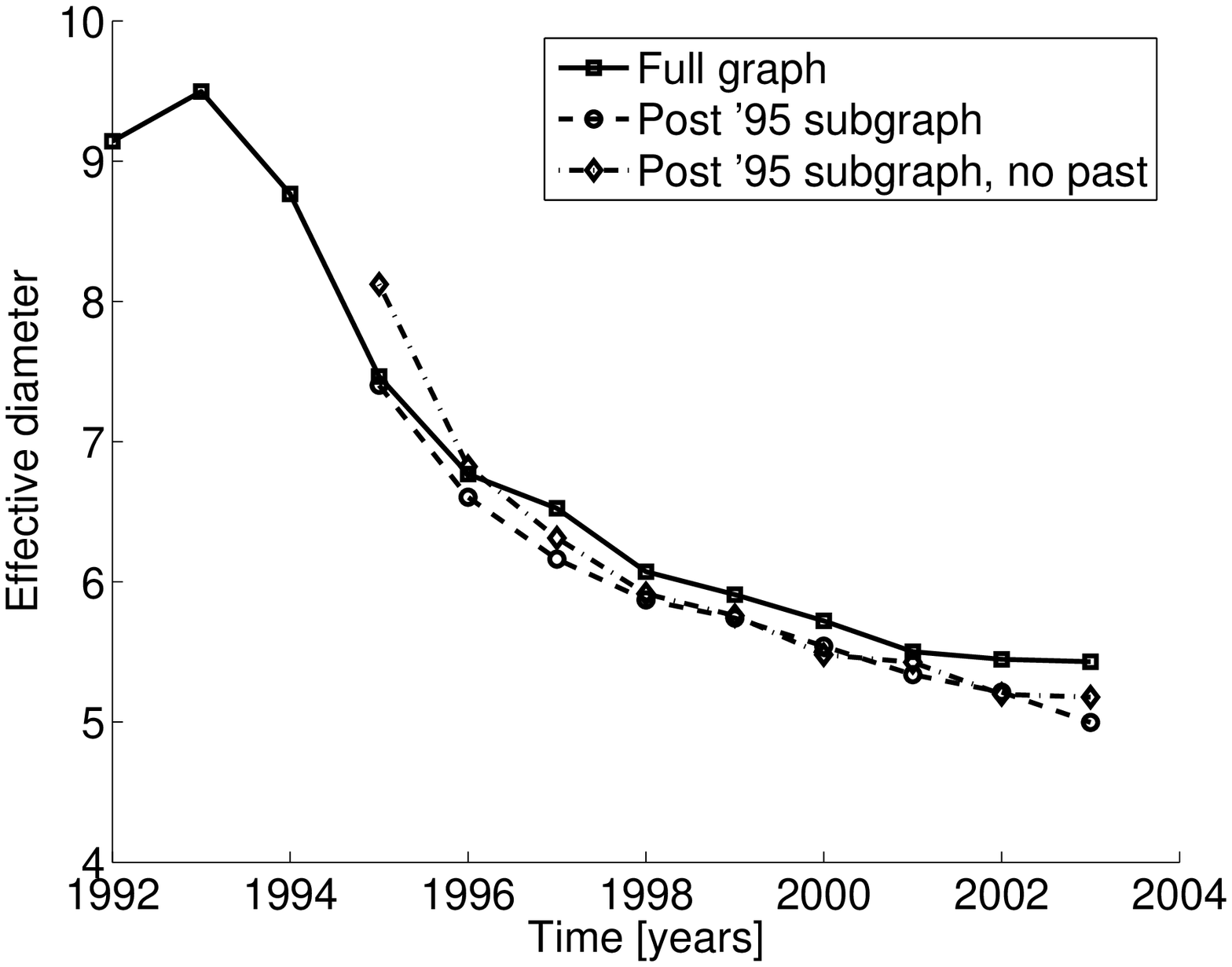, width=2.2in} &
    \epsfig{file=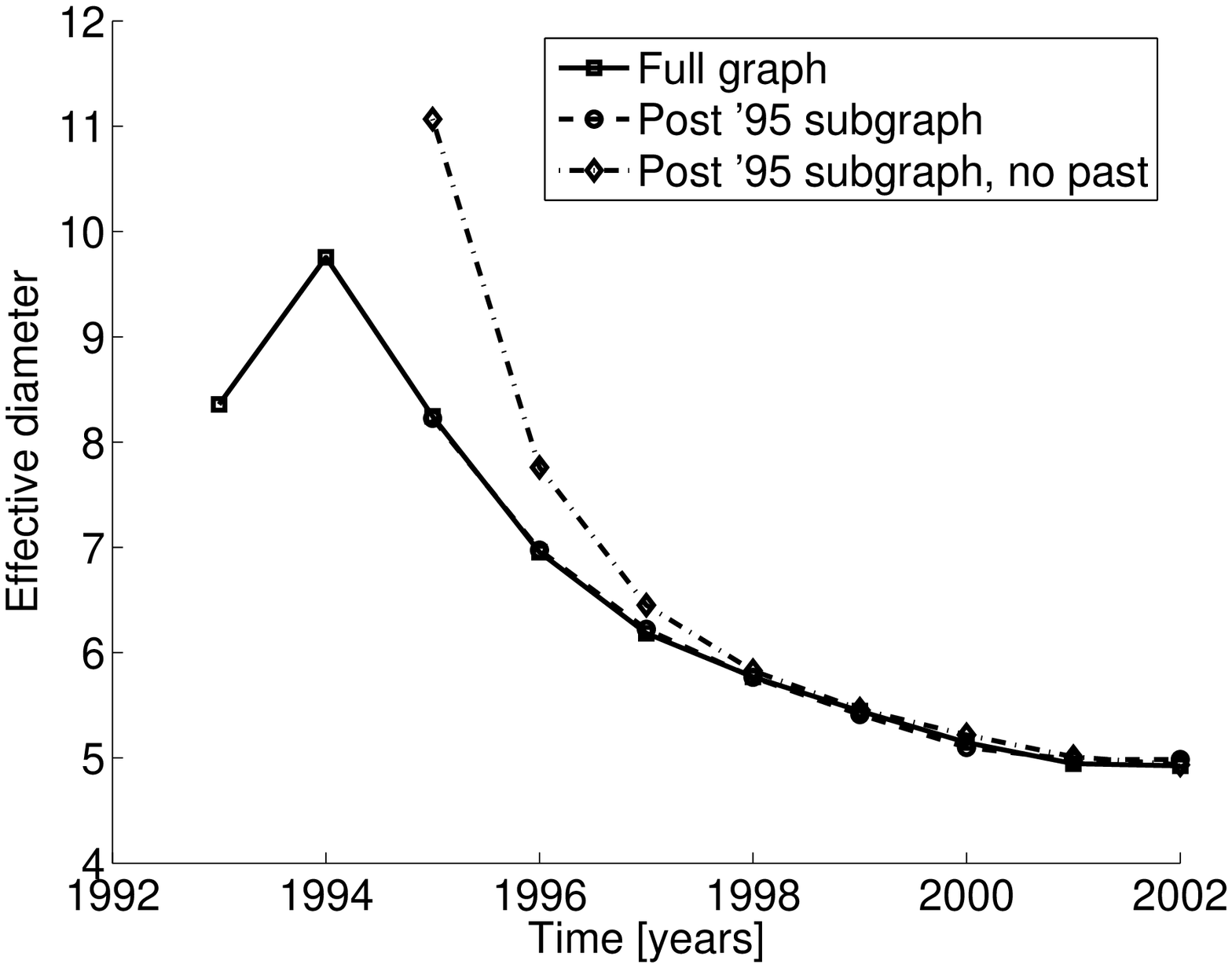 , width=2.2in} \\
    (a) arXiv citation graph & (b) Affiliation network \\
    \epsfig{file=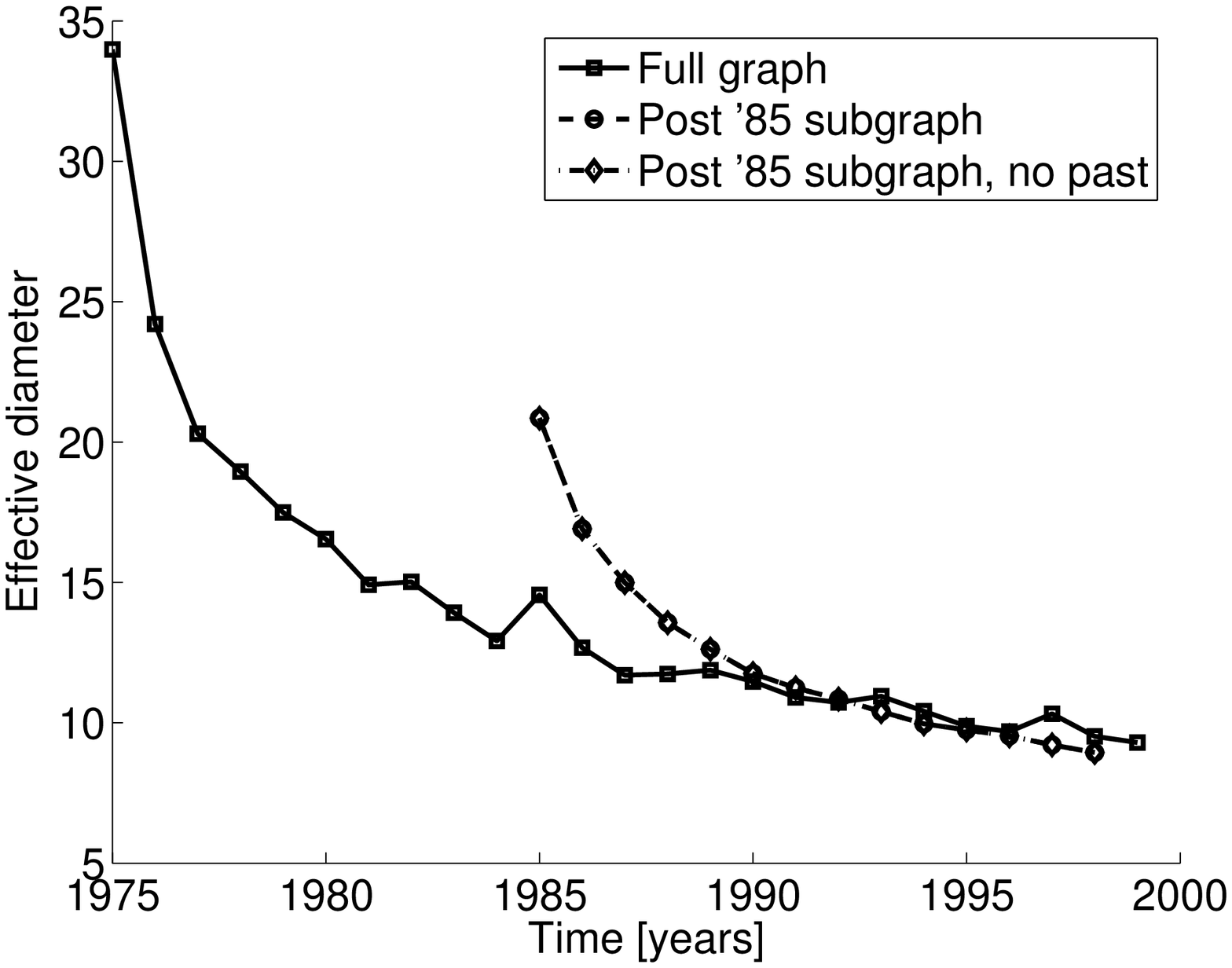, width=2.2in} &
    \epsfig{file=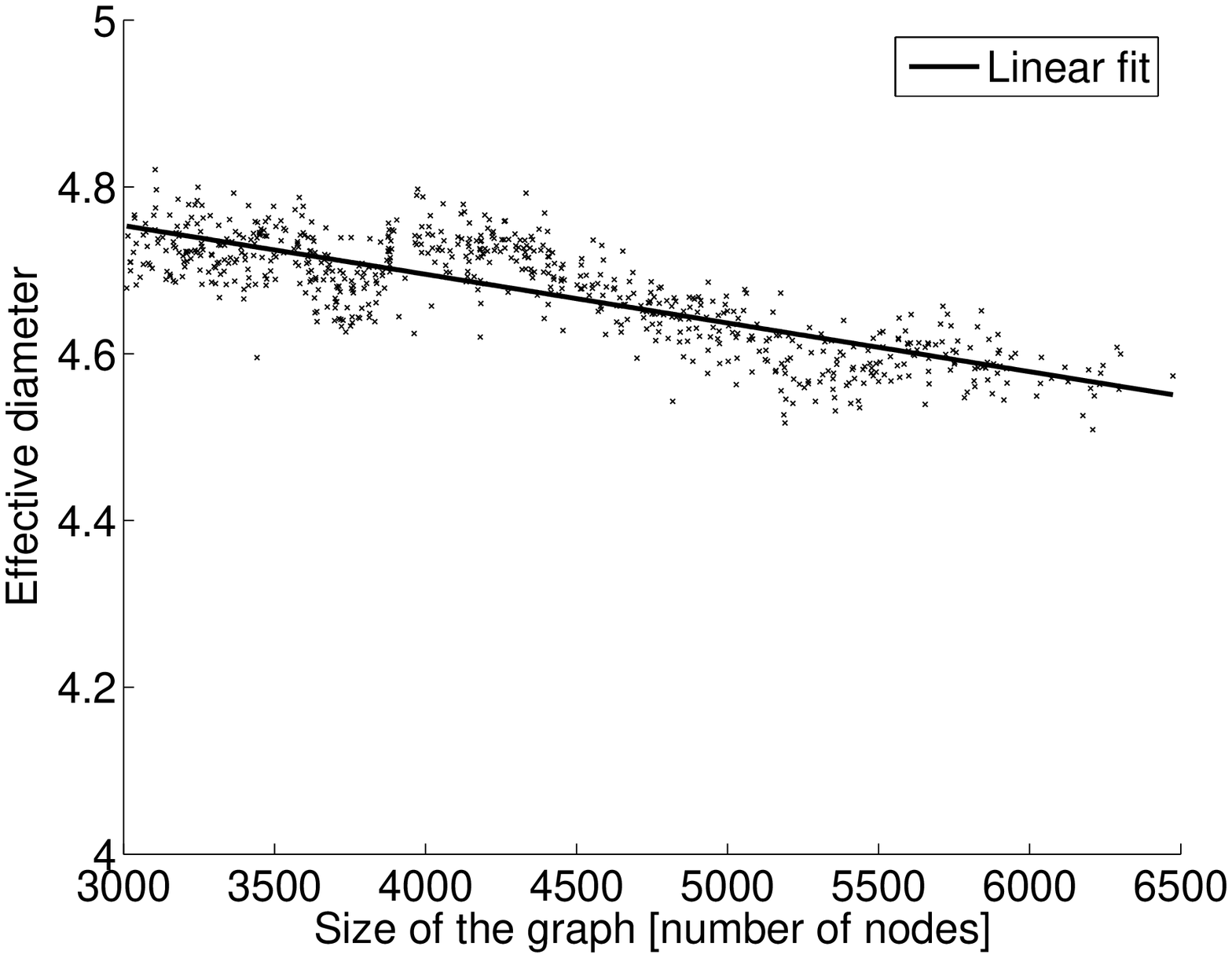, width=2.2in} \\
    (c) Patents citation graph & (d) Autonomous Systems \\
    \epsfig{file=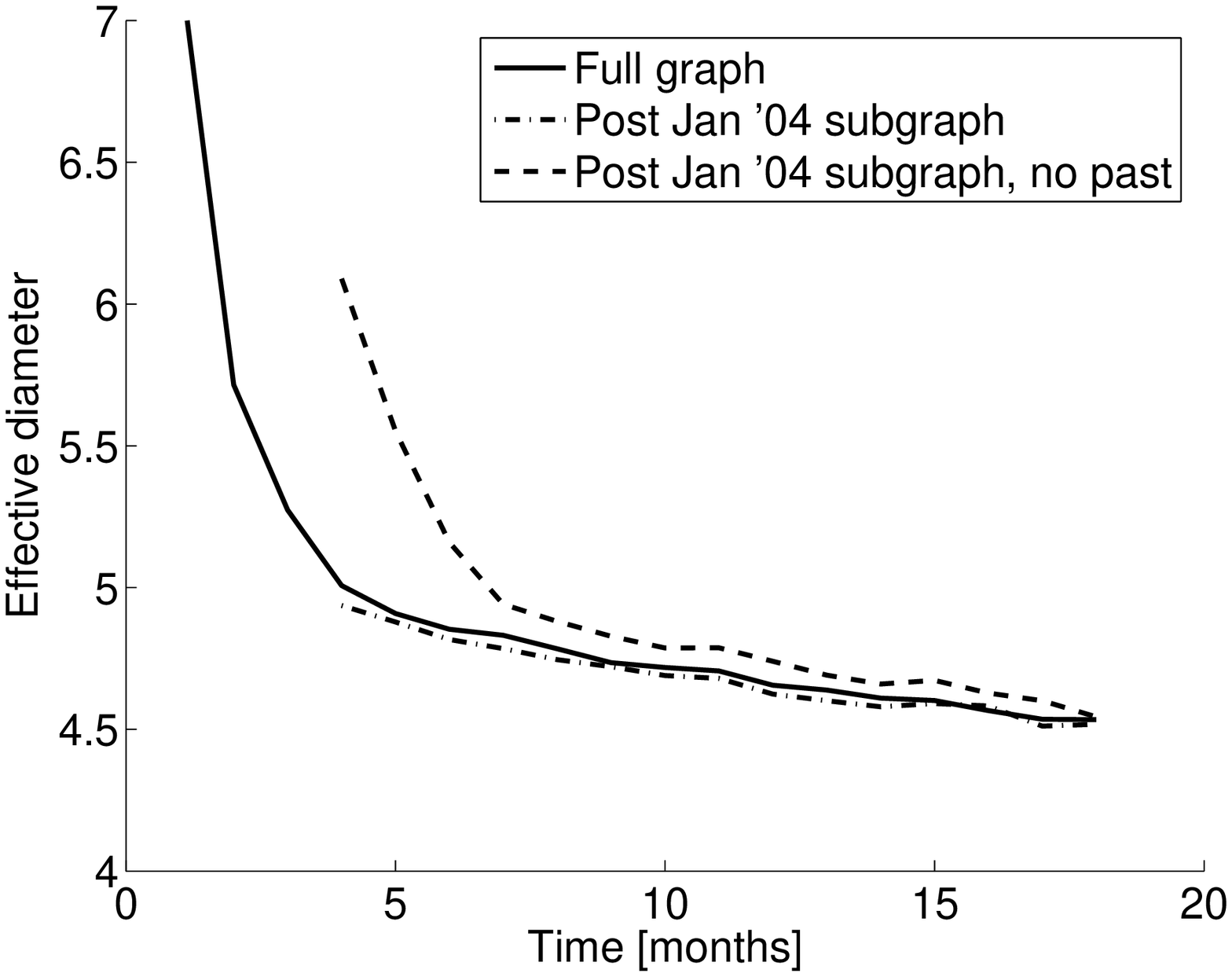, width=2.2in} &
    \epsfig{file=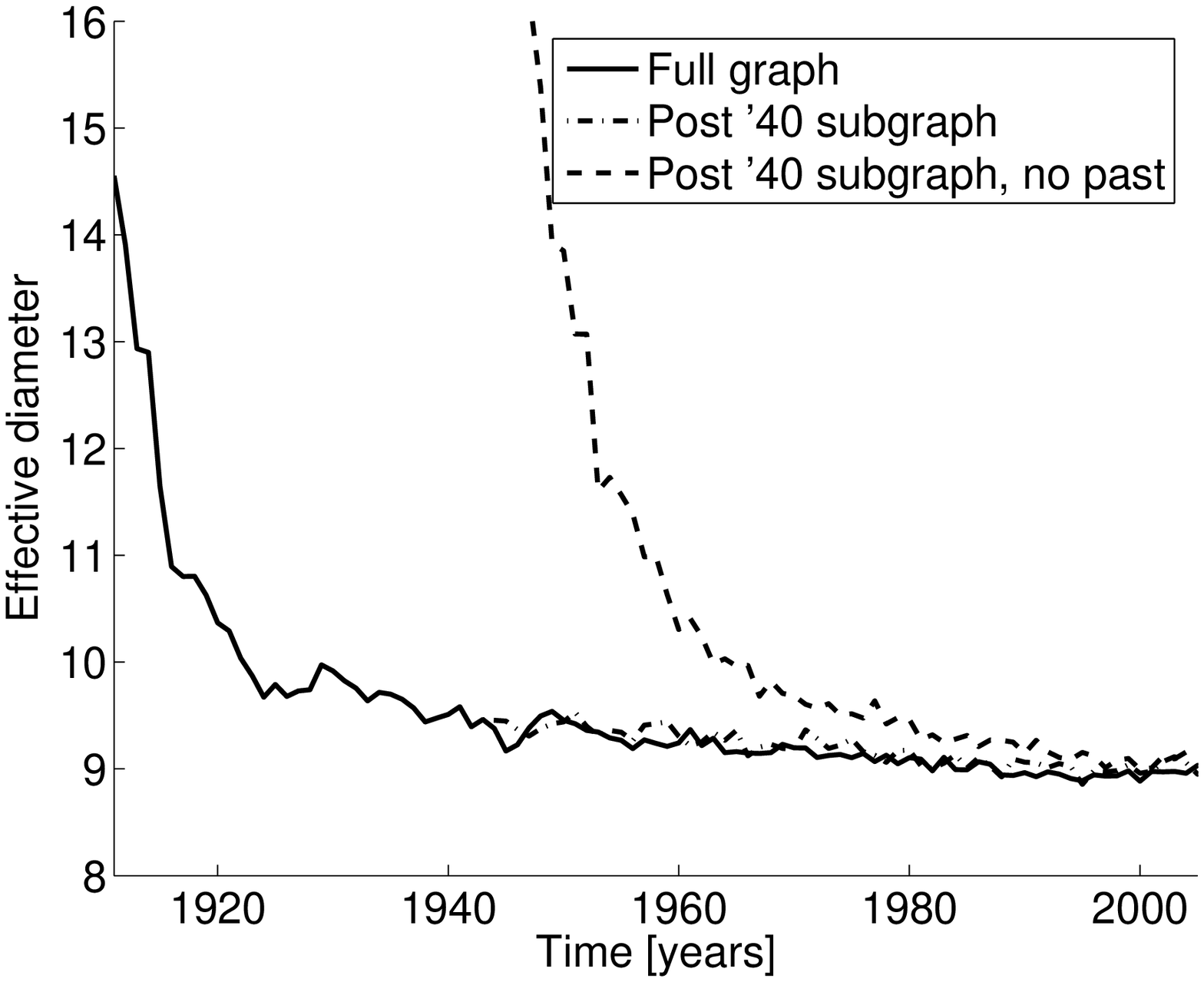, width=2.2in} \\
    (e) Email network & (f) IMDB actors to movies network\\
  \end{tabular}
  \caption{The effective diameter over time for 6 different
  datasets. Notice consistent decrease of the diameter over time.}
  \label{fig:Diameter}
\end{center}
\end{figure}

We now discuss the behavior of the effective diameter over time, for
this collection of network datasets. Following the conventional wisdom
on this topic, we expected the underlying question to be whether we
could detect the differences among competing hypotheses concerning the
growth rates of the diameter --- for example, the difference between
logarithmic and sub-logarithmic growth. Thus, it was with some surprise
that we found the effective diameters to be actually {\em decreasing}
over time (Figure~\ref{fig:Diameter}).

Let us define the necessary concepts underlying the observations. We
say that two nodes in a network are {\em connected} if there is an
undirected path between them; for each natural number $d$, let
$g(d)$ denote the fraction of connected node pairs whose shortest
connecting path has length at most $d$. The {\em hop-plot} for the
network is the set of pairs $(d,g(d))$; it thus gives the cumulative
distribution of distances between connected node pairs. We extend
the hop-plot to a function defined over all positive real numbers by
linearly interpolating between the points $(d,g(d))$ and
$(d+1,g(d+1))$ for each $d$, and we define the {\em effective
diameter} of the network to be the value of $d$ at which the
function $g(d)$ achieves the value $0.9$.

\begin{definition}
Graph $G$ has the {\em diameter} $d$ if the maximum length of
undirected shortest path over all {\em connected} pairs of nodes is
$d$. The length of the path is the number of segments (edges, links, hops) it contains.
\end{definition}

We also use {\em full diameter} to refer to this quantity. Notice
the difference between the usual and our definition of the diameter.
For a disconnected graph the diameter as usually defined to be
infinite, here we avoid this problem by considering only pairs of nodes that are connected. Also note we ignore the directionality of an edge if the graph is directed.

\begin{definition}
For each natural number $d$, let $g(d)$ denote the {\em fraction} of
connected node pairs whose undirected shortest connecting path in a graph $G$
has length at most $d$. And let $D$ be an integer for which $g(D-1)
< 0.9$ and $g(D) \ge 0.9$. Then the graph $G$ has the {\em integer
effective diameter} $D$~\cite{tauro01conceptual-topology}.
\end{definition}

In other words, the {\em integer effective diameter} is the smallest
number of hops $D$ at which at least 90\% of all connected pairs of
nodes can be reached.

Last we give the definition of the {\em effective diameter} as
considered in this paper. Originally we defined $g(d)$, a fraction
of connected pairs of nodes at distance at most $d$, only for
natural numbers $d$. Now we extend the definition of $g$ to all
positive reals $x$ by linearly interpolating the function value
between $g(d)$ and $g(d+1)$ ($d \le x < d+1$): $g(x)=g(d) +
(g(d+1)-g(d))(x-d)$.

\begin{definition}
Let $D$ be a value where $g(D) = 0.9$, then graph $G$ has the {\em
effective diameter} $D$.
\end{definition}

This definition varies slightly from an alternate definition of the
effective diameter used in earlier work: the minimum integer value $d$ such
that at least $90\%$ of the connected node pairs are at distance at
most $d$. Our variation smooths this definition by allowing it to
take {\em non-integer} values.

The effective diameter is a more robust quantity than the diameter
(defined as the maximum distance over all connected node pairs),
since the diameter is prone to the effects of degenerate structures
in the graph (e.g. very long chains). However, our experiments show
that the effective diameter and diameter tend to exhibit
qualitatively similar behavior. Note that under these definitions
the effective diameter and the diameter are well defined even if the
graph is disconnected.

We follow the same procedure as in case of Densification Power Law measurements. For each time $t$, we create a graph
consisting of nodes up to that time, and compute the effective
diameter of the undirected version of the graph.

Figure~\ref{fig:Diameter} shows the effective diameter over time;
one observes a decreasing trend for all the graphs. We performed a
comparable analysis to what we describe here for all 11 graph
datasets in our study, with very similar results. For the citation
networks in our study, the decreasing effective diameter has the
following interpretation: Since all the links out of a node are
``frozen'' at the moment it joins the graph, the decreasing distance
between pairs of nodes appears to be the result of subsequent papers
acting as ``bridges'' by citing earlier papers from disparate areas.
Note that for other graphs in our study, such as the AS dataset, it
is possible for an edge between two nodes to appear at an arbitrary
time after these two nodes join the graph.

We note that the effective diameter of a graph over time is
necessarily bounded from below, and the decreasing patterns of the
effective diameter in the plots of Figure~\ref{fig:Diameter} are
consistent with convergence to some asymptotic value. However,
understanding the full ``limiting behavior'' of the effective
diameter over time, to the extent that this is even a well-defined
notion, remains an open question.

\subsubsection{Validating the shrinking diameter conclusion}

Given the unexpected nature of this result, we wanted to verify that
the shrinking diameters were not attributable to artifacts of our
datasets or analyses. We explored this issue in a number of ways,
which we now summarize; the conclusion is that the shrinking
diameter appears to be a robust, and intrinsic, phenomenon.
Specifically, we performed experiments to account for (a) possible
sampling problems, (b) the effect of disconnected components, (c)
the effect of the ``missing past''(as in the previous subsection),
and (d) the dynamics of the emergence of the giant component.

\begin{itemize}
\item {\em Possible sampling problems:} Computing shortest paths among
all node pairs is computationally prohibitive for graphs of our
scale. We used several different approximate methods, obtaining
almost identical results from all of them. In particular, we applied
the Approximate Neighborhood Function (ANF) approach
\cite{palmer02anf} (in two different implementations), which can
estimate effective diameters for very large graphs, as well as a
basic sampling approach in which we ran exhaustive breadth-first
search from a subset of the nodes chosen uniformly at random. The
results using all these methods were essentially identical.

Plots on figure~\ref{fig:Diameter} were created by averaging over
100 runs of the ANF, the approximate diameter algorithm. For all
datasets the standard error is less than 10\%. For clarity of
presentation we do not show the error bars.

\item {\em Disconnected components:} One can also ask about the effect
of small disconnected components. All of our graphs have a single
{\em giant component} -- a connected component (or a weakly connected
component in the case of directed graphs, ignoring the direction of
the edges) that accounts for a significant fraction of all nodes.
For each graph, we computed effective diameters for both the entire
graph and for just the giant component; again, our results are
essentially the same using these two methods.

\item {\em ``Missing Past'' effects:} A third issue is the problem of
the ``missing past,'' the same general issue that surfaced in the
previous subsection when we considered densification. In particular,
we must decide how to handle citations to papers that predate our
earliest recorded time. (Note that the missing past is not an issue
for the AS network data, where the effective diameter also
decreases.)

To understand how the diameters of our networks are affected by this
unavoidable problem, we perform the following test. We pick some
positive time $t_0 > 0$, and determine what the diameter would look
like
as a function of time, {\em if this were the beginning of our data}.
We then put back in the nodes and the edges from before time $t_0$,
and
study how much the diameters change. If this change is small --- or at
least if it does not affect the qualitative conclusions --- then it
provides evidence that the missing past is not influencing the overall
result.

Specifically, we set this cut-off time $t_0$ to be the beginning of
$1995$ for the arXiv (since we have data from $1993$), and to be
$1985$ for the patent citation graph (we have data from $1975$). For
Email network we set the cut-off time to January 2004 and for IMDB
to 1940 (we also experimented with 1920 and 1960 and findings were
consistent). We then compared the results of three measurements:

\begin{itemize}
\item[$-$] {\em Diameter of full graph.} For each time $t$ we compute the effective diameter of the graph's giant component.

\item[$-$] {\em Post-$t_0$ subgraph.} We compute the effective
diameter of the post-$t_0$ subgraph using all nodes and edges. This
means that for each time $t$ ($t > t_0$) we create a graph using all
nodes dated before $t$. We then compute the effective diameter of
the subgraph of the nodes dated between $t_0$ and $t$. To compute
the effective diameter we can use all edges and nodes (including
those dated before $t_0$). This means that we are measuring
distances {\em only} among nodes dated between $t_0$ and $t$ while
also using nodes and edges before $t_0$ as ``shortcuts'' or
``bypasses''. The experiment measures the diameter of the graph if
we knew the full (pre-$t_0$) past --- the citations of the papers
which we have intentionally excluded for this test.

\item[$-$] {\em Post-$t_0$ subgraph, no past.} We set $t_0$ the same
way as in previous experiment, but then for all nodes dated before
$t_0$ we delete all their out-links. This creates the graph we would
have gotten if we had started collecting data only at time $t_0$.
\end{itemize}

In Figure~\ref{fig:Diameter}, we superimpose the effective diameters
using the three different techniques. If the missing past does not
play a large role in the diameter, then all three curves should lie
close to one another. We observe this is the case for the arXiv
citation graphs. For the arXiv paper-author affiliation graph, and
for the patent citation graph, the curves are quite different right
at the cut-off time $t_0$ (where the effect of deleted edges is most
pronounced), but they quickly align with one another. As a result,
it seems clear that the continued decreasing trend in the effective
diameter as time runs to the present is not the result of these
boundary effects.

\item {\em Emergence of giant component:} A final issue is the
dynamics by which the giant component emerges. For example, in the
standard Erd\"os-Renyi random graph model (which has a substantially
different flavor from the growth dynamics of the graphs here), the
diameter of the giant component is quite large when it first
appears, and then it shrinks as edges continue to be added. Could
shrinking diameters in some way be a symptom of emergence of giant
component?

\begin{figure}[!tp]
\begin{center}
  \begin{tabular}{cc}
    \epsfig{file=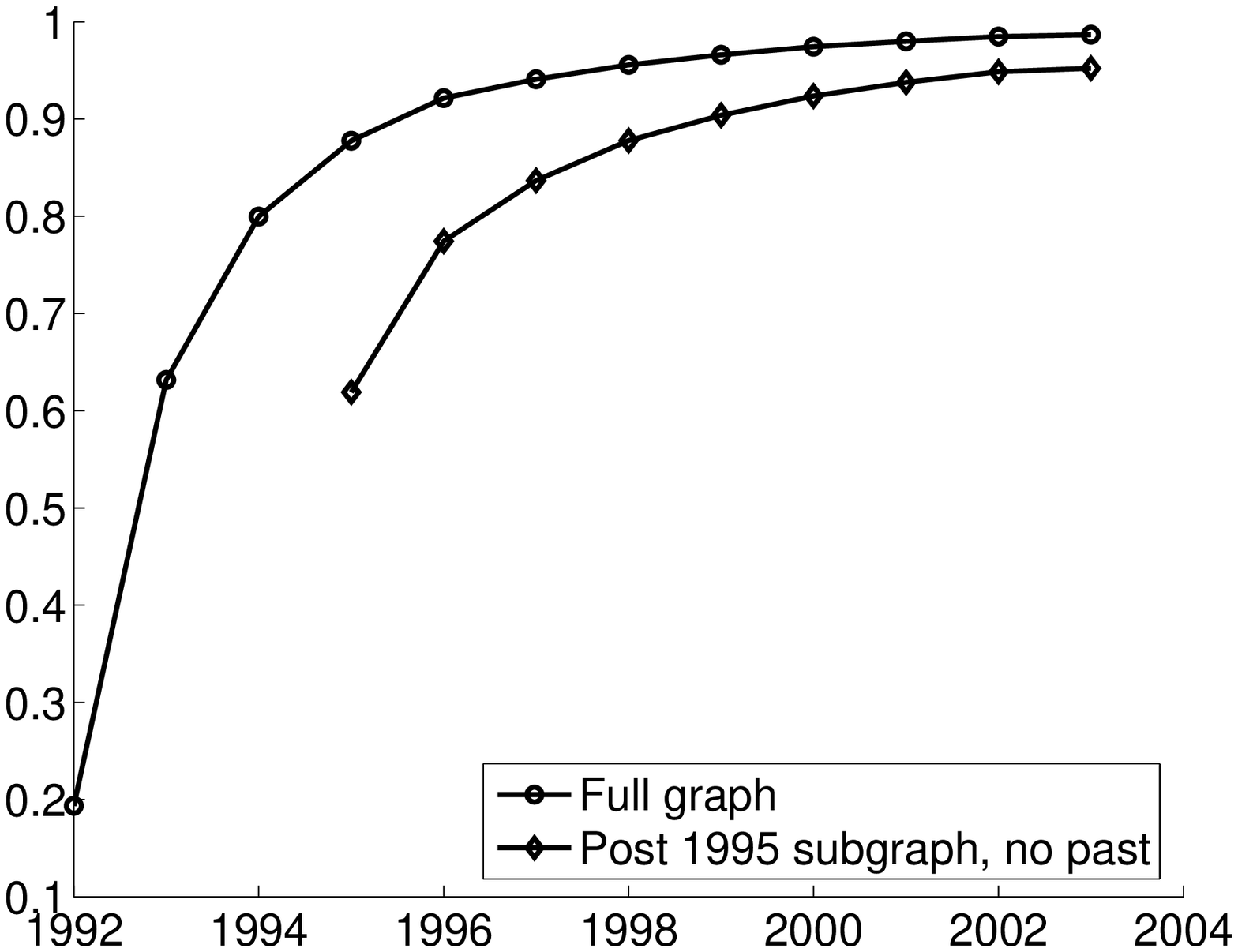, width=2.2in} &
    \epsfig{file=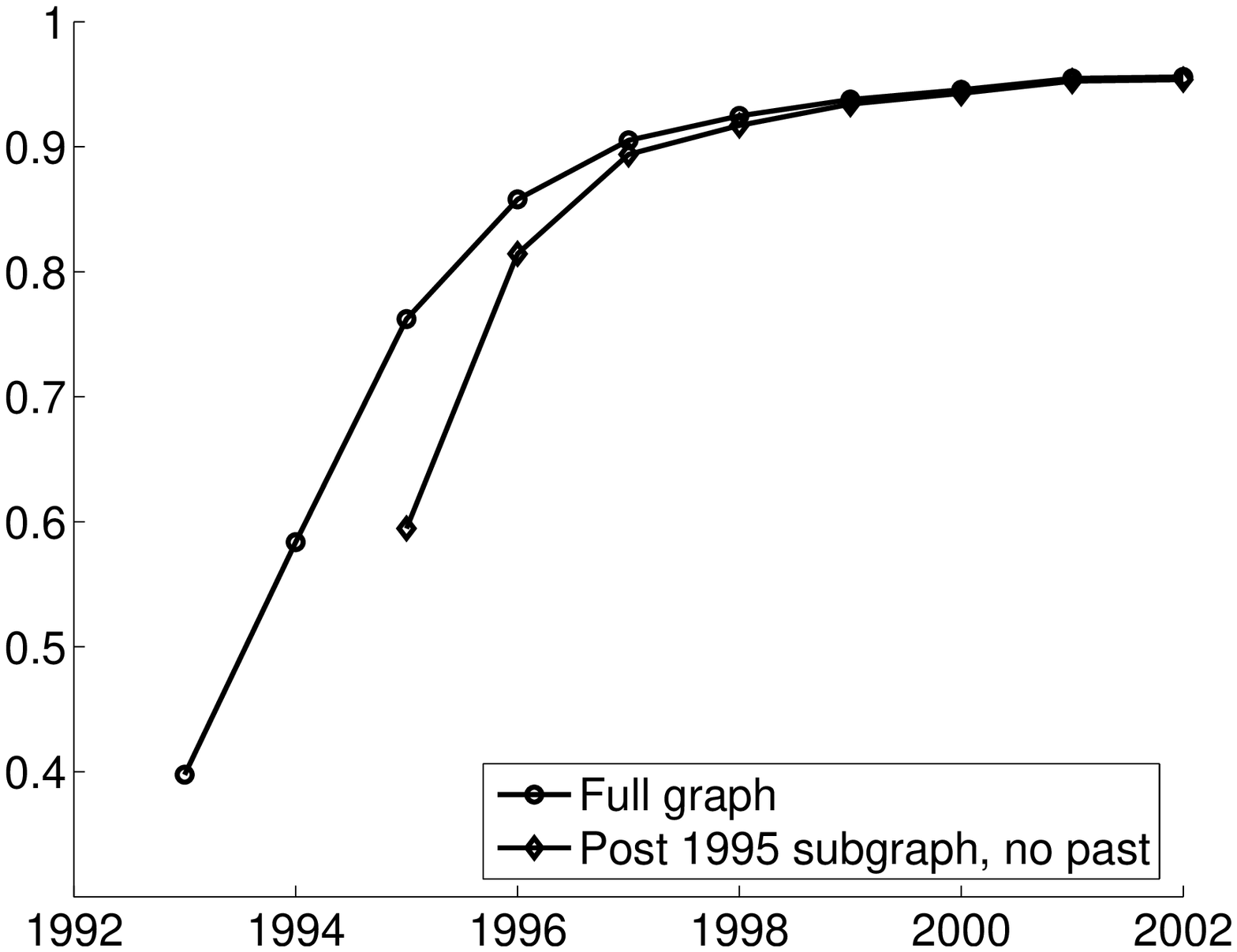 , width=2.2in} \\
    (a) arXiv citation graph & (b) Affiliation network \\
  \end{tabular}
  \caption{The fraction of nodes that are part of the giant
  connected component over time. We see that after 4 years the 90\% of
  all nodes in the graph belong to giant component.}
  \label{fig:gccSize}
\end{center}
\end{figure}

It appears fairly clear that this is not the case.
Figure~\ref{fig:gccSize} shows the fraction of all nodes that are
part of the largest connected component (GCC) over time. We plot the
size of the GCC for the full graph and for a graph where we had no
past --- i.e., where we delete all out-links of the nodes dated
before the cut-off time $t_0$. Because we delete the out-links of
the pre-$t_0$ nodes the size of GCC is smaller, but as the graph
grows the effect of these deleted links becomes negligible.

We see that within a few years the giant component accounts for
almost all the nodes in the graph. The effective diameter, however,
continues to steadily decrease beyond this point. This indicates
that the decrease is happening in a ``mature'' graph, and not
because many small disconnected components are being rapidly glued
together.
\end{itemize}

Based on all this, we believe that the decreasing diameters in our
study reflect a fundamental property of the underlying networks.
Understanding the possible causes of this property, as well as the
causes of the densification power laws discussed earlier, will be
the subject of the next section.

\section{Proposed Models}
\label{sec:explanations}

We have now seen that densification power laws and shrinking
effective diameters are properties that hold across a range of
diverse networks. Moreover, existing models do not capture these
phenomena. We would like to find some simple, local model of
behavior, which could naturally lead to the macroscopic phenomena we
have observed. We present increasingly sophisticated models, all of
which naturally  achieve the observed densification; the last one
(the ``Forest Fire'' model) also exhibits shrinking diameter and all
the other main patterns known (including heavy-tailed in- and
out-degree distributions).

\subsection{\CGA}

What are the underlying principles that drive all our observed graphs
to
obey a densification power law,
without central control or coordination? We seek a model
in which the densification exponent arises from intrinsic features of
the process that generates nodes and edges.  While one could clearly
define a graph model in which $e(t) \propto n(t)^a$ by simply having
each node, when it arrives at time $t$, generate $n(t)^{a-1}$ out-links
--- the equivalent of positing that each author of a paper in a
citation
network has a rule like, ``Cite $n^{a-1}$ other documents,'' hard-wired
in his or her brain --- such a model would not provide any insight into
the origin of the exponent $a$, as the exponent is unrelated to the
operational details by which the network is being constructed. Instead,
our goal is to see how underlying properties of the network evolution
process itself can affect the observed densification behavior.

We take the following approach. Power laws often appear in
combination with {\em self-similar} structures. Intuitively, a
self-similar object consists of miniature replicas of
itself~\cite{Schroeder91Fractals}. Our approach involves two steps,
both of which are based on self-similarity.

We begin by searching for self-similar, recursive structures. In
fact, we can easily find several such recursive sets: For example,
computer networks form tight groups (e.g., based on geography),
which consist of smaller groups, and so on, recursively. Similarly
for patents: they also form conceptual groups (``chemistry'',
``communications'', etc.), which consist of sub-groups, and so on
recursively. Several other graphs feature such ``communities within
communities'' patterns.

For example, it has been argued (see e.g.~\cite{watts02identity} and
the references therein) that social structures exhibit
self-similarity, with individuals organizing their social contacts
hierarchically. Moreover, pairs of individuals belonging to the same
small community form social ties more easily than pairs of
individuals who are only related by membership in a larger
community. In a different domain, Menczer studied the frequency of
links among Web pages that are organized into a topic hierarchy such
as the Open Directory~\cite{menczer02growing}. He showed that link
density among pages decreases with the height of their least common
ancestor in the hierarchy. That is, two pages on closely related
topics are more likely to be hyperlinked than are two pages on more
distantly related topics.

This is the first, qualitative step in our explanation for the
Densification Power Law. The second step is quantitative. We will
need a numerical measure of the difficulty in crossing communities.
The extent to which it is indeed difficult to form links across
communities will be a property of the domain being studied. We call
this the {\em {\CCDC}}, and we define it more precisely below.

\begin{table}
\begin{center}
  \begin{tabular}{c|l}
    {\sc Symbol} & {\sc Description}\\\hline
    $a$ & {\em Densification Exponent}\\
    $c$ & {\em \CCDC}\\
    $f(h)$ & {\em \CCDF}\\
    $n(t)$ & number of nodes at time $t$\\
    $e(t)$ & number of edges at time $t$\\
    $b$ & community branching factor\\
    $\dbar$ & expected average node out-degree\\
    $H$ & height of the tree\\
    $h(v,w)$ & least common ancestor height of leaves $v, w$\\
    $\fwdprob$ & forest fire forward burning probability\\
    $\bckprob$ & forest fire backward burning probability\\
    $\backratio$ & ratio of backward and forward probability,
    $\backratio=\fwdprob/\bckprob$\\
    $\ddslope$ & power-law degree distribution exponent\\
  \end{tabular}
\caption{Table of symbols}
\label{tab:symbols}
\end{center}
\end{table}

\subsubsection{The Basic Version of the Model}
\label{sec:basicModel}

We represent the recursive structure of
communities-within-communities as a tree $\Gamma$, of height $H$. We
shall show that even a simple, perfectly balanced tree of constant
fanout $b$ is enough to lead to a densification power law, and so we
will focus the analysis on this basic model.

The nodes $V$ in the graph we construct will be the leaves of the
tree; that is, $n = |V|$. (Note that $n = b^H$.) Let $h(v,w)$ define
the standard tree distance of two leaf nodes $v$ and $w$: that is,
$h(v, w)$ is the height of their least common ancestor (the height
of the smallest sub-tree containing both $v$ and $w$).

We will construct a random graph on a set of nodes $V$ by specifying
the probability that $v$ and $w$ form an edge as a function $f$ of
$h(v,w)$. We refer to this function $f$ as the {\em \CCDF}. What
should be the form of $f$? Clearly, it should decrease with $h$; but
there are many forms such a decrease could take.

The form of $f$ that works best for our purposes comes from the
self-similarity arguments we made earlier: We would like $f$ to be
scale-free; that is, $f(h) / f(h-1)$ should be level-independent and
thus constant. The only way to achieve level-independence is to
define $f(h) = f(0) c^{-h}$. Setting $f(0)$ to 1 for simplicity, we
have:
\begin{equation}
    f(h) = c^{-h}
    \label{eq:CCLP}
\end{equation}

where $c \ge 1$. We refer to the constant $c$ as the {\em \CCDC}.
Intuitively, cross-communities links become harder to form as $c$
increases.

This completes our development of the model, which we refer to as
{\em \CGA}: If the nodes of a graph belong to
communities-within-communities, and if the cost for cross-community
edges is scale-free (Eq.~(\ref{eq:CCLP})), the \GPL\ follows
naturally. No central control or exogenous regulations are needed to
force the resulting graph to obey this property. In short,
self-similarity itself leads to the \GPL.

\begin{theorem}
In the \CGA\ random graph model just defined, the expected average
out-degree $\dbar$ of a node is proportional to:

\begin{eqnarray}
  \dbar & = & n^{1-\log_{b}(c)} ~~~~~~ \textrm{if} ~~ 1 \le c \le b
  \nonumber \\
        & = & \log_{b}(n)     ~~~~~~~~ \textrm{if} ~~ c = b \nonumber
        \\
        & = & constant          ~~~~~~ \textrm{if} ~~ c > b \nonumber
\end{eqnarray}
\label{th:basic}
\end{theorem}

\begin{proof}
For a given node $v$, the expected out-degree (number of links) $\dbar$
of the node is proportional to

\begin{equation}
\dbar  =  \sum_{x \neq v}{f(h(x, v))}
       =  \sum_{j=1}^{\log_{b}(n)}{(b-1)b^{j-1} c^{-j}}
       =  {b-1 \over c} \sum_{j=1}^{\log_{b}(n)}
             {{\left({b \over c} \right)} ^{j-1}}.
  \label{eq:numEdges}
\end{equation}

There are three different cases: if $1 \le c < b$ then by summing the
geometric series we obtain

\begin{eqnarray*}
\dbar  & =  & {b-1 \over c} \cdot
  {{{\left({b \over c} \right)}^{\log_{b}(n)}-1}
  \over { {\left({b \over c} \right)} - 1 }}
       =  \left({b-1 \over b - c }\right) {(n^{1-\log_{b}(c)} - 1)} \\
       & = &  \Theta(n^{1-\log_{b}(c)}).
\end{eqnarray*}

In the case when $c = b$ the series sums to

\begin{eqnarray*}
  \dbar & = & \sum_{x \neq v}{f(h(x, v))}
  = {b-1 \over b} \sum_{j=1}^{\log_{b}(n)} {{\left({b \over b}
  \right)}^{j-1}}
  = {b-1 \over b} \log_{b}(n) \\
  & = &  \Theta(\log_{b}(n)).
\end{eqnarray*}

The last case is when {\CCDC} $c$ is greater than branching factor $b$
($c > b$), then the sum in Eq.~(\ref{eq:numEdges}) converges to a
constant even if carried out to infinity, and so we obtain
$\dbar = \Theta(1)$.
\end{proof}

Note that when $c < b$, we get a densification law with exponent
greater
than $1$: the expected out-degree is $n^{1 - \log_b(c)}$, and so the
total number of edges grows as $n^a$ where $a = 2 - \log_b (c)$.
Moreover, as $c$ varies over the interval $[1,b)$, the exponent $a$
ranges over all values in the interval $(1,2]$.

\begin{corollary}
  If the \CCDF\ is scale-free ($f(h)$ $=$ $c^{-h}$, with $1 < c < b$),
  then the \CGA\ obeys the \GPL\ with exponent
  $$a = 2 - \log_b(c)$$
\end{corollary}

\subsubsection{Dynamic \CGA}

So far we have discussed a model in which nodes are first organized
into
a nested set of communities, and then they start forming links. We now
extend this to a setting in which nodes are added over time, and the
nested structure deepens to accommodate them. We will assume that a
node
only creates out-links at the moment it is added (and hence, only to
nodes already present); this is natural for domains like citation
networks in which a paper's citations are written at the same time as
the paper itself.

Specifically, the model is as follows. Rather than having graph
nodes reside only at the leaves of the tree $\Gamma$, there will now
be a graph node corresponding to every internal node of $\Gamma$ as
well. Initially, there is a single node $v$ in the graph, and our
tree $\Gamma$ consists just of $v$. In time step $t$, we go from a
complete $b$-ary tree of depth $t-1$ to one of depth $t$, by adding
$b$ new leaves as children of each current leaf.  Each of these new
leaves will contain a new node of the graph.

Now, each new node forms out-links according to a variant of the
process in which all graph nodes are leaves. However, since a new
node has the ability to link to internal nodes of the existing tree,
not just to other leaves, we need to extend the model to incorporate
this. Thus, we define the {\em tree-distance} $d(v,w)$ between nodes
$v$ and $w$ to be the length of a path between them in $\Gamma$ ---
this is the length of the path from $v$ up to the least common
ancestor of $v$ and $w$, plus the length of the path from this least
common ancestor down to $w$.  Note that if $v$ and $w$ are both
leaves, then $d(v,w) = 2h(v,w)$, following our definition of
$h(v,w)$ from above.

The process of forming out-links is now as follows: For a constant
$c$, node $v$ forms a link to each node $w$, independently, with
probability $c^{-d(v,w)/2}$. (Note that dividing by $2$ in the
exponent means this model gives the same probability as basic model
in the case when both $v$ and $w$ are leaves.)

Like the first model, this process produces a densification law with
exponent $a = 2 - \log_b(c)$ when $c < b$. However, for $c < b^2$,
it also yields a heavy-tailed distribution of in-degrees ---
something that the basic model did not produce. We describe this in
the following theorem.

\begin{theorem}
The Dynamic Community Guided Attachment model just defined has
the following properties.
\begin{itemize}
        \item When $c < b$, the average node degree is $n^{1 -
\log_b(c)}$ and the in-degrees follow a Zipf distribution with
exponent $\frac12 \log_b(c)$.
        \item When $b < c < b^2$, the average node degree is
        constant, and the in-degrees follow a Zipf distribution with
        exponent
        $1 - \frac12\log_b(c)$.
        \item When $c > b^2$, the average node degree is constant
and the probability of an in-degree exceeding any constant bound $k$
decreases exponentially in $k$.
\end{itemize}
\end{theorem}

\begin{proof}
In the proof, all logarithms will be expressed in base $b$ unless
specified otherwise.

We begin with the following basic facts. If a node is at height $h$
in the tree, then the number of nodes at distance $d \leq h$ from it
is $\T{b^d}$. Nodes at distance $d > h$ can be reached by going up
for $j$ steps, and then down for $d - j$ steps (if $d - j \leq h +
j$). This is maximized for $j = (d - h)/2$, and so the total number
of nodes reachable at distance $d$ is $\T{b^{(d + h)/2}}$.

\bigskip {\bf Case 1: $c < b$}
In this case, the expected out-degree for a leaf node is

$$
  \sum_{d = 0}^{2 \log n} \T{\frac{b^{d/2}}{c^{d/2}}} =
  \T{\frac{b^{\log n}}{c^{\log n}}} = \T{\frac{n}{c^{\log n}}} =
  \T{n^{1 - \log c}}.
$$

Since the expected out-degree values for other nodes are smaller,
and since a constant fraction of all nodes are leaves, it follows
that the expected value of the out-degree taken over all nodes is
$\T{n^{1 - \log c}}$ as well.

Now we compute the expected in-degree of a node at height $h$. This
is

$$
  \sum_{d \leq h} \T{\frac{b^d}{c^{d/2}}} + \sum_{d > h}
  \T{\frac{b^{(d + h)/2}}{c^{d/2}}} = \sum_{d \leq h}
  \T{\frac{b^{d/2}}{c^{d/2}}} b^{d/2} + \sum_{d > h}
  \T{\frac{b^{d/2}}{c^{d/2}}} b^{h/2}.
$$

The largest term in this sum is the last, for $d = 2 \log n - h$.
Here it takes the value

$$
  \T{\frac{b^{\log n}}{c^{\log n - (h/2)}}} =
  \T{\frac{b^{\log n}}{c^{\log n}}} c^{h/2} =
  \T{n^{1 - \log c}c^{h/2}}.
$$

The maximum expected in-degree $z$ is achieved for $h = \log n$,
when we get

$$
  z = \T{n^{1 - \log c} c^{.5 \log n}} = \T{n^{1 - .5 \log c}}.
$$

So for a node at depth $t = \log n - h$, we get an expected
in-degree of

$$
  \T{n^{1 - \log c} c^{(\log n - t)/2}} = \T{z c^{-t/2}}.
$$

Hence, to compute a Zipf exponent, we see that a node of degree rank
$r = {b^t}$ has depth $t$, so it has degree

$$
  \T{\frac{z}{c^{t/2}}} = \T{\frac{z}{r^{.5 \log c}}}.
$$

\bigskip {\bf Case 2: $b < c < b^2$}
In this case, the expected out-degree for a leaf node is

$$
  \sum_{d = 0}^{2 \log n} \T{\frac{b^{d/2}}{c^{d/2}}} = \Theta(1).
$$

Since the expected out-degree values for other nodes are smaller, it
follows that the expected value of the out-degree taken over all
nodes is $\T{1}$ as well.

Now we compute the expected in-degree of a node at height $h$. This
is

$$
  \sum_{d \leq h} \T{\frac{b^d}{c^{d/2}}} + \sum_{d > h}
  \T{\frac{b^{(d + h)/2}}{c^{d/2}}} = \sum_{d \leq h}
  \T{\frac{b^{d/2}}{c^{d/2}}} b^{d/2} + \sum_{d > h}
  \T{\frac{b^{d/2}}{c^{d/2}}} b^{h/2}.
$$

Since $b < c < b^2$, these terms increase geometrically up to $d =
h$, then decrease. Thus, the largest term is for $d = h$, where it
is $\T{b^h c^{-h/2}}$.

Thus the maximum degree is $z = \T{n^{1 - .5 \log c}}$, and for
depth $t = \log n - h$, we get a degree of

$$
  \T{\pfrac{b}{c^{1/2}}^{\log n} \pfrac{b}{c^{1/2}}^{-t}} =
  \T{z \pfrac{b}{c^{1/2}}^{-t}}.
$$

Now, $b/c^{1/2} = b^{1 - .5 \log c}$, so a node of degree rank $r =
b^t$ (at depth $t$) has degree $\T{z / r^{1 - .5 \log c}}$.

\bigskip {\bf Case 3: $c > b^2$}
The expected out-degrees here are only smaller than they are in the
previous case, and hence the expected value of the out-degree taken
over all nodes is $\T{1}$.

The node whose in-degree is most likely to exceed a fixed bound $k$
is the root, at height $h = \log n$. The in-degree of the root is a
sum $X$ of independent $0$-$1$ random variables $X_v$, where $X_v$
takes the value $1$ if node $v$ links to the root, and $X_v$ takes
the value $0$ otherwise. We have

$$
  EX = \sum_v EX_v = \sum_{d \leq \log n} \T{\frac{b^{d}}{c^{d/2}}} =
\Theta(1),
$$

and hence by Chernoff bounds, the probability that it exceeds a
given value $k > EX$ decreases exponentially in $k$.
\end{proof}

Thus, the dynamic \CGA\ model exhibits three qualitatively different
behaviors as the parameter $c$ varies: densification with
heavy-tailed in-degrees; then constant average degree with
heavy-tailed in-degrees; and then constant in- and out-degrees with
high probability. Note also the interesting fact that the Zipf
exponent is maximized for the value of $c$ right at the onset of
densification.

Finally, we have experimented with versions of the dynamic \CGA\
model in which the tree is not balanced, but rather deepens more on
the left branches than the right (in a recursive fashion). We have
also considered versions in which a single graph node can ``reside''
at two different nodes of the tree $\Gamma$, allowing for graph
nodes to be members of different communities. Experimental results
and overall conclusions were all the time the same and consistent
regardless of the particular version (modification) of the dynamic
\CGA\ model used.

\subsection{The Forest Fire Model}

\CGA\ and its extensions show how densification can arise naturally,
and even in conjunction with heavy-tailed in-degree distributions.
However, it is not a rich enough class of models to capture all the
properties in our network datasets. In particular, we would like to
capture both the shrinking effective diameters that we have
observed, as well as the fact that real networks tend to have
heavy-tailed out-degree distributions (though generally not as
skewed as their in-degree distributions). The \CGA\ models do not
exhibit either of these properties.

Specifically, our goal is as follows. Given  a (possibly empty) initial
graph ${G}$, and a sequence of new nodes $v_1$ $\ldots$ $v_k$, we want
to design a simple randomized process to successively link $v_i$ to
nodes of ${G}$ ($i=1, \ldots k$) so that the resulting graph
${G}_{final}$ will obey all of the following patterns: heavy-tailed
distributions for in- and out-degrees, the \GPL, and shrinking
effective
diameter.

We are guided by the intuition that such a graph generator may arise
from a combination of the following components:

\begin{itemize}
  \item some type of ``rich get richer'' attachment process,
  to lead to heavy-tailed in-degrees;
  \item some flavor of the ``copying'' model~\cite{kumar00stochastic},
  to lead to communities;
  \item some flavor of \CGA, to produce a version of the \GPL;
  \item and a yet-unknown ingredient, to lead to shrinking diameters.
\end{itemize}

Note that we will {\em not be assuming a community hierarchy on
nodes}, and so it is not enough to simply vary the \CGA\ model.

Based on this, we introduce the {\em Forest Fire Model}, which is
capable of producing all these properties. To set up this model, we
begin with some intuition that also underpinned \CGA: nodes arrive in
over time; each node has a ``center of gravity'' in some part of the
network; and its probability of linking to other nodes decreases
rapidly
with their distance from this center of gravity. However, we add to
this
picture the notion that, occasionally, a new node will produce a very
large number of out-links. Such nodes will help cause a more skewed
out-degree distribution; they will also serve as
``bridges'' that connect formerly disparate parts of the network,
bringing the diameter down.

\subsubsection{The Basic Forest Fire Model}

Following this plan, we now define the most basic version of the model.
Essentially, nodes arrive one at a time and form out-links to some
subset of the earlier nodes; to form out-links, a new node $v$ attaches
to a node $w$ in the existing graph, and then begins ``burning'' links
outward from $w$, linking with a certain probability to any new node it
discovers. One can view such a process as intuitively corresponding to
a
model by which an author of a paper identifies references to include in
the bibliography.  He or she finds a first paper to cite, chases a
subset of the references in this paper (modeled here as random), and
continues recursively with the papers discovered in this way. Depending
on the bibliographic aids being used in this process, it may also be
possible to chase back-links to papers that cite the paper under
consideration. Similar scenarios can be considered for social networks:
a new computer science (CS) graduate student arrives at a university, meets
some older CS students, who introduce him/her to their friends (CS or
non-CS), and the introductions may continue recursively.

We formalize this process as follows, obtaining the Forest Fire Model.
To begin with, we will need two parameters, a {\em forward burning
probability} $\fwdprob$, and a {\em backward burning ratio}
$\backratio$, whose roles will be described below. Consider a node $v$
joining the network at time $t > 1$, and let $G_t$ be the graph
constructed thus far.  ($G_1$ will consist of just a single node.) Node
$v$ forms out-links to nodes in $G_t$ according to the following
process.

\begin{itemize}
  \item[(i)] $v$ first chooses an {\em ambassador node} $w$ uniformly
  at
  random, and forms a link to $w$.
  \item[(ii)] We generate two random numbers: $x$ and $y$ that
  are geometrically distributed with means $\fwdprob/(1 - \fwdprob)$
  and $\backratio\fwdprob/(1 - \backratio\fwdprob)$ respectively.
  Node $v$ selects $x$ out-links and $y$ in-links of $w$ incident to nodes that were not yet visited. Let $w_1, w_2, \ldots, w_{x+y}$ denote
  the other ends of these selected links. If not enough in- or
  out-links are available, $v$ selects as many as it can.
  \item[(iii)] $v$ forms out-links to $w_1, w_2, \ldots, w_{x+y}$,
  and then applies step (ii) recursively to each of $w_1, w_2,
  \ldots, w_{x+y}$. As the process continues, nodes cannot be visited a
  second time, preventing the construction from cycling.
\end{itemize}

Thus, the ``burning'' of links in Forest Fire model begins at $w$,
spreads to $w_1, \ldots,$ $w_{x+y}$, and proceeds recursively until
it dies out. In terms of the intuition from citations in papers, the
author of a new paper $v$ initially consults $w$, follows a subset
of its references (potentially both forward and backward) to the
papers $w_1, \ldots, w_{x+y}$, and then continues accumulating
references recursively by consulting these papers. The key property
of this model is that certain nodes produce large
``conflagrations,'' burning many edges and hence forming many
out-links before the process ends.

Despite the fact that there is no explicit hierarchy in the Forest
Fire Model, as there was in \CGA, there are some subtle similarities
between the models. Where a node in \CGA\ was the child of a parent
in the hierarchy, a node $v$ in the Forest Fire Model also has an
``entry point'' via its chosen ambassador node $w$. Moreover, just
as the probability of linking to a node in \CGA\ decreased
exponentially in the tree distance, the probability that a new node
$v$ burns $k$ successive links so as to reach a node $u$ lying $k$
steps away is exponentially small in $k$. (Of course, in the Forest
Fire Model, there may be many paths that could be burned from $v$ to
$u$, adding some complexity to this analogy.)

In fact, our Forest Fire Model combines the flavors of several older
models, and produces graphs qualitatively matching their properties.
We establish this by simulation, as we describe below, but it is
also useful to provide some intuition for why these properties
arise.

\begin{itemize}
\item {\em Heavy-tailed in-degrees.} Our model has a ``rich get
richer'' flavor:  highly linked nodes can easily be reached by a
newcomer, no matter which ambassador it starts from.

\item {\em Communities.} The model also has a ``copying''
flavor: a newcomer copies several of the neighbors of his/her
ambassador (and then continues this recursively).

\item {\em Heavy-tailed out-degrees.}  The recursive nature of
link formation provides a reasonable chance for a new node to burn
many edges, and thus produce a large out-degree.

\item {\em \GPL.} A newcomer will have a lot of links near the
community of his/her ambassador; a few links beyond this, and
significantly fewer farther away. Intuitively, this is analogous to
the \CGA, although without an explicit set of communities.

\item {\em Shrinking diameter.} It is not a priori clear why the
Forest Fire Model should exhibit a shrinking diameter as it grows.
Graph densification is helpful in reducing the diameter, but it is
important to note that densification is certainly not enough on its
own to imply shrinking diameter.  For example, the \CGA \ model
obeys the \GPL, but our experiments also show that the diameter
slowly increases (not shown here for brevity).
\end{itemize}

\begin{figure}[!h]
\begin{center}
  \begin{tabular}{cc}
    \epsfig{file=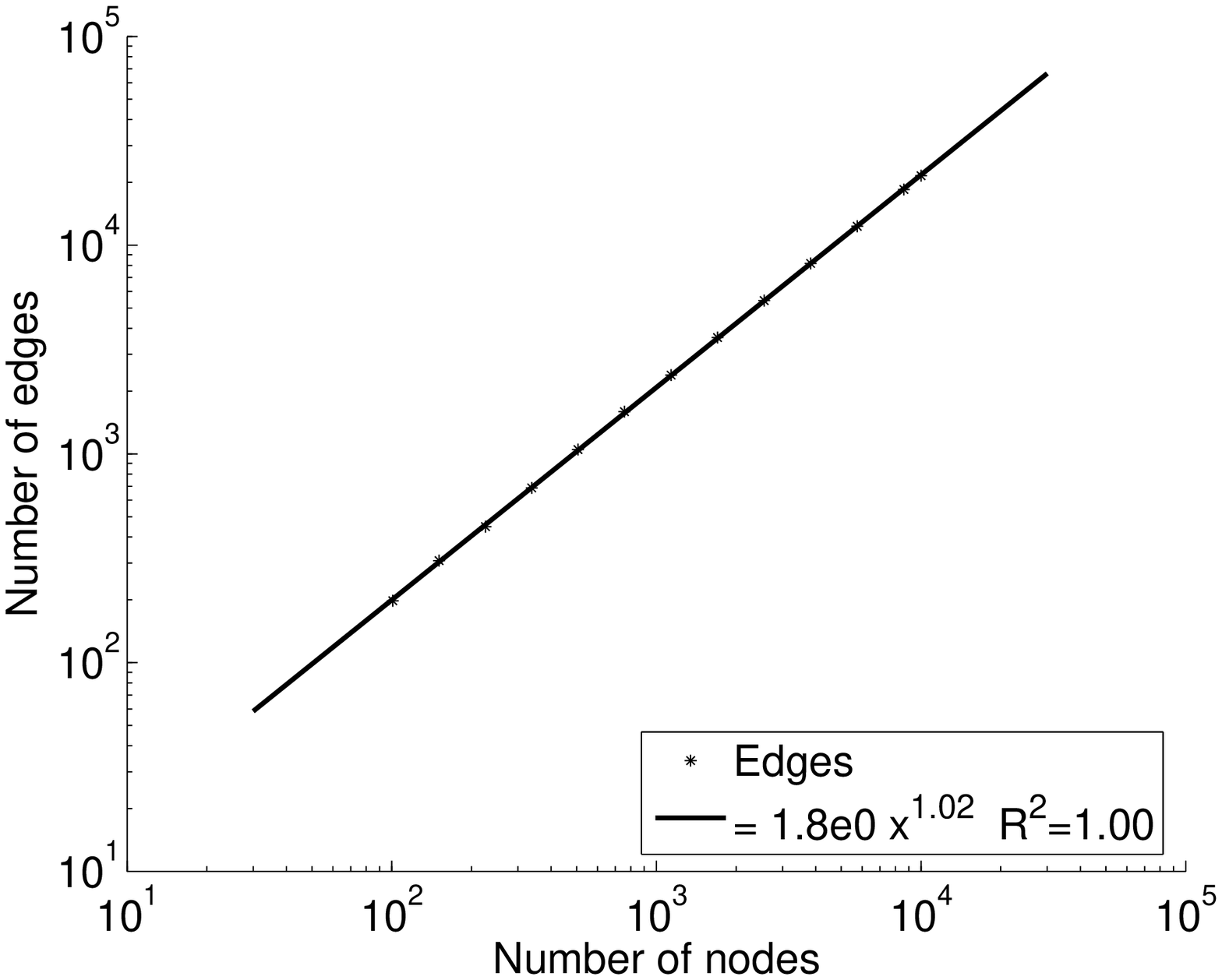 , width=2.0in} &
    \epsfig{file=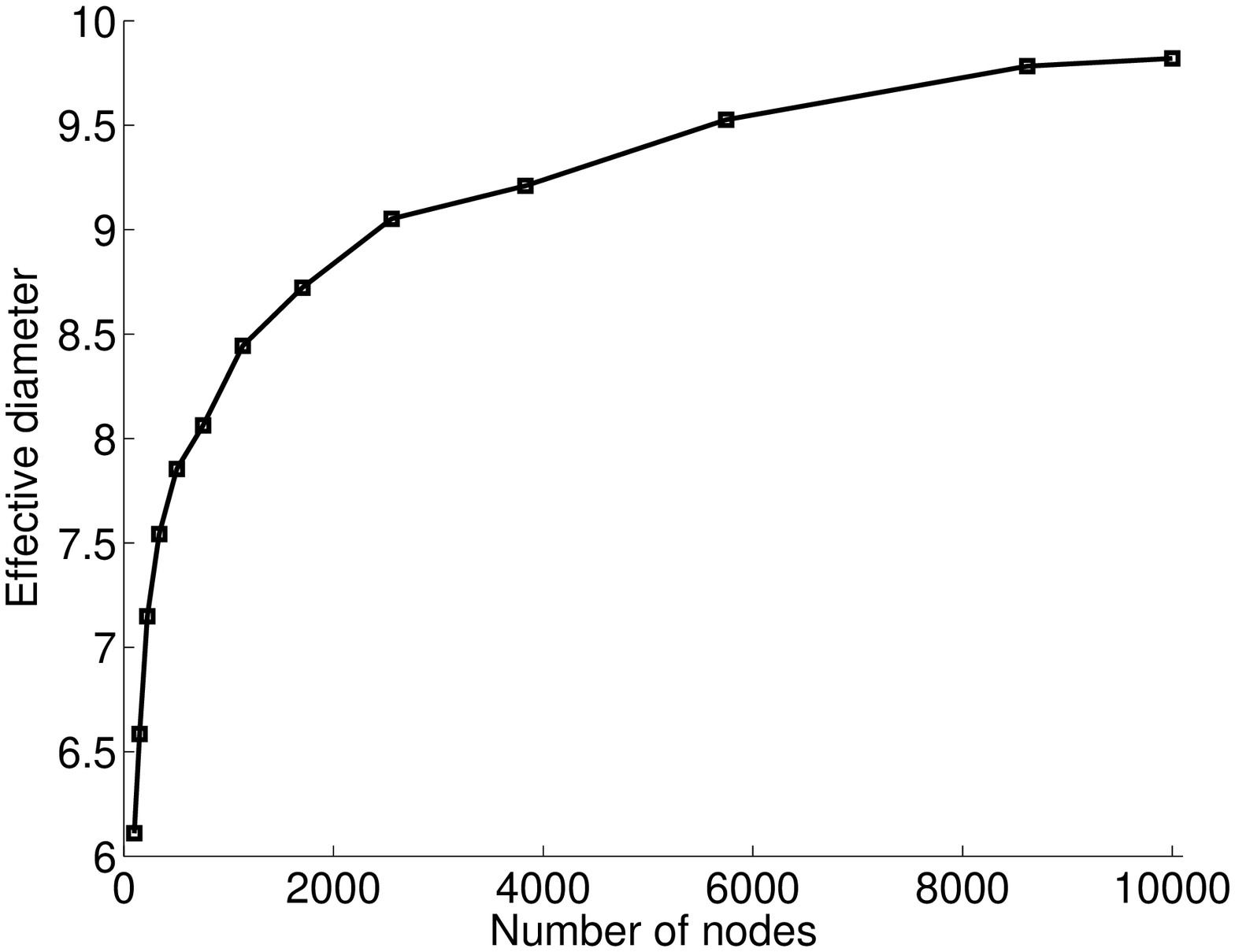, width=2.0in} \\
    \epsfig{file=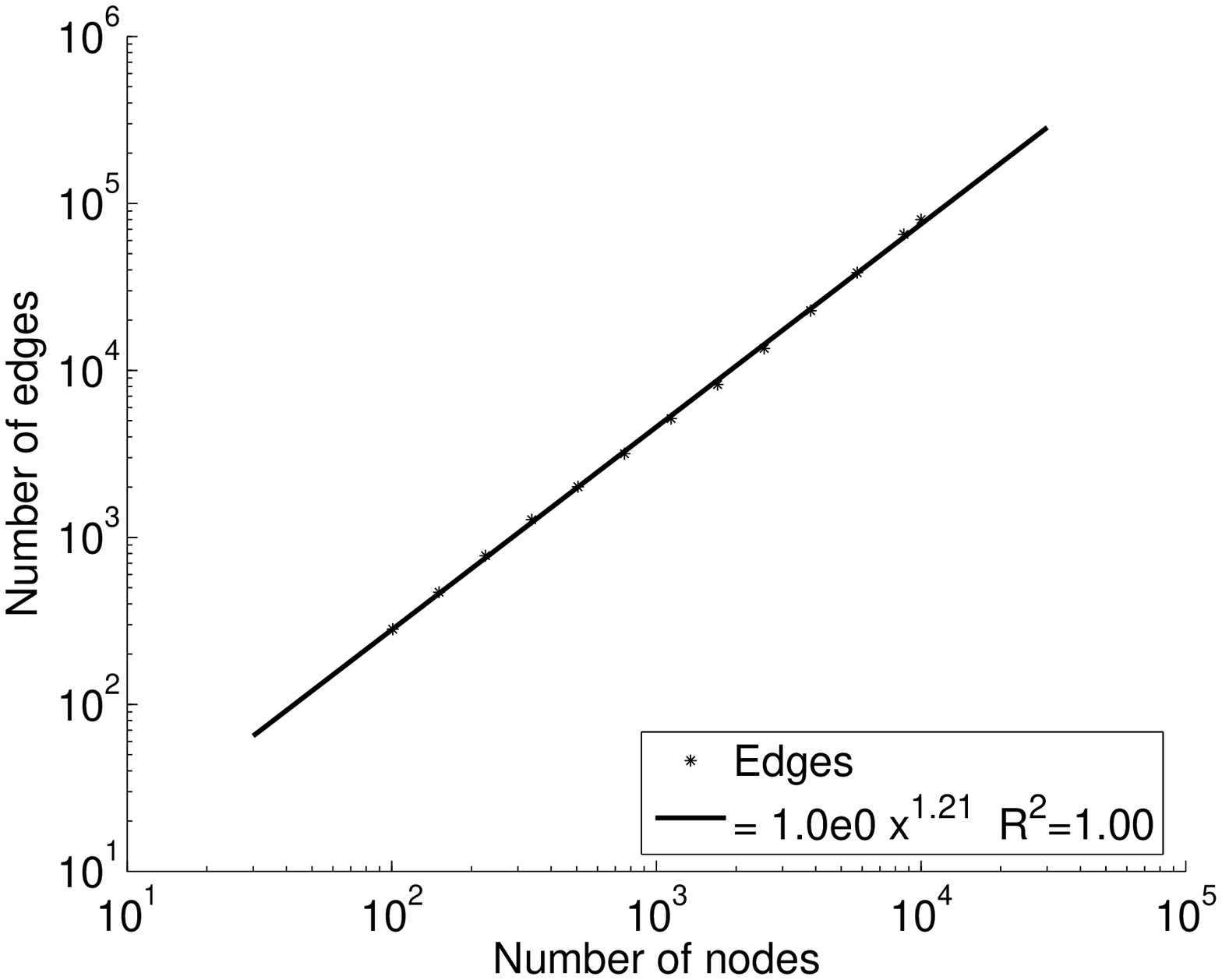 , width=2.0in} &
    \epsfig{file=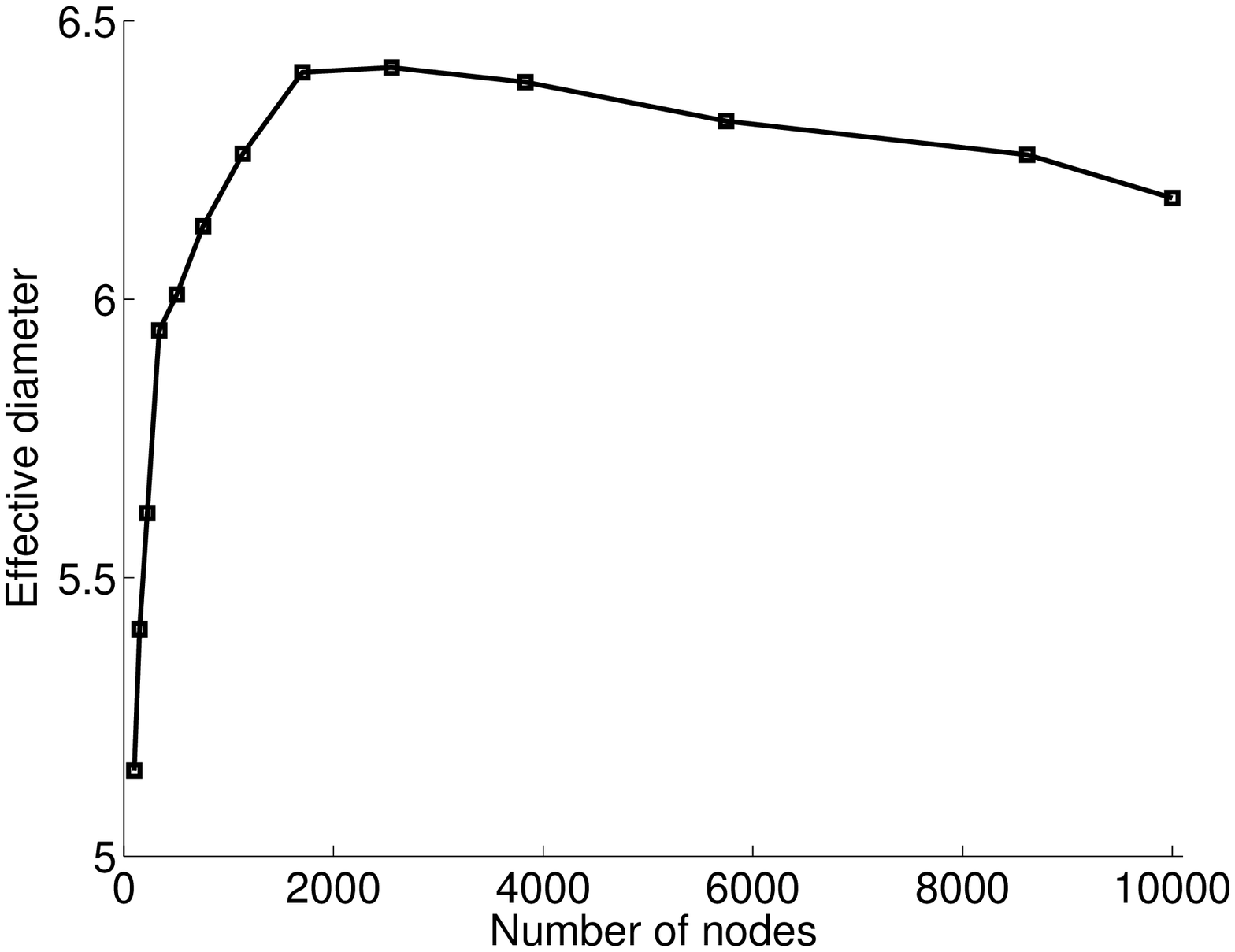, width=2.0in} \\
    \epsfig{file=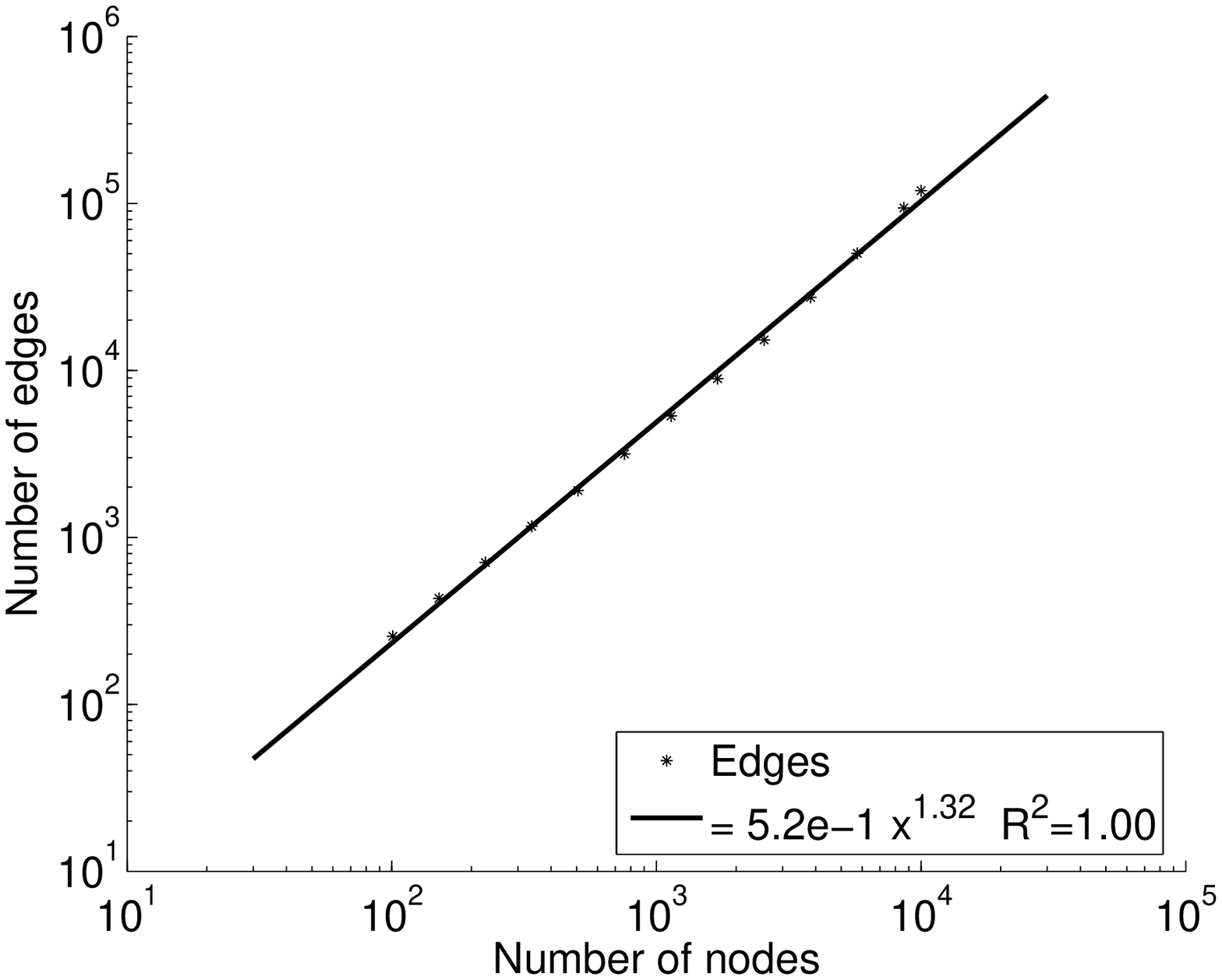 , width=2.0in} &
    \epsfig{file=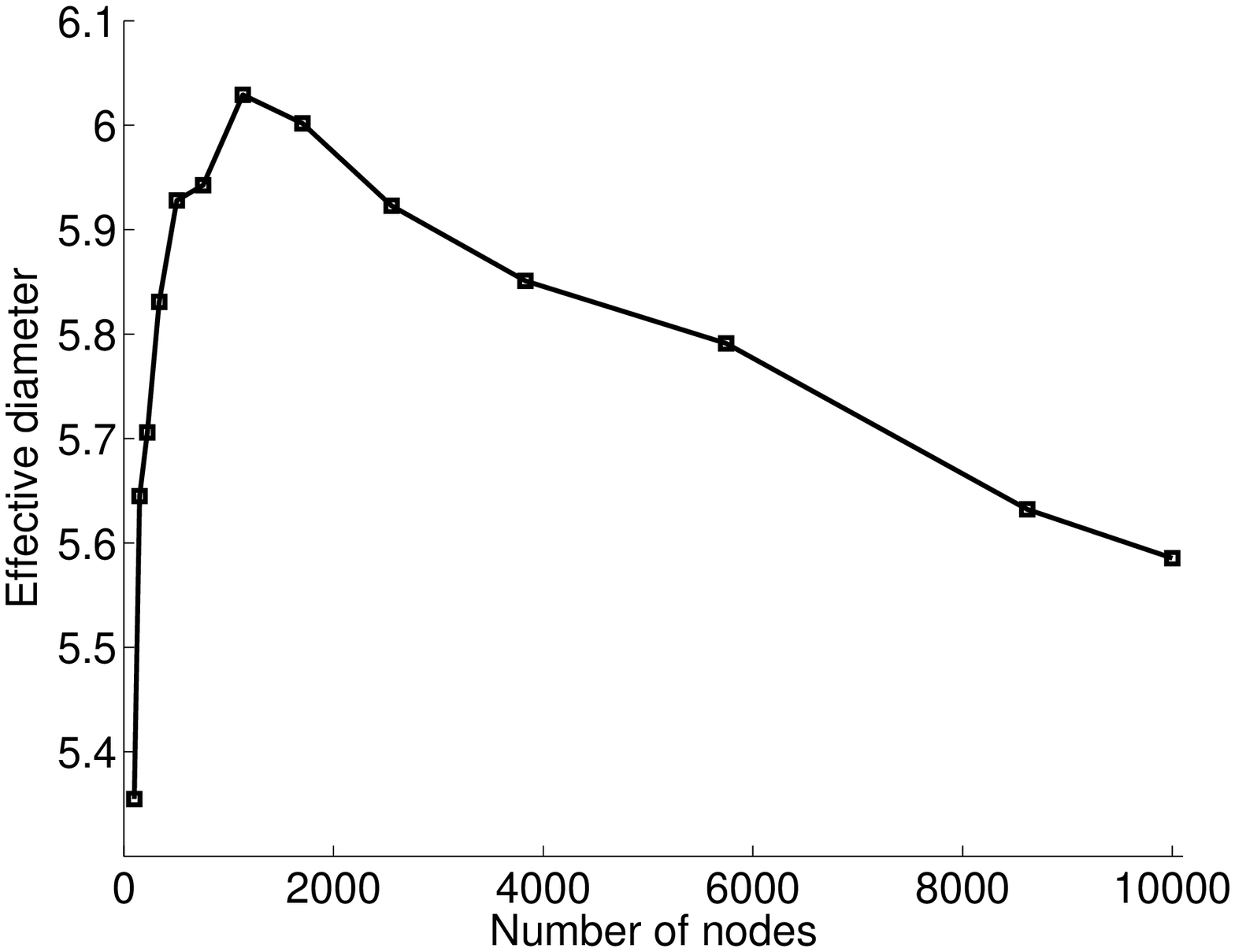, width=2.0in} \\
    \epsfig{file=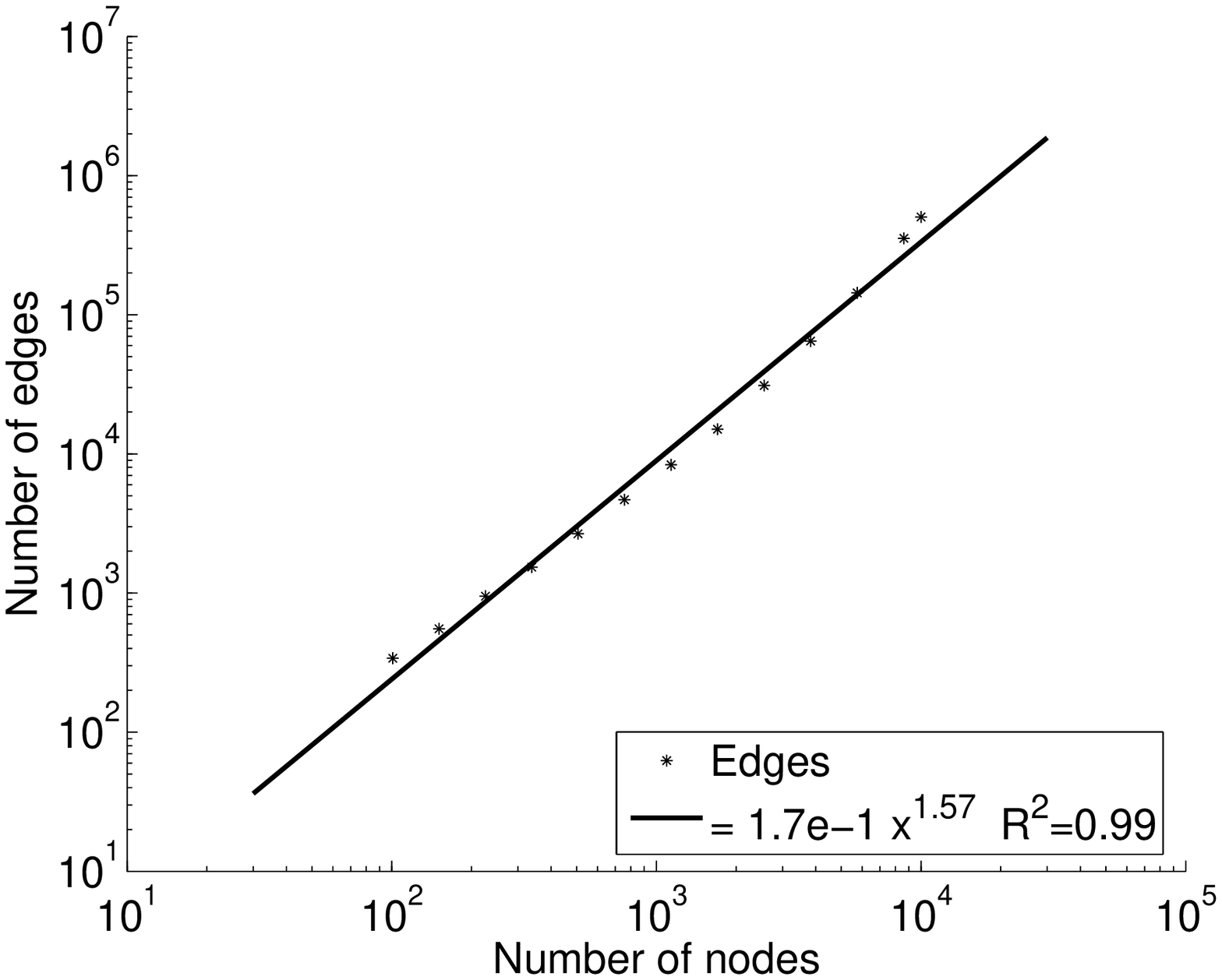 , width=2.0in} &
    \epsfig{file=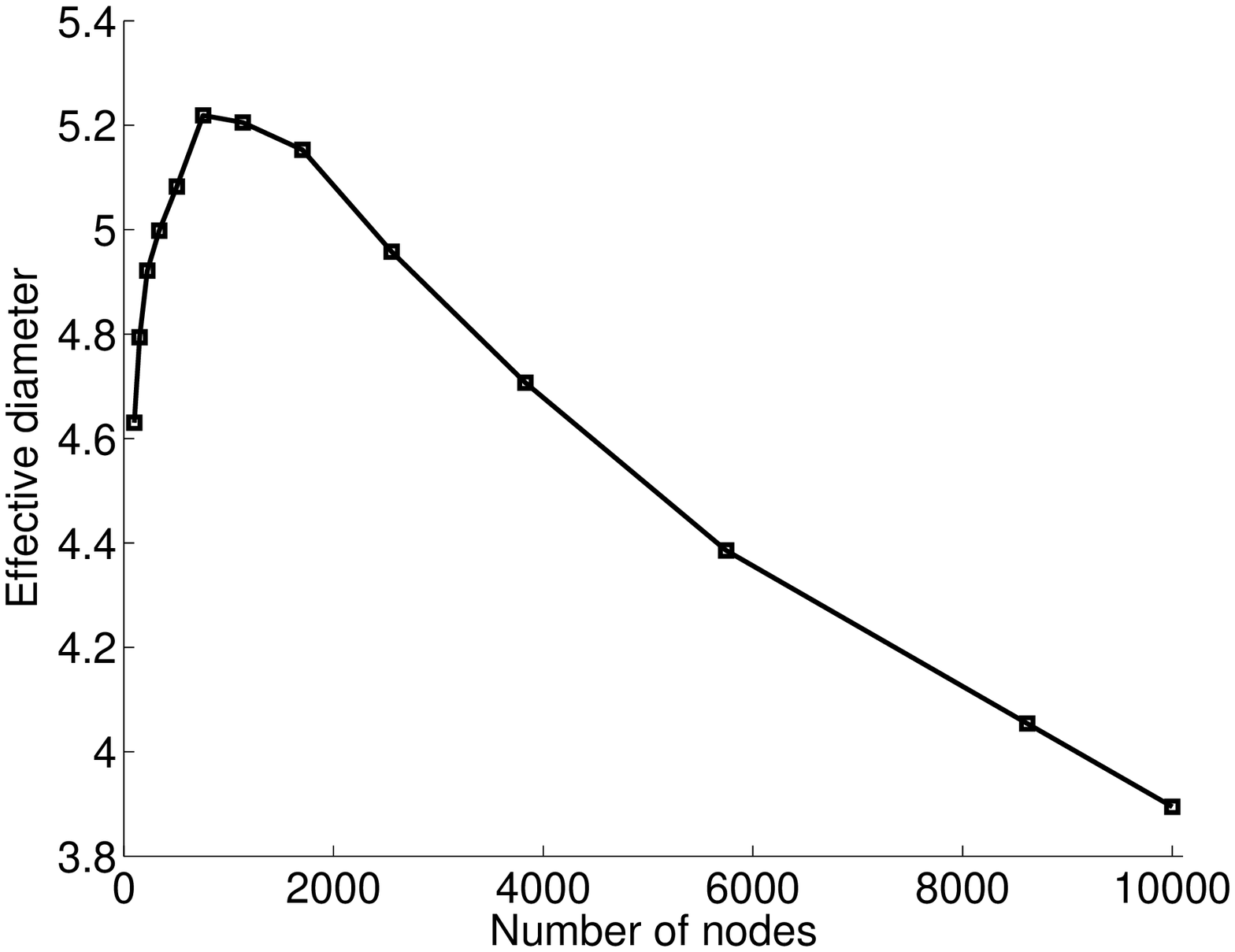, width=2.0in} \\
    \GPL & Effective diameter  \\
  \end{tabular}
  \caption{The \GPLplot\ and the diameter for Forest Fire model. Row 1:
  sparse graph ($a=1.01<2$), with increasing diameter (forward
  burning probability: $p= 0.35$, backward probability: $\bckprob=
  0.20$). Row 2: (most realistic case:) densifying graph
  ($a=1.21<2$) with slowly decreasing diameter ($p= 0.37$,
  $\bckprob=0.32$). Row 3: densifying graph (($a=1.32<2$) with
  decreasing
  diameter ($p= 0.37$, $\bckprob= 0.33$). Row 4: dense graph with
  densification exponent close to 2 ($a=1.57$) and decreasing
  diameter ($p= 0.38$, $\bckprob= 0.35$).} \label{fig:gplDiam3538}
  \vspace{-0.4in}
\end{center}
\end{figure}

\begin{figure}[t]
\begin{center}
  \begin{tabular}{cc}
    \epsfig{file=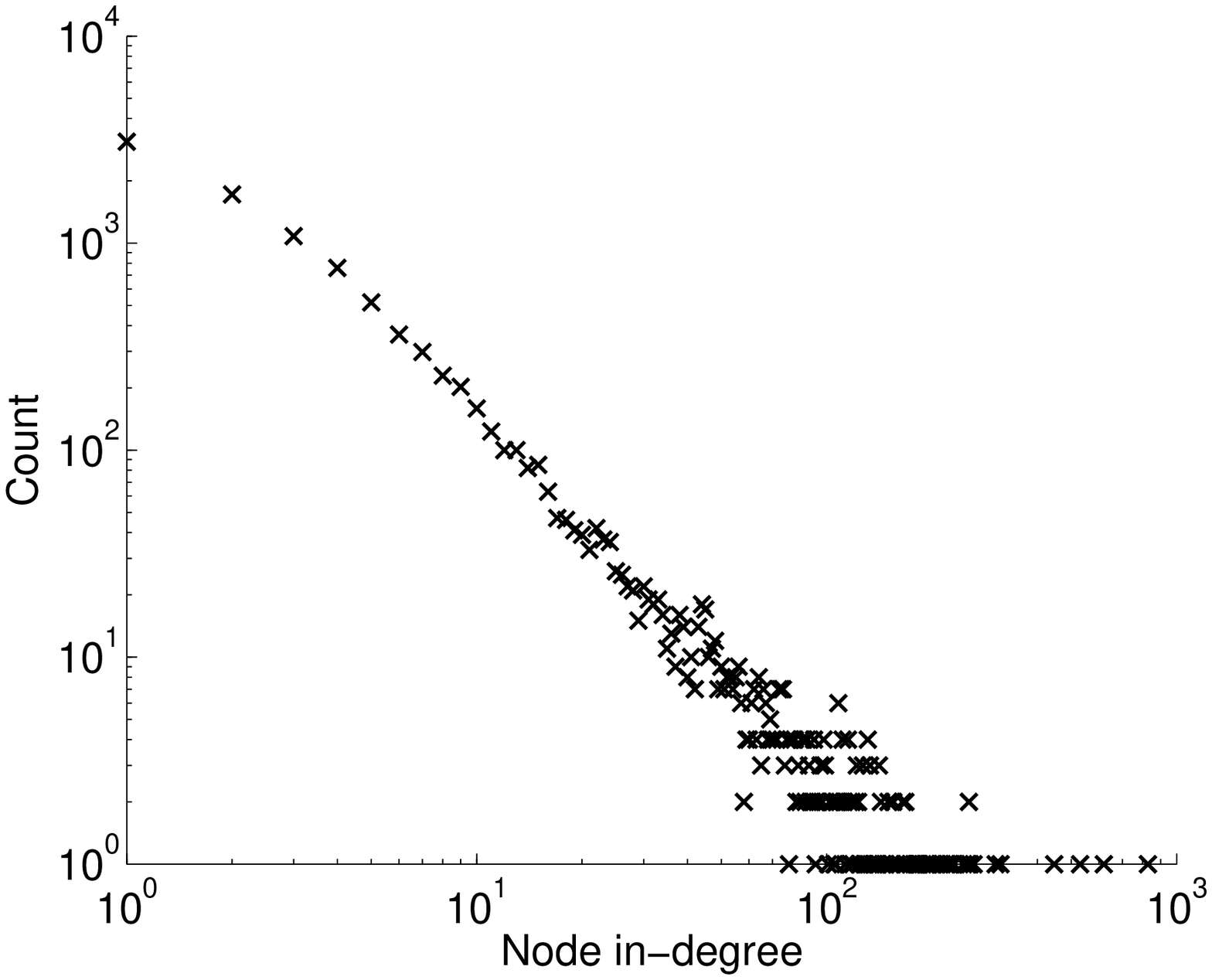, width=2.2in} &
    \epsfig{file=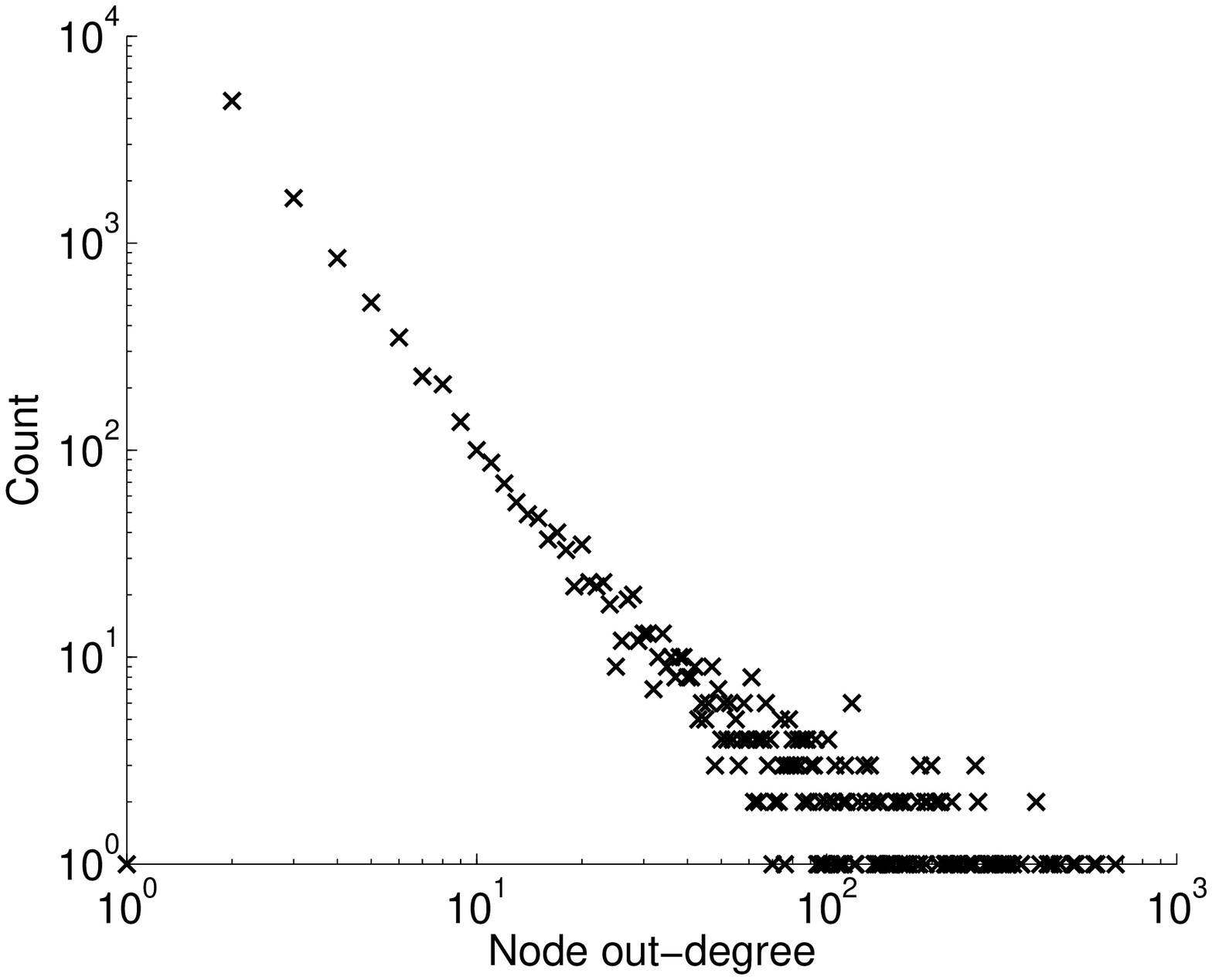, width=2.2in} \\
    In-degree & Out-degree  \\
  \end{tabular}
  \caption{Degree distribution of a sparse graph with decreasing
  diameter
  (forward burning probability: 0.37, backward probability: 0.32).}
  \label{fig:inOutDeg}
\end{center}
\end{figure}

Rigorous analysis of the Forest Fire Model appears to be quite
difficult. However, in simulations, we find that by varying just the
two parameters $\fwdprob$ and $\backratio$, we can produce graphs
that densify ($a > 1 $), exhibit heavy-tailed distributions for both
in- and out-degrees (Fig.~\ref{fig:inOutDeg}), and have diameters
that decrease. This is illustrated in Figure~\ref{fig:gplDiam3538},
which shows plots for the effective diameter and the Densification
Power Law exponent as a function of the number of nodes for some
selections of $p$ and $r$.

We see that depending on the forward and backward burning parameters
the Forest Fire Model is capable of generating sparse or dense
graphs with effective diameters that either increase or decrease,
while also producing power-law in- and out-degree distributions
(figure~\ref{fig:inOutDeg}). Informally, a dense graph has close to
a linear number of edges incident to each node, while a sparse graph
has significantly fewer than a linear number of edges incident to
each node.

Also notice the high sensitivity of the parameter space. We fix the
forward burning probability $p$, and by increasing the backward
burning probability $p_b$ for only a few percent we move from an
increasing to a slowly and then to more rapidly decreasing effective
diameter (figure~\ref{fig:gplDiam3538}).

\begin{figure}[t]
\begin{center}
  \begin{tabular}{cc}
    \epsfig{file=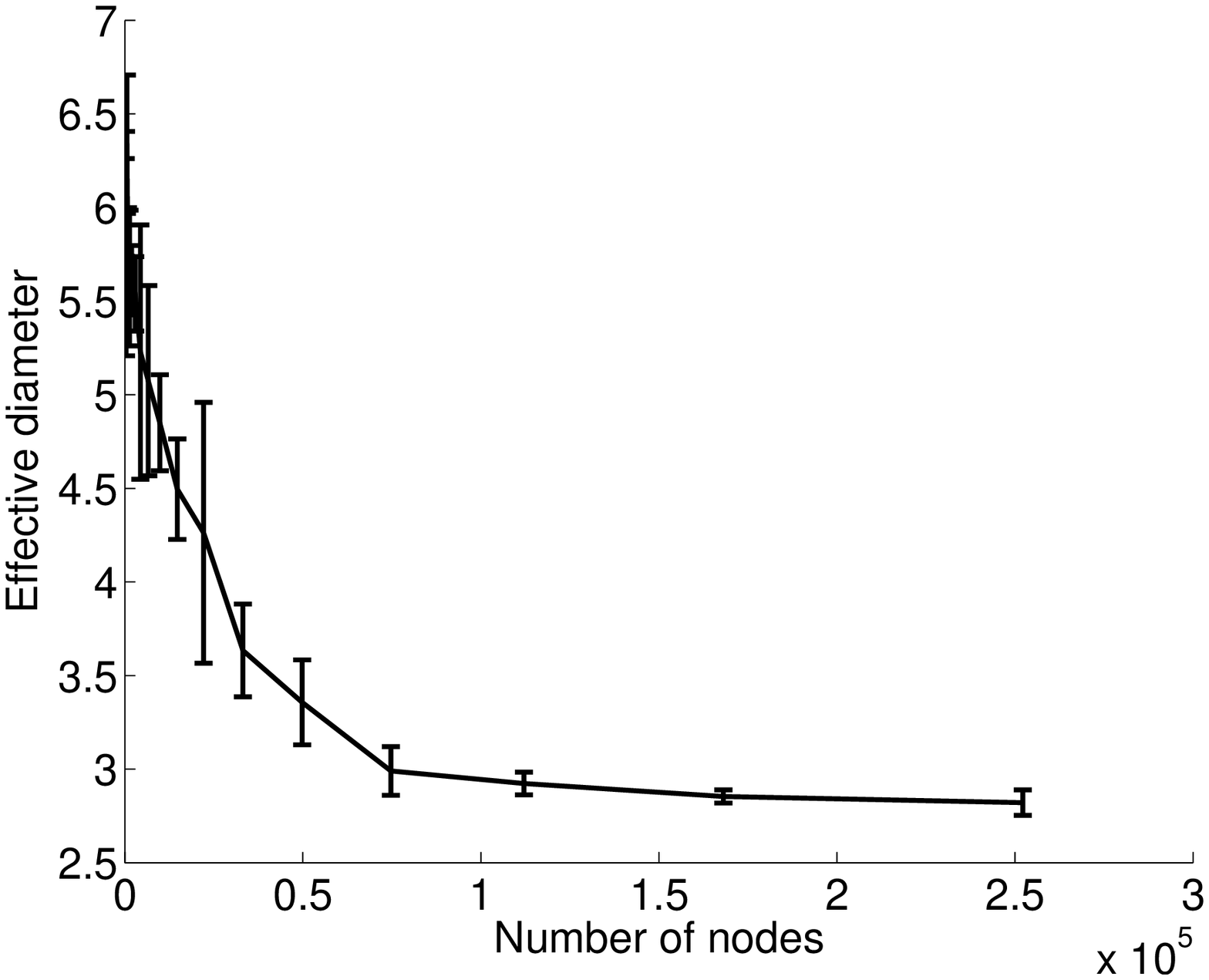, width=2.2in} &
    \epsfig{file=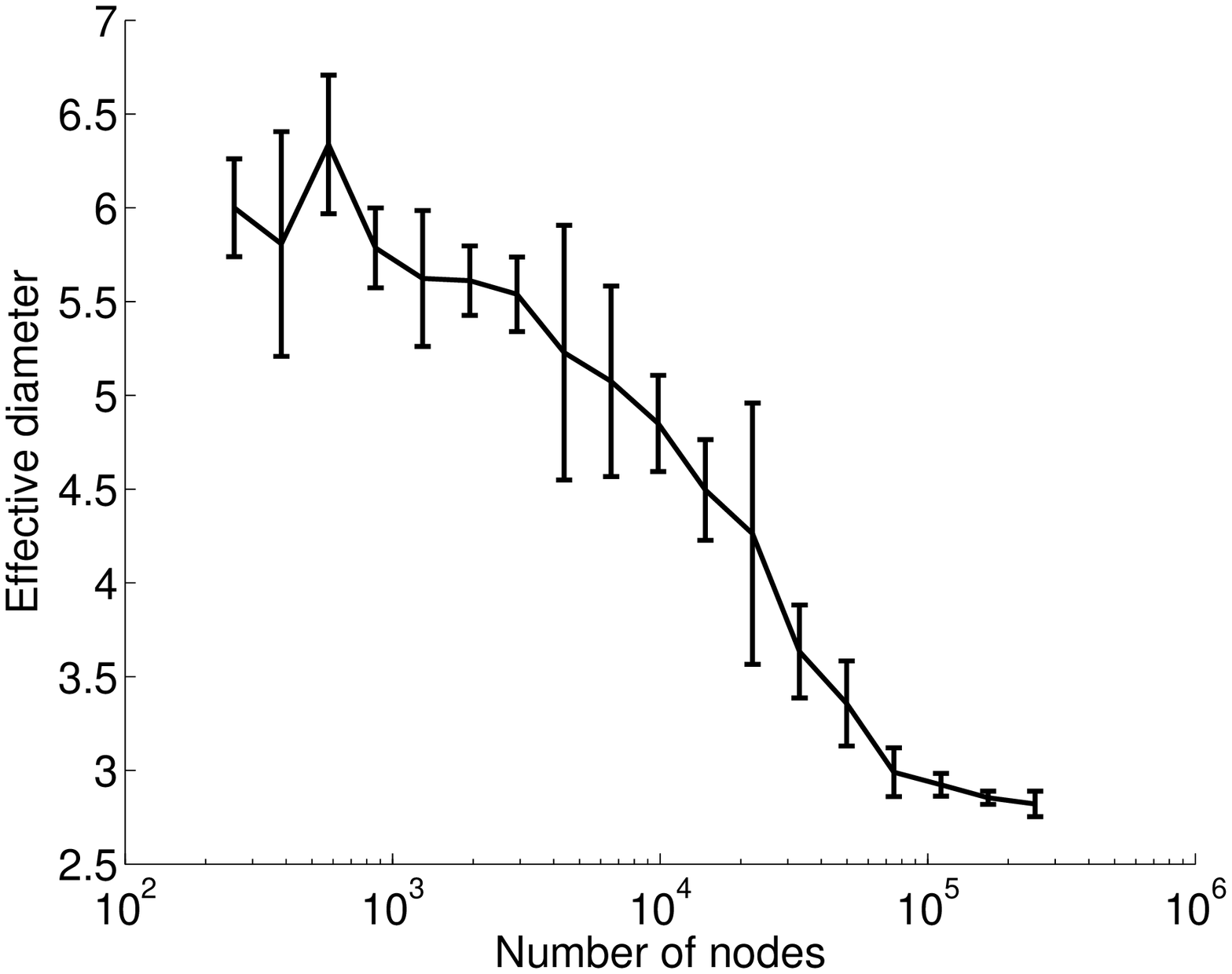, width=2.2in} \\
    (a) Effective diameter & (b) Effective diameter (log nodes) \\
    $p=0.37$, $p_b=0.34$ & $p=0.37$, $p_b=0.34$ \\
  \end{tabular}
  \caption{Evolution of effective diameter of Forest Fire
  model while generating a large graph. Both
  plots show the same data; left one plots on linear scales and the
  right one plots on log-linear scales (effective diameter vs. log
  number of nodes). Error bars show the confidence interval of the
  estimated effective diameter. Notice that the effective diameter
  shrinks and then slowly converges.}
  \label{fig:ffDiam}
\end{center}
\end{figure}

Figure~\ref{fig:ffDiam} plots the evolution of the effective
diameter of Forest Fire. We generated a single large graph on
$250,000$ nodes and measured the effective diameter over time. Error
bars present 1 standard deviation of the estimated effective
diameter over 10 runs. Both plots show the same data. The left figure plots the
number of nodes on linear while the right plots the log number of
nodes. Notice the convergence of the effective diameter. At first it
shrinks more rapidly and then slowly converges to a low value.

\subsubsection{Extensions to the Forest Fire Model}

Our basic version of the Forest Fire Model exhibits rich structure
with just two parameters. By extending the model in natural ways, we
can fit observed network data even more closely. We propose two
natural extensions: ``{\em orphans}'' and {multiple ambassadors}.

``{\em Orphans}'': In both the patent and arXiv citation graphs,
there are many isolated nodes, that is, documents with no citations
into the corpus. For example, many papers in the arXiv only cite
non-arXiv papers. We refer to them as {\em orphans}. Our basic model
does not produce orphans, since each node always links at least to
its chosen ambassador. However, it is easy to incorporate orphans
into the model in two different ways. We can start our graphs with
$n_0 > 1$ nodes at time $t=1$; or we can have some probability $q>0$
that a newcomer will form no links (not even to its ambassador) and
so become an orphan.

We find that such variants of the model have a more pronounced
decrease in the effective diameter over time, with large distances
caused by groups of nodes linking to different orphans gradually
diminishing as further nodes arrive to connect them together.

{\em Multiple ambassadors}: We experimented with allowing newcomers
to choose more than one ambassador with some positive probability.
That is, rather than burning links starting from just one node,
there is some probability that a newly arriving node burns links
starting from two or more. This extension also accentuates the
decrease in effective diameter over time, as nodes linking to
multiple ambassadors serve to bring together formerly far-apart
parts of the graph.

{\em Burning a fixed percentage of neighbors:} We also considered a
version of Forest Fire where the fire burns a fixed percentage of
node's edges, i.e. the number of burned edges is proportional to the
node's degree. When a fire comes into a node, for each unburned
neighbor we {\em independently} flip a biased coin, and the fire
spreads to nodes where the coin came up heads. This process continues recursively until no nodes are burned. In case of forward
and backward burning probabilities we have two coins, one for out-
and one for in-edges.

The problem with this version of the model is that, once there is a
single large fire that burns a large fraction of the graph, many
subsequent fires will also burn much of the graph.  This results in
a bell-shaped, non-heavy-tailed degree distribution and gives two
regimes of densification --- slower densification before the first
big fire, and quadratic ($a$ = 2) densification afterwards.

We also experimented with the model where burning probability
decayed exponentially as the fire moves away from the ambassador
node.

\subsubsection{Phase plot}

In order to understand the densification and the diameter properties
of graphs produced by the Forest Fire Model, we explored the full
parameter space of the basic model in terms of the two underlying
parameters: the forward burning probability $\fwdprob$ and the
backward burning ratio $\backratio$.

Note, there are two equivalent ways to parameterize the Forest Fire
model. We can use the forward burning probability $\fwdprob$ and the
backward burning ratio $\backratio$; or the forward burning
probability $\fwdprob$ and the backward burning probability
$\bckprob$ ($\bckprob = \backratio \fwdprob$). We examine both and
show two cuts through the parameter space.

\begin{figure}[!tp]
\begin{center}
  \begin{tabular}{cc}
    \epsfig{file=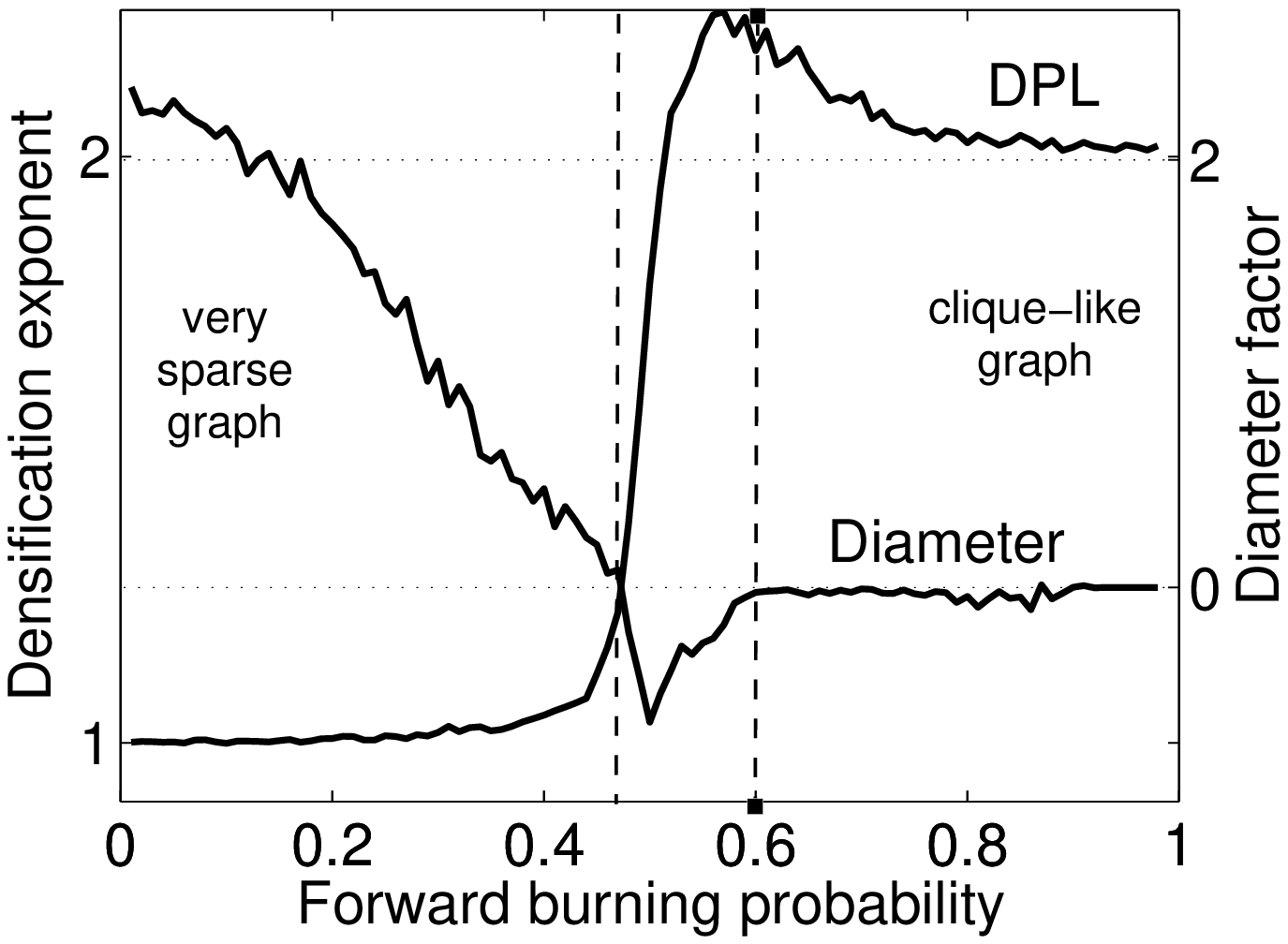 , width=2.4in} &
    \epsfig{file=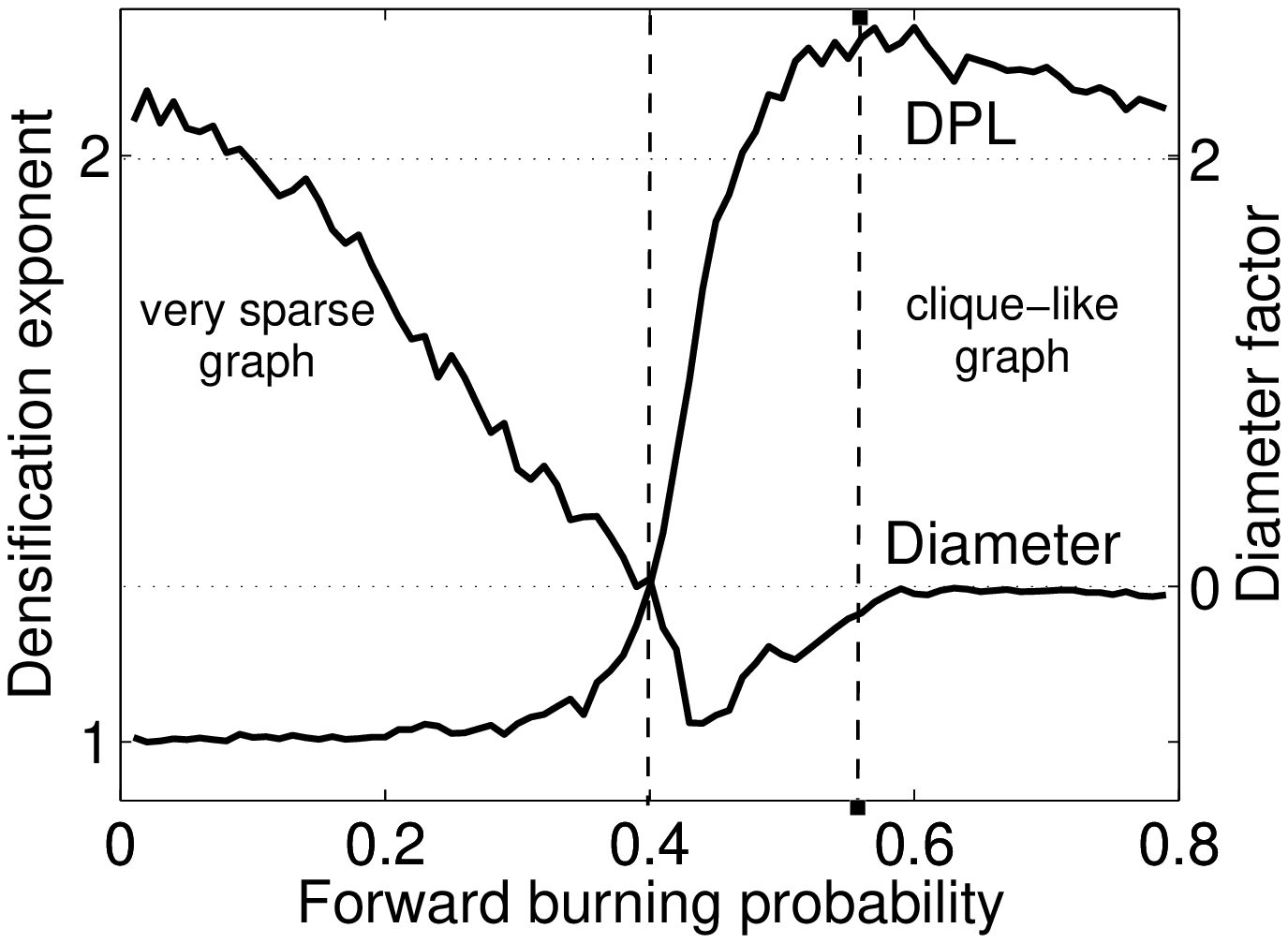, width=2.4in} \\
    (a) We fix burning ratio, $r=0.5$ &
    (b) We fix backward burning prob., $p_b=0.3$ \\
    and vary forward burning probability $p$ &
    and vary forward burning probability $p$ \\
  \end{tabular}
  \caption{We vary the forward burning probability while fixing
  burning ratio (a) or backward burning probability (a). The plot
  gives a very precise cut through Forest Fire parameter space.
  Notice that each plot has {\em two} vertical axes: DPL exponent on
  the left, and the diameter log-fit factor on the right. Observe a
  very sharp transition in DPL exponent and a narrow region,
  indicated by vertical dashed lines, where Forest Fire produces
  slowly densifying graphs with decreasing effective diameter.}
  \label{fig:phaseCut}
\end{center}
\end{figure}

Figure~\ref{fig:phaseCut} shows how the densification exponent and
the effective diameter depend on forward burning probability
$\fwdprob$. In the left plot of figure~\ref{fig:phaseCut} we fix the
backward burning probability $\bckprob=0.3$, and in the right plot
we fix the backward burning ratio $\backratio=0.5$. We vary forward
burning probability, and plot the \GPL\ exponent. The densification
exponent $a$ is computed as in Section~\ref{sec:observations}, by
fitting a relation of the form $e(t) \propto n(t)^a$. Notice the
very sharp transition between the regimes with no densification and
those with very high densification.

On the same plot we also show the {\em Effective diameter log-fit
factor} $\alpha$. We fit a logarithmic function of the form
$diameter = \alpha \log t + \beta$ (where $t$ is the current time,
and hence the current number of vertices) to the last half of the
effective diameter plot; we then report the factor $\alpha$. Thus,
Diameter Factor $\alpha < 0$ corresponds to decreasing effective
diameter over time, and $\alpha > 0$ corresponds to increasing
effective diameter.

Going back to Figure~\ref{fig:phaseCut}, notice that at low values
of forward burning probability $\fwdprob$, we observe increasing
effective diameter and no densification ($a=1$). As $\fwdprob$
increases, the effective diameter grows slower and slower. For a
narrow band of $\fwdprob$ we observe {\em decreasing effective
diameter}, negative $\alpha$ (the small valley around
$\fwdprob=0.45$). With high values of $\fwdprob$ the effective
diameter is constant ($\alpha \approx 0$), which means that the
generated graph is effectively a clique with effective diameter
close to 1 and DPL exponent $a \approx 2$. Also notice that the
sharp transition in the DPL exponent and the decreasing effective
diameter are very well aligned.

This simulations indicate that even the basic Forest Fire Model is
able to produce sparse and slowly densifying (with densification
exponent near $1$) graphs in which the effective diameter decreases.

\begin{figure}[!tp]
\begin{center}
  \begin{tabular}{cc}
    \epsfig{file=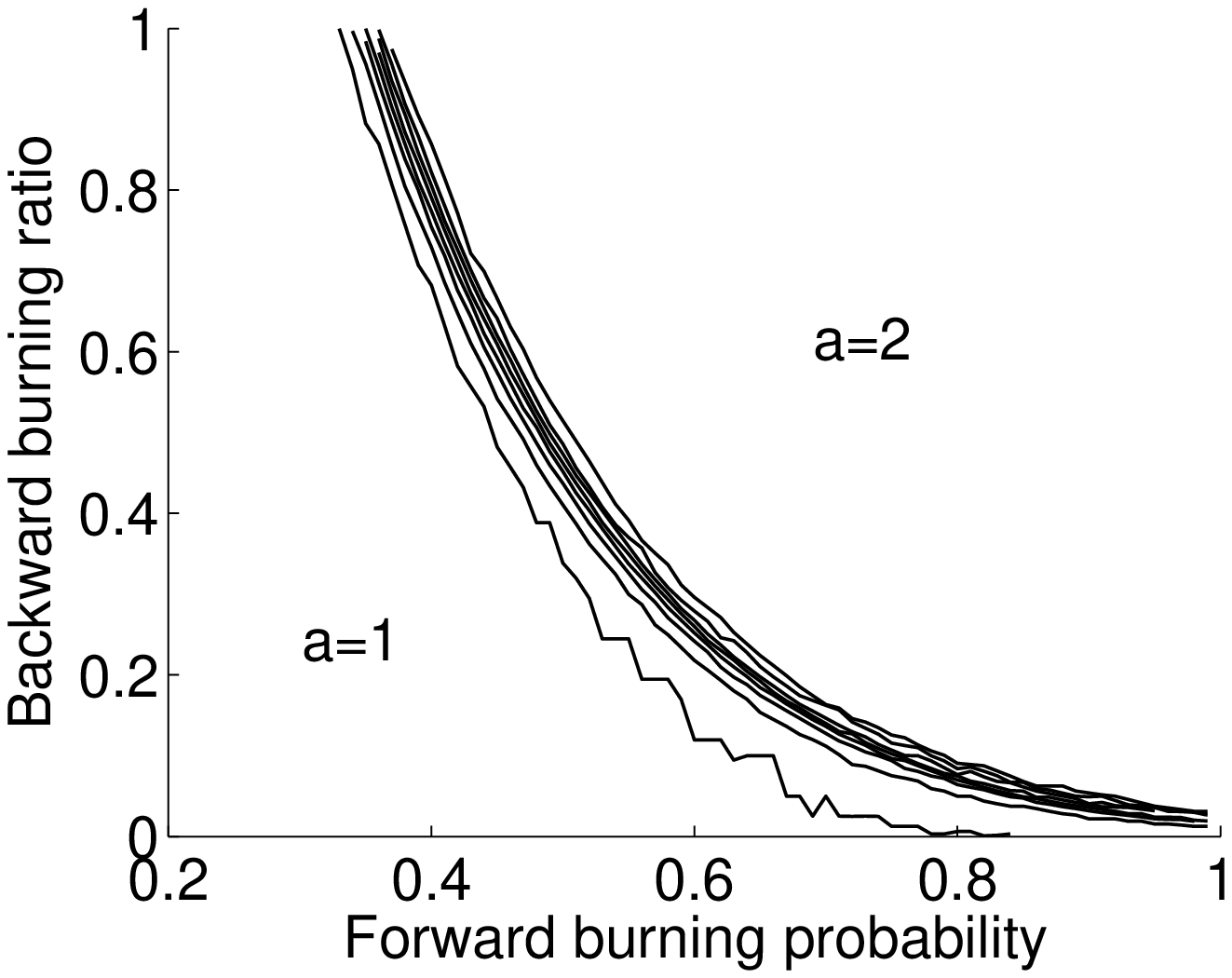, width=2.4in} &
    \epsfig{file=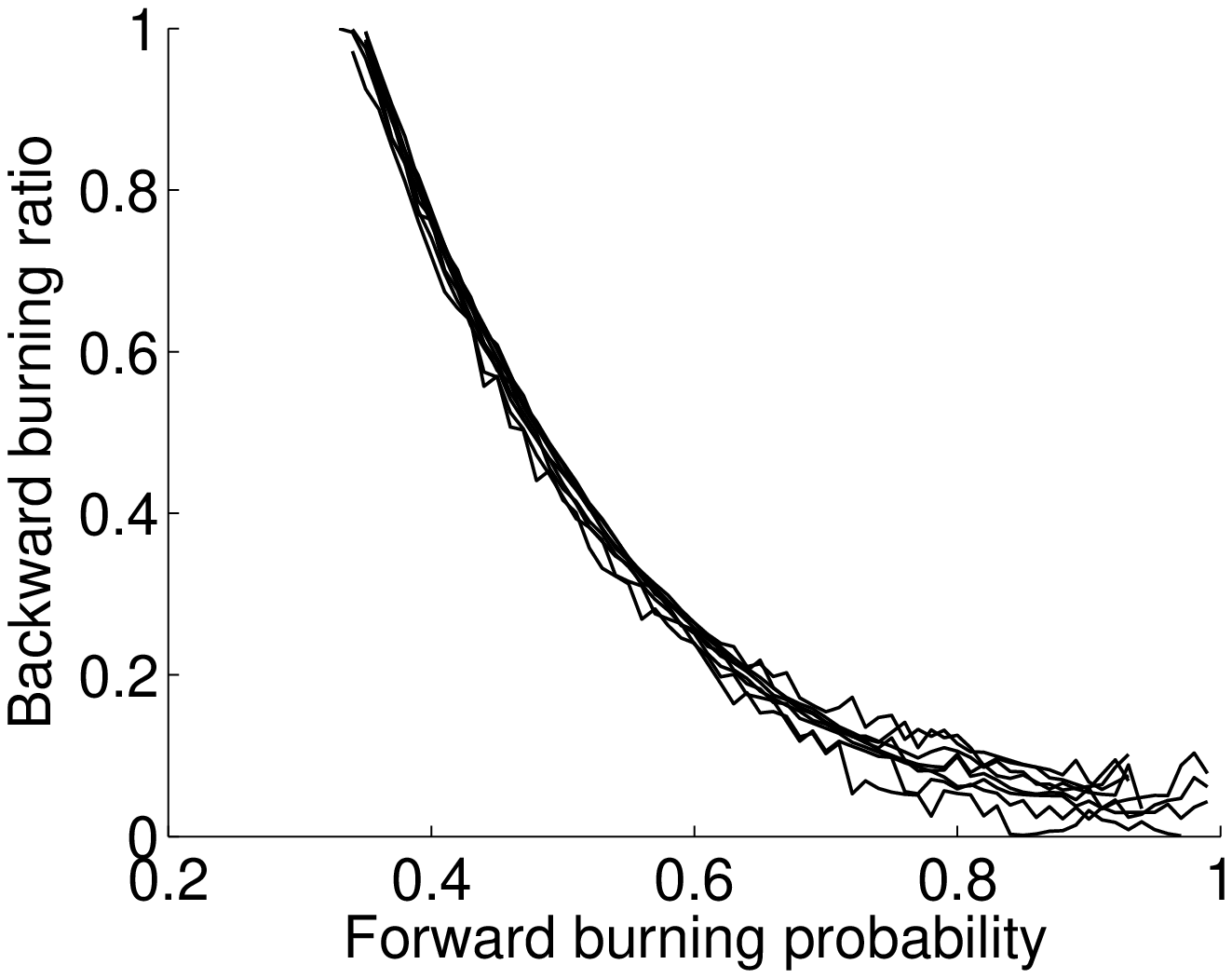, width=2.4in} \\
     (a) Densification exponent & (b) Diameter log-fit factor \\
  \end{tabular}
  \caption{Contour plots: The \GPL\ exponent $a$ (left) and the
  effective diameter log-fit factor $\alpha$ (right) over the
  parameter space (forward-burning probability and backward burning
  ratio) of the Forest Fire model.}
  \label{fig:phase}
\end{center}
\end{figure}

Figure~\ref{fig:phase} shows how the densification exponent and the
effective diameter depend on the values of the Forest Fire
parameters $\fwdprob$ and $\backratio$.

Figure~\ref{fig:phase}(a) gives the contour plot of the
densification exponent $a$. The lower left part corresponds to $a=1$
(the graph maintains constant average degree), and in the upper
right part $a=2$ -- the graph is ``dense'', that is, the number of
edges grows quadratically with the number of nodes, as, e.g., in the
case of a clique. The contours in-between correspond to $0.1$
increase in DPL exponent: the left-most contour corresponds to
$a=1.1$ and the right-most contour corresponds to $a=1.9$ The
desirable region is in-between; we observe that it is very narrow:
$a$ increases dramatically along a contour line, suggesting a sharp
transition.

Figure~\ref{fig:phase}(b) gives the contour plot for the Effective
diameter log-fit factor $\alpha$ as defined above. Each contour
correspond to diameter factor $\alpha$. We vary $\alpha$ in range
$-0.3 \le \alpha \le 0.1$, with step-size $0.05$. Notice, the
boundary in parameter space between decreasing and increasing effective diameter is very narrow.

\begin{figure}[!tp]
\begin{center}
  \begin{tabular}{cc}
    \epsfig{file=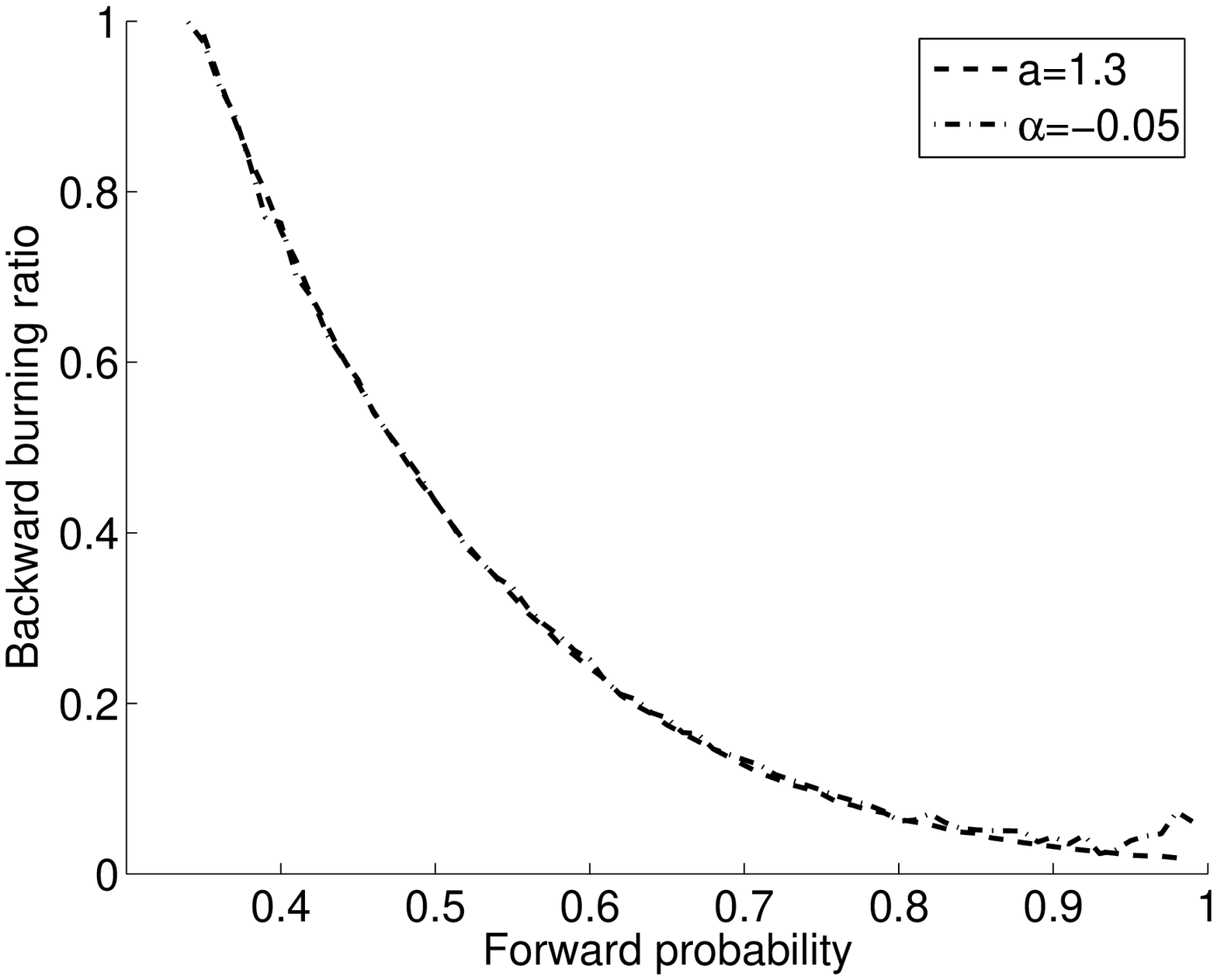, width=2.2in} &
    \epsfig{file=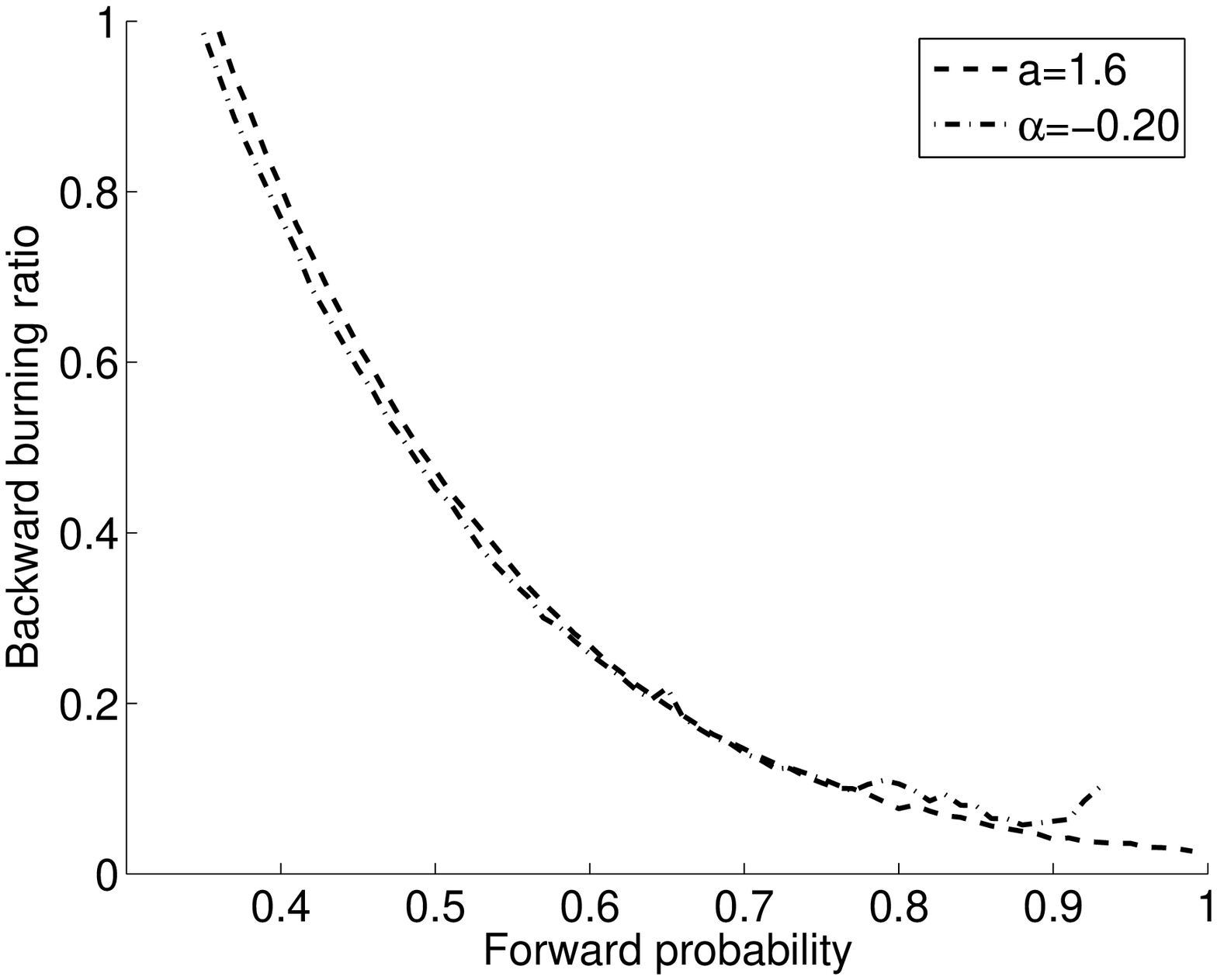, width=2.2in} \\
    (a) Densification exponent $a=1.3$ & (b) Densification exponent
    $a=1.6$ \\
     Diameter factor $\alpha=-0.05$ & Diameter factor $\alpha=-0.20$ \\
  \end{tabular}
  \caption{We superimpose the \GPL\ exponent $a$ and diameter
  log-fit $\alpha$ factor over the Forest Fire Model parameter space.
  Notice that the shape of transition boundary of the densification
  and the shrinking diameter very much follow the same shape.}
  \label{fig:phImpose}
\end{center}
\end{figure}

Do contour plots of \GPL\ and Shrinking Diameters from
Figure~\ref{fig:phase} follow the same shape? More exactly, does the
boundary between decreasing and increasing diameters follow the same
shape as the transition in the densification exponent?

We answer this question on figure~\ref{fig:phImpose}, where we
superimpose phase contours of DPL and the effective diameter over
the Forest Fire parameter space. The left plot superimposes phase
contours for the \GPL\ exponent $a=1.3$ and the diameter log-fit
factor $\alpha=-0.05$. The right plot superimposes contours for
$a=1.6$ and $\alpha=-0.30$. In both cases we observe very good
alignment of the two phase lines which suggests the same shape of
the transition boundary for the \GPL\ exponent and the Effective
Diameter.

We also observe similar behavior with orphans and multiple
ambassadors. These additional features in the model help further
separate the diameter decrease/increase boundary from the
densification transition, and so widen the region of parameter space
for which the model produces reasonably sparse graphs with
decreasing effective diameters.

\section{Densification and the degree distribution over time}
\label{sec:ddOverTm}

Many real world graphs exhibit power-law degree
distributions~\cite{barabasi99emergence,faloutsos99powerlaw}. As we
saw in section~\ref{sec:observations} the average degree increases
over time, and the graphs densify following the power-law relationship
between the number of nodes and the number of edges. Here we analyze
the relation between the densification and the power-law degree
distribution over time, and find evidence that some of the real world
graphs obey the relations we find. A similar analysis was also
performed by Dorogovtsev and Mendes~\cite{dorogovtsev02growth}
although without specific measurements or comparison to empirical
data.

We analyze the following two cases: If the degree distribution of a time evolving graph is power-law, and it maintains {\em constant} power-law exponent $\ddslope$ over time, then we show that for $1 < \ddslope < 2$
\GPL\ with exponent

\[
a = 2/\ddslope.
\]

arises. In this case the \GPL\ is the consequence of the fact that a power-law distribution with exponent $\ddslope < 2$ has no finite expectation~\cite{newman05power}, and thus the average degree grows as degree exponent is constant.

Our second result is for the case when temporally evolving graph
densifies with densification exponent $a$, and follows a power-law
degree distribution with exponent $\ddslope > 2$ that we alow to {\em change} over time. We show that in this case for a given densification exponent $a$, the
power-law degree exponent $\ddslope_n$ has to evolve with the size of the graph $n$ as

\[
\ddslope_n = \frac{4 n^{a-1} - 1}{2 n^{a-1} - 1}
\]

This shows that \GPL\ and the degree distribution are related and
that one implies the other.

\subsection{Constant degree exponent over time}

First, we analyze the case where the graph over time maintains
power-law degree distribution with a constant exponent
$\ddslope$. Power law distribution $p(x) = cx^{-\ddslope}$ with
exponent $\ddslope < 2$ has infinite
expectation~\cite{newman05power}, i.e. as the number of samples
increases, the mean also increases. Assuming that the exponent
(slope) of the degree distribution does {\em not change over time},
a natural question to ask is: {\em what is the relation between the
\GPL\ exponent and the degree distribution over time?}  The
following theorem answers the question:

\begin{theorem}
In a temporally evolving graph with a power-law degree
distribution having constant degree exponent $\ddslope$ over time,
the \GPL\ exponent $a$ is:
\begin{eqnarray}
      a & = & 1 ~~~~~~~           \textrm{if} ~~ \ddslope > 2 \\
        & = & 2/\ddslope ~~~~  \textrm{if} ~~ 1 \le \ddslope \le 2 \\
        & = & 2  ~~~~~~~         \textrm{if} ~~ \ddslope < 1
\end{eqnarray}
\label{th:dpldd}
\end{theorem}

\begin{proof}
Assume that at any time $t$ the degree distribution of an undirected
graph $G$ follows a power law. This means the number of nodes $D_d$
with degree $d$ is $D_d = c d^{-\ddslope}$, where $c$ is a constant.
Now assume that at some point in time the maximum degree in the graph
is $\dmax$. Later as the graph grows we will let $\dmax
\rightarrow \infty$. Using the previous power-law relation, we can
calculate the number of nodes $n$ and the number of edges $e$ in the
graph:

\begin{eqnarray*}
  n  & =  & \sum_{d=1}^{\dmax} c d^{-\ddslope}
    \approx \int_{d=1}^{\dmax} d^{-\ddslope}
    = c \frac{\dmax^{1-\ddslope} - 1}{1-\ddslope} \\
  e   & = & \frac{1}{2} \sum_{d=1}^{\dmax} c d^{1-\ddslope}
    \approx \int_{d=1}^{\dmax} d^{1-\ddslope}
    = c \frac{\dmax^{2-\ddslope} - 1}{2-\ddslope} \\
\end{eqnarray*}

Now, we let the graph grow, so $\dmax \rightarrow \infty$. Then the
\GPL\ exponent $a$ is:

\begin{eqnarray*}
  a = \lim_{\dmax \rightarrow \infty}\frac{\log(e)}{\log(n)} =
    \frac{\ddslope\log(\dmax)+\log(|\dmax^{2-\ddslope} - 1|) -
    \log(|2-\ddslope|)}
         {\ddslope\log(\dmax)+\log(|\dmax^{1-\ddslope} - 1|) -
         \log(|1-\ddslope|)}
\end{eqnarray*}

Note, that the degree distribution exponent is $\ddslope$, so we
also have the relation $\log(c) = \ddslope \log(\dmax)$. Now, we
have 3 cases:

{\bf Case 1}: $\ddslope > 2$. No densification:

\begin{eqnarray*}
  a = \frac{\ddslope\log(\dmax) + o(1)} {\ddslope\log(\dmax) + o(1)} = 1
\end{eqnarray*}

{\bf Case 2}: $1 < \ddslope < 2$ is the interesting case where  densification arises:

\begin{eqnarray*}
  a = \frac{\ddslope\log(\dmax)+(2-\ddslope)\log(\dmax) + o(1)}
  {\ddslope\log(\dmax)+ o(1)} = \frac{2}{\ddslope}
\end{eqnarray*}

{\bf Case 3:} $1 < \ddslope $. Maximum densification -- the
graph is basically a clique and the number of edges grows
quadratically with the number of nodes:

\begin{eqnarray*}
  a = \frac{\ddslope\log(\dmax)+(2-\ddslope)\log(\dmax) + o(1)}
  {\ddslope\log(\dmax)+ (1-\ddslope)\log(\dmax) + o(1)} = 2
\end{eqnarray*}
\end{proof}

This shows that for cases when graph evolves by maintaining the constant power-law degree exponent $\ddslope > 2$ over time it does not densify. However, for cases when $ddslope < 2$ we observe densification. This can easily be explained. The densification means that the number of edges grows faster than the number of nodes. So, for densification to appear the tail of the degree distribution has to grow, {\em i.e.} has to accumulate more mass over time. Here, this is the case since power-law distributions with exponent $\ddslope < 2$ have no finite expectation. In the case of degree distribution this means that the expected node degree grows as the graph accumulates more nodes.

\subsection{Evolving degree distribution}

There also exist graphs with degree distribution $\ddslope > 2$ which can also densify. Now, we allow the degree distribution to change over time. In fact, the degree distribution has to flatten over time to accumulate more mass in the tail as more nodes are added to allow for densification. This is what we explore next.

In the previous section we assumed that the exponent $\ddslope$ of
the power-law degree distribution remains constant over time, and
then found the range for power-law degree exponent $\ddslope$ where
it leads to densification. Now, we assume \GPL\ with exponent $a$, allow degree distribution to change over time, and ask {\em How should the power-law degree exponent $\ddslope$ change over time to allow for densification?}  We show the following result:

\begin{theorem}
Given a time evolving graph on $n$ nodes that evolves according to
\GPL\ with exponent $a > 1$ and has a Power-Law degree distribution
with exponent $\ddslope_n > 2$, then the degree exponent
$\ddslope_n$ evolves with the number of nodes $n$ as

\begin{eqnarray}
      \ddslope_n = \frac{4n^{a-1}-1}{2n^{a-1}-1}
      \label{eq:dplddVsTm}
\end{eqnarray}
\label{th:dplddVsTm}
\end{theorem}

\begin{proof}
An undirected graph $G$ on $n$ nodes has $e=\frac{1}{2}n\dbar$
edges, where $\dbar$ is the average degree in graph $G$. Then the
\GPL\ exponent $a$ is

\begin{eqnarray}
  a = \frac{\log(e)}{\log(n)}
    = \frac{\log(n) + \log(\dbar) - \log(2)}{\log(n)}
  \label{eq:ddOverTmA}
\end{eqnarray}

In a graph with power-law degree distribution, $p(x)=x^{-\ddslope}$,
with exponent $\ddslope > 2$, the average degree $\dbar$ is

\begin{eqnarray}
  \dbar \approx \int_{1}^{\infty}x p(x)\ \textrm{d}x
    = c \int_{1}^{\infty} x^{-\ddslope+1} \textrm{d}x
    = \frac{c}{2-\ddslope} x^{-\ddslope+2} \biggr\vert_{1}^{\infty}
    = \frac{\ddslope-1}{\ddslope -2}.
  \label{eq:avgD}
\end{eqnarray}

Now, substituting $\dbar$ in equation~\ref{eq:ddOverTmA} with the
result of equation~\ref{eq:avgD}, and solving for $\ddslope$, we
obtain:

\begin{eqnarray}
  \ddslope_n = \frac{4n^{a-1}-1}{2n^{a-1}-1}
  \label{eq:ddSlopeOverTm}
\end{eqnarray}
\end{proof}

Here we found the evolution pattern that degree distribution with exponent $\ddslope > 2$ has to follow in order to allow for densification. As theorem~\ref{eq:ddOverTmA} shows the degree distribution has to flatten over time, so that the expected node degree increases, which is the result of densification.

\subsection{Measurements on real networks}

Next, given the analysis from the previous section, we went back to
the data and checked if graphs we analyzed before behave according
to the results of theorems~\ref{th:dpldd} and~\ref{th:dplddVsTm}.

First, we show an example of a graph where the evolution of the
degree distribution and the \GPL\ exponent follow the results of
theorem~\ref{th:dpldd}. Using the email network described in
section~\ref{sec:email} we found that the degree distribution
follows a power-law with exponent $\ddslope$ that remains constant over
time.

\begin{figure}[t]
\begin{center}
  \begin{tabular}{cc}
    \epsfig{file=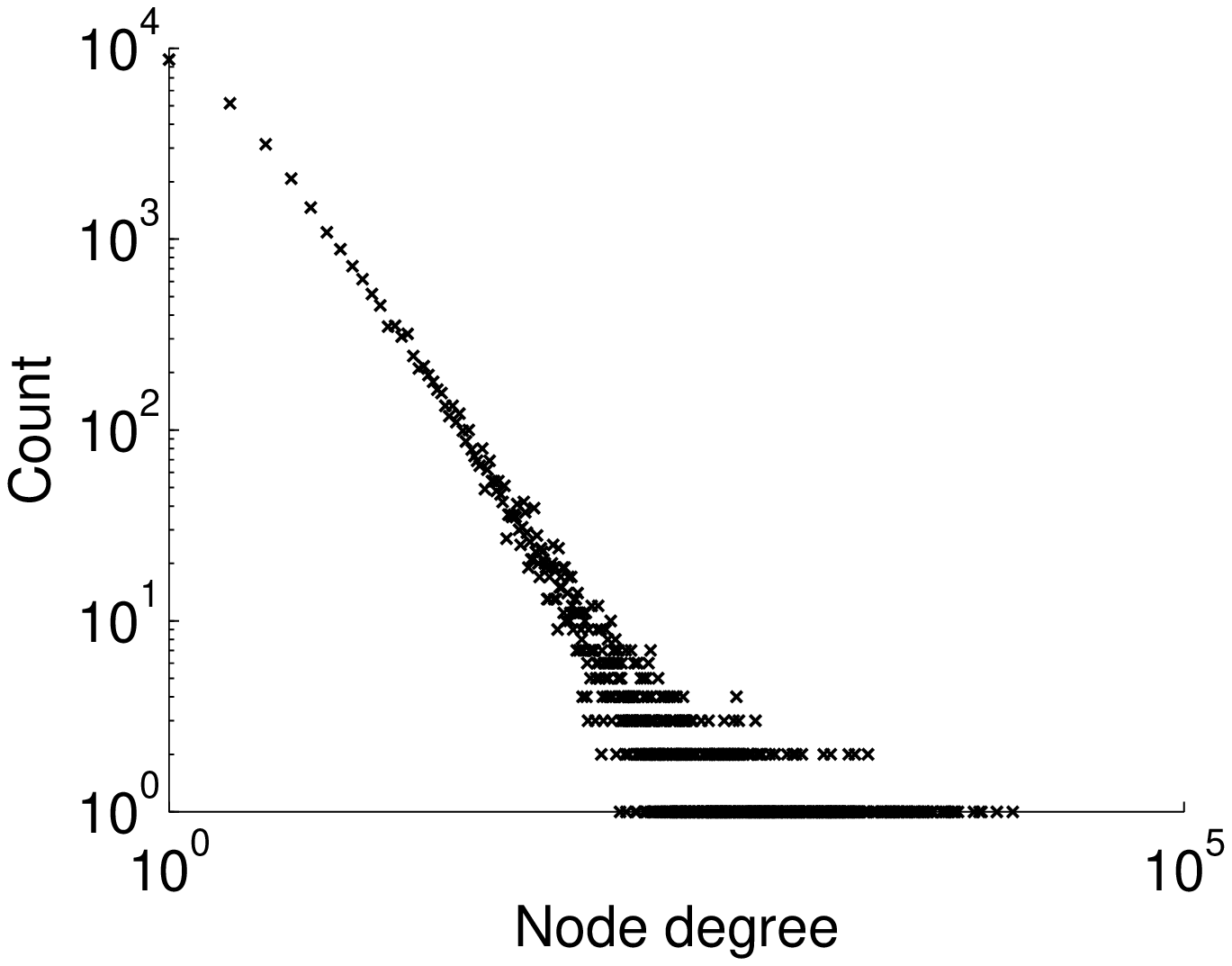, width=2.4in} &
    \epsfig{file=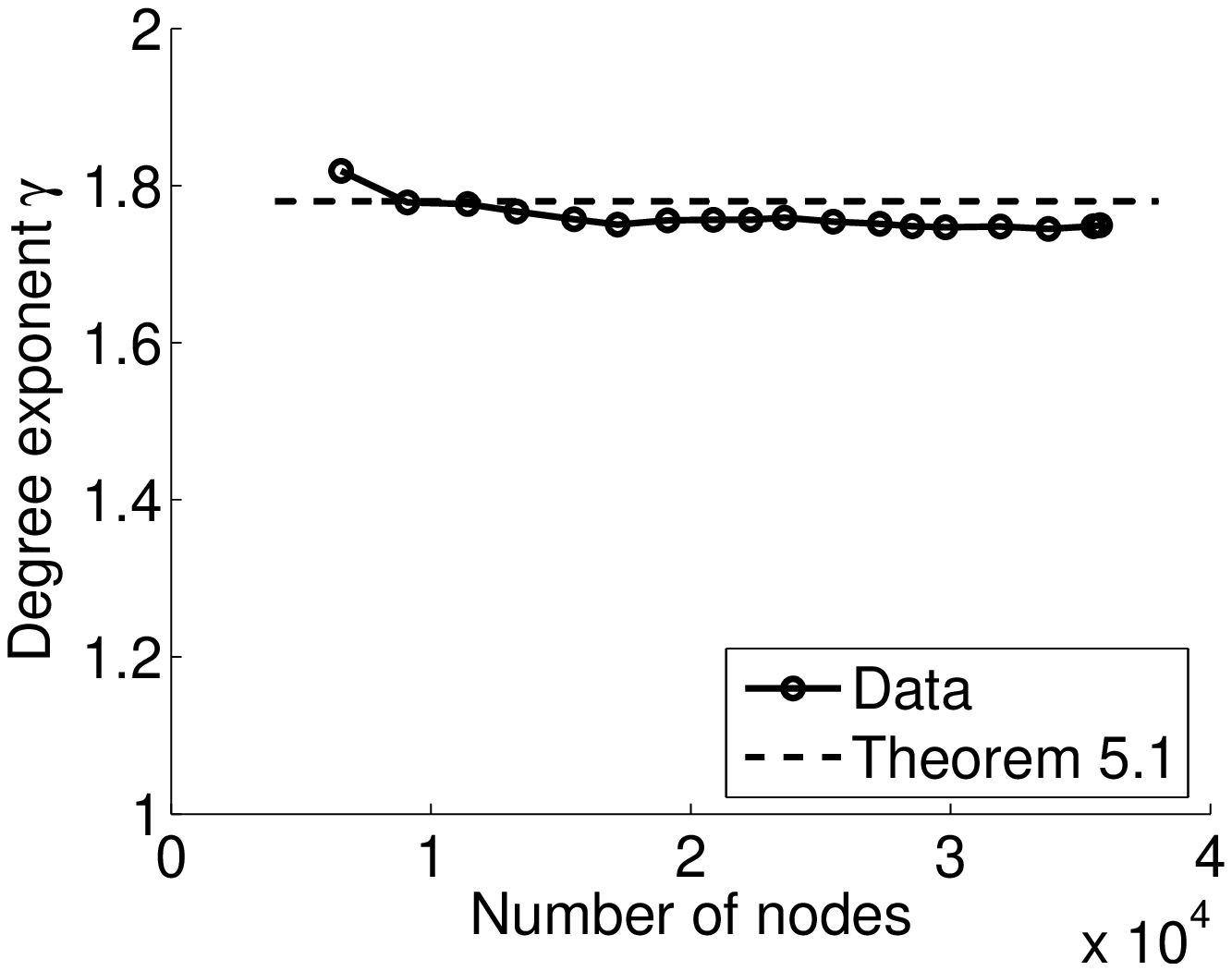, width=2.4in} \\
    (a) Degree distribution & (b) Degree exponent over time\\
    \end{tabular}
  \caption{Degree distribution (a) and the degree exponent $\ddslope$
  over time (b) for the email network. The network maintains constant
  slope $\ddslope$ of degree distribution over time.
  Notice that $\ddslope<2$. We observe a remarkably good agreement
  between the result of Theorem~\ref{th:dpldd}
  (\GPL\ exponent $a=1.13$), and our measurements (\GPL\ exponent
  $a=1.11$) in figure~\ref{fig:powerGrowth}(e).}
  \label{fig:emailDdCf}
\end{center}
\end{figure}

Figure~\ref{fig:emailDdCf}(a) shows the degree distribution of the
the email network for last snapshot of the network, i.e. last 2
months of the data. We create the networks by using a 2 month sliding window. We fit the power-law degree exponent $\ddslope$
using Maximum Likelihood Estimation (MLE), and plot its evolution
over time in figure~\ref{fig:emailDdCf}(b). Notice $\ddslope$
remains practically constant over time, which is also in agreement with
observations reported in~\cite{kossinets06email}. Also notice that
the power-law degree exponent $\ddslope = 1.76 < 2$.
Given the degree exponent $\ddslope$, and using
theorem~\ref{th:dpldd} we obtain the theoretical value of the \GPL\
exponent $a = 2/1.76 \approx 1.13$. The value of DPL exponent we
measured in section~\ref{sec:observations}
figure~\ref{fig:powerGrowth}(e) is $a=1.11$, which is a remarkably
good agreement. This shows that there exist graphs in the real world
that densify and have decreasing diameter while maintaining constant
degree exponent over time.

\begin{figure}
\begin{center}
    \begin{tabular}{cc}
    \epsfig{file=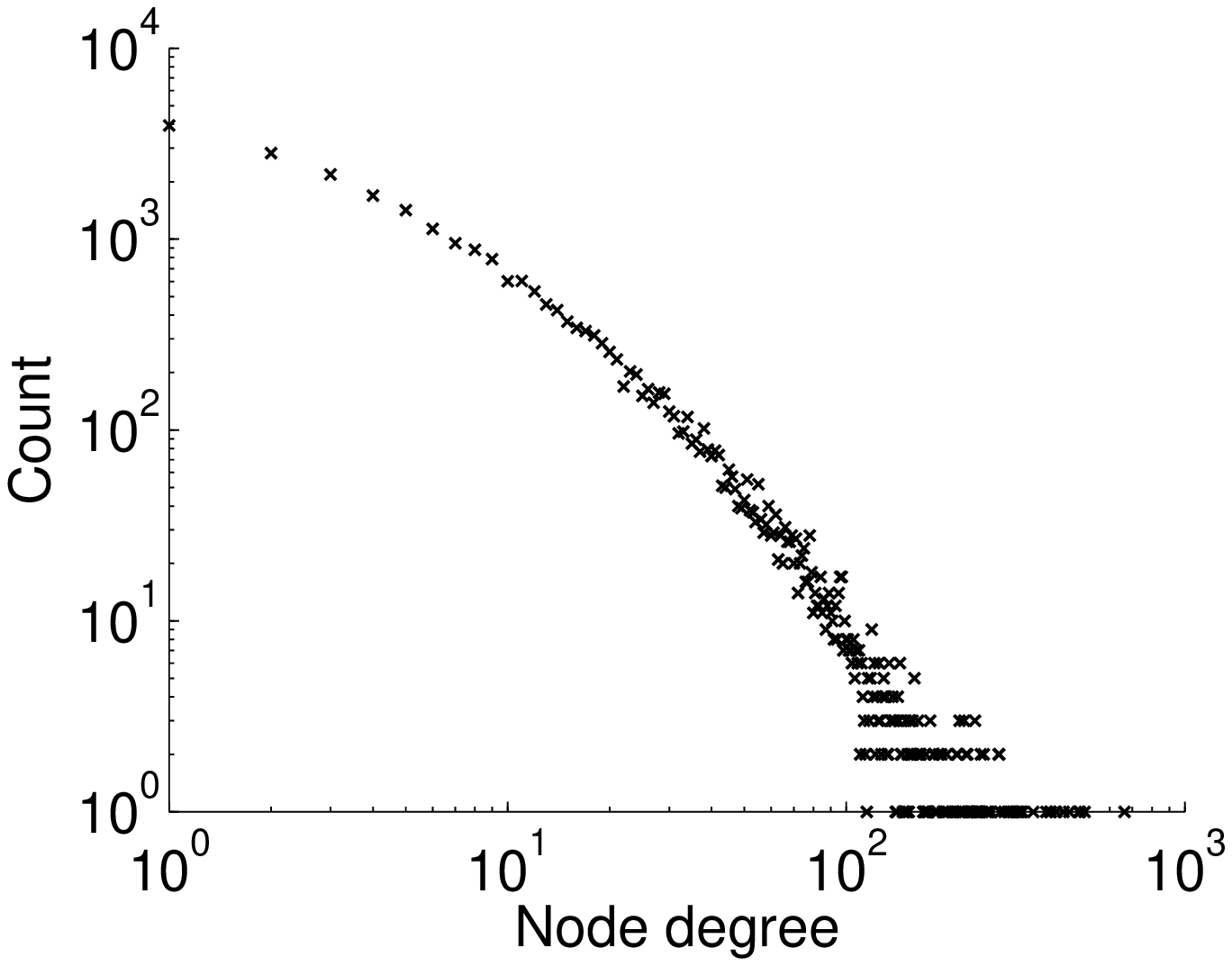, width=2.4in} &
    \epsfig{file=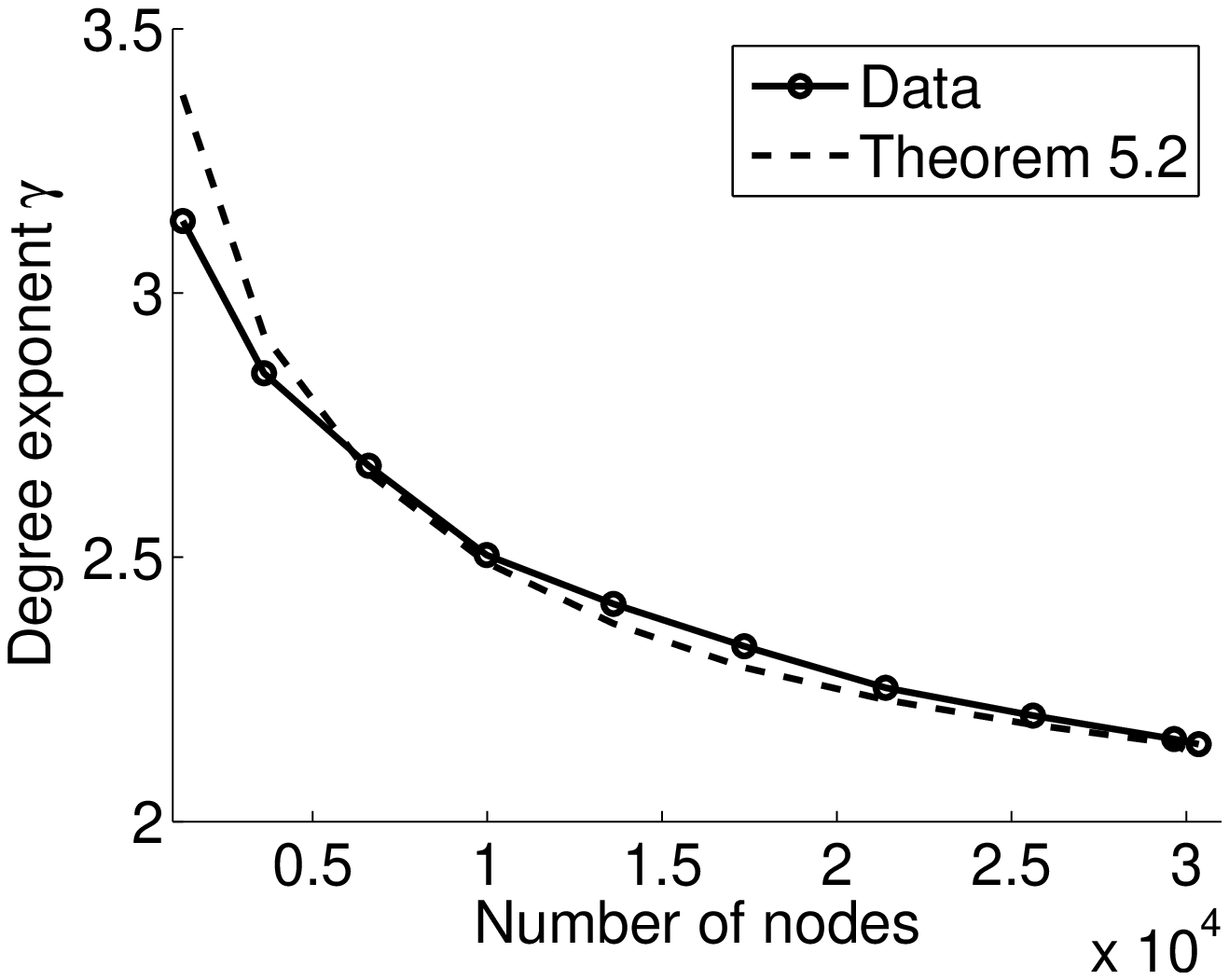, width=2.4in} \\
    (a) Degree distribution & (b) Degree exponent over time\\
  \end{tabular}
  \caption{Degree distribution (a) and the degree exponent over time (b)
  for the HEP--PH citation network. The network follows power-law degree
  distribution only in the tail. Degree distribution
  exponent $\ddslope$ is decreasing over time. Notice a good
  agreement of degree distribution evolution (solid line) as predicted
  by the theorem~\ref{th:dplddVsTm} (dashed line).}
  \label{fig:hepphDdCf}
\end{center}
\end{figure}

Last, we show an example of a temporally evolving graph that
densifies, and has the power-law degree exponent $\ddslope$ changing over time.

Figure~\ref{fig:hepphDdCf}(a) plots the degree distribution of the
full HEP--PH citation network from section~\ref{sec:arxiv}.  In this
case the degree distribution only follows a power-law in the tail of
the distribution, so we applied the following procedure. For every
year $y$, $1992 \le y \le 2002$ we create a citation graph and
measure the exponent of the power-law degree distribution. We apply
logarithmic binning and fit the power-law degree distribution using
MLE on the tail of the degree distribution starting at minimum
degree 10. We plot the resulting degree exponent $\ddslope$ over
time as a function of the size of the graph in
figure~\ref{fig:hepphDdCf}(b).

Using dashed-lines we also plot the degree exponent $\ddslope$ as
obtained by theorem~\ref{th:dplddVsTm}. Since the graph does not
exhibit power-law degree distribution on the entire range, and due
to missing past effects, we had to appropriately scale time axis
with a manually chosen value. Regardless of the manual scaling we think
this result indicates that for a class of temporally evolving graphs
the degree distribution flattens over time as given by the
theorem~\ref{th:dplddVsTm}. This seems to be the case for HEP--PH
citation network where the evolution of the degree exponent
qualitatively follows the result of theorem~\ref{th:dplddVsTm}.

\begin{figure}[t]
\begin{center}
  \begin{tabular}{cc}
    \epsfig{file=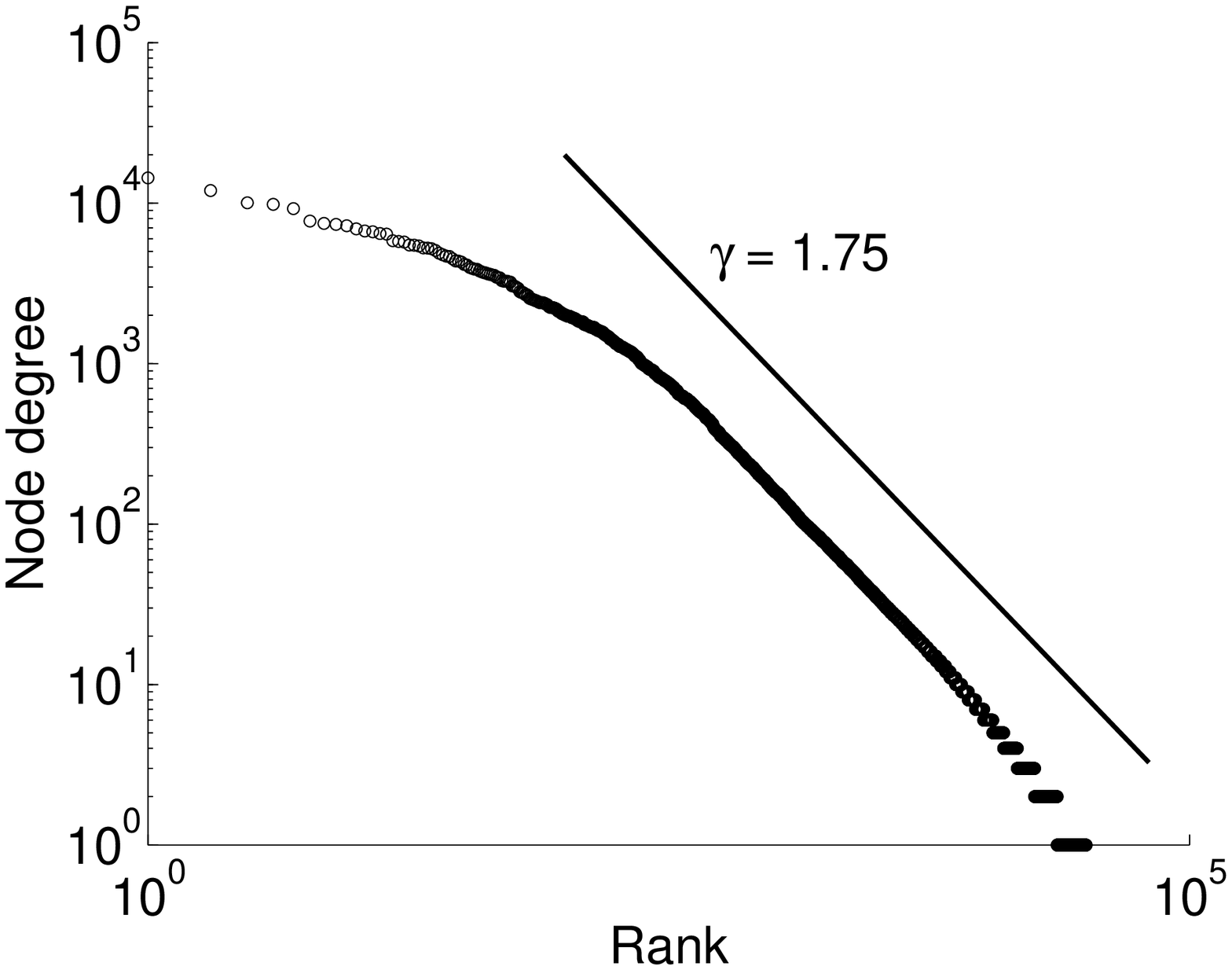, width=2.4in} &
    \epsfig{file=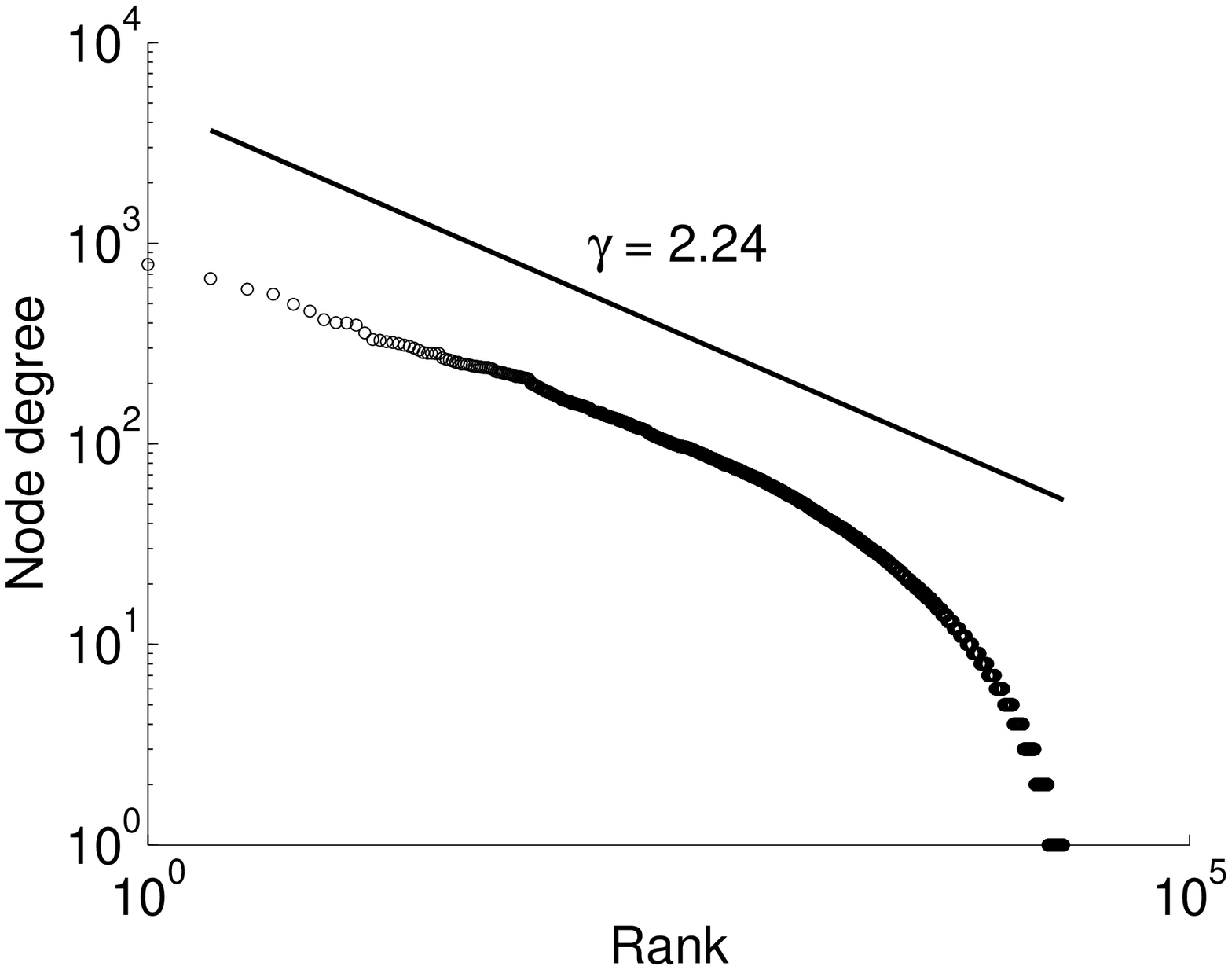, width=2.4in} \\
    (a) Email network & (b) HEP--PH citation network\\
    \end{tabular}
  \caption{Rank Degree plot for the degree distribution of the
  email and HEP--PH datasets. We use the same data as in
  figures~\ref{fig:emailDdCf}(a) and~\ref{fig:hepphDdCf}(a) but plot
  node degree vs. rank using the log-log scales.
  The solid lines present the power-law decay with exponent
  $\ddslope = 1.75$ and $\ddslope = 2.24$, respectively.}
  \label{fig:rankDeg}
\end{center}
\end{figure}

Figure~\ref{fig:rankDeg} further investigates the degree
distribution of the email and HEP--PH networks. We use the same data
as in figures~\ref{fig:emailDdCf}(a) and~\ref{fig:hepphDdCf}(a), and
plot the number of nodes of a certain degree against the rank. The
solid lines present the power-law decay with exponents $\ddslope =
1.75$ and $\ddslope = 2.24$, respectively. The actual slope of the
plotted line is $1/(\ddslope -1)$, which is the relation between the
power-law exponent $\ddslope$ and the slope of the rank degree plot
(see~\cite{ladaRanking} for more details on these relationships).

In both plots of figure~\ref{fig:rankDeg} we observe linearity which
suggests a power-law relationship for a part of the degree
distribution. For the email network we observe linearity in the
tail, and for the HEP--PH citation network in the first part of the
distribution. These two plots show that in our two datasets the power-law degree distribution does not hold for the entire range. However, we still
observe a significant range where power-law relationship seems to
hold. Regardless of these irregularities there is still very good
agreement of the data with the results of theorems~\ref{th:dpldd}
and~\ref{th:dplddVsTm}, which suggests that there exists graphs that
densify by maintaining constant power law degree exponent
(theorem~\ref{th:dpldd}), and also graphs that densify by degree exponent flattening over time (theorem~\ref{th:dplddVsTm}).

\section{Conclusion}
\label{sec:conclusion}

Despite the enormous recent interest in large-scale network data, and
the range of interesting patterns identified for static snapshots of
graphs (e.g. heavy-tailed distributions, small-world phenomena), there
has been relatively little work on the properties of the time evolution
of real graphs. This is exactly the focus of this work. The main
findings and contributions follow:

\begin{itemize}
  \item The \GPL: In contrast to the standard modeling
  assumption that the average out-degree remains constant over time, we
  discover that real graphs have out-degrees that grow over time,
  following a natural pattern (Eq.~(\ref{eq:gpl})).

  \item  Shrinking diameters: Our experiments also show that the
  standard assumption of slowly growing diameters does not hold in a
  range of real networks; rather, the diameter may actually exhibit a
  gradual decrease as the network grows.

  \item We show that our \CGA\ model leads to the \GPL, and
  that it needs only one parameter to achieve it.

  \item We give the Forest Fire Model, based on only
  two parameters, which is able to capture patterns observed both in
  previous work and in the current study: heavy-tailed in- and
  out-degrees, the \GPL, and a shrinking diameter.

  \item We notice that the Forest Fire Model exhibits a sharp transition
  between sparse graphs and graphs that are densifying. Graphs with
  decreasing effective diameter are generated around this
  transition point.

  \item Finally, we find a fundamental relation between the temporal
  evolution of the graph's power-law degree distribution and the \GPL\ exponent.
  We also observe that real datasets exhibit this type of relation.
\end{itemize}

Our work here began with an investigation of the time-evolution of a
set of large real-world graphs across diverse domains. It resulted
in the finding that real-world graphs are becoming denser as they
grow, and that in many cases their effective diameters are
decreasing. This challenges some of the dominant assumptions in
recent work on random graph models, which assumes constant (or at
most logarithmic) node degrees, and diameters that increase slowly
in the number of nodes. Building on these findings, we have proposed
a set of simple graph generation processes, capable of producing
graphs that exhibit densification and exhibit decreasing effective
diameter.

Our results have potential relevance in multiple settings, including
'what if' scenarios; in forecasting of future parameters of computer
and social networks; in anomaly detection on monitored graphs; in
designing graph sampling algorithms; and in realistic graph
generators.

\section{Acknowledgments}
\label{sec:ack}

We thank Panayiotis Tsaparas of HIIT for helpful
comments, Michalis Faloutsos and George Siganos of UCR, for help
with the data and for early discussions on the Autonomous System
dataset, and Sergey Dorogovtsev and Alexei Vazquez for pointing us
to references.

This material is based upon work supported by the National Science
Foundation under Grants No.
   IIS-0209107 
   SENSOR-0329549 
   EF-0331657
   IIS-0326322 
   IIS-0534205, 
  CCF-0325453, IIS-0329064, CNS-0403340, CCR-0122581, 
a David and Lucile Packard Foundation Fellowship, and also by the
Pennsylvania Infrastructure Technology Alliance (PITA), a
partnership of Carnegie Mellon, Lehigh University and the
Commonwealth of Pennsylvania's Department of Community and Economic
Development (DCED). Additional funding was provided by a generous
gift from Hewlett-Packard. Jure Leskovec was partially supported by
a Microsoft Research Graduate Fellowship.

Any opinions, findings, and conclusions or recommendations expressed
in this material are those of the author(s) and do not necessarily
reflect the views of the National Science Foundation, or other
funding parties


\end{document}